\documentclass[twocolumn, aps, pra, longbibliography,  
 superscriptaddress]{revtex4-1}
\usepackage[english]{babel}
\usepackage[dvipsnames]{xcolor}
\usepackage{tikz}
\usetikzlibrary{quantikz}
\usepackage{tikz-cd}
\usepackage{graphicx}
\usepackage{dcolumn}
\usepackage{bm}
\usepackage{amsmath}
\usepackage{amssymb}
\usepackage{dsfont}
\usepackage{lipsum}  
\usepackage{color}
\usepackage{multirow}
\usepackage{stmaryrd}
\usepackage{upgreek}
\usepackage{bbm}
\usepackage{xcolor}
\usepackage{ragged2e}
\usepackage{siunitx}
\usepackage[colorlinks=true,urlcolor=blue,citecolor=blue,linkcolor=blue]{hyperref}
\usepackage{braket}
\usepackage{mathtools}
\usepackage[mathlines]{lineno}
\usepackage[normalem]{ulem}
\usepackage{scalerel}
\usepackage{ bbold }
\usepackage{inputenc}
\usepackage[T1]{fontenc}
\usepackage{tikz}
\usepackage{cleveref}
\usepackage{amsfonts}
\usepackage{natbib}
\usepackage{booktabs}
\usepackage{algorithm} 
\usepackage{algpseudocode} 
\usepackage{blkarray}

\usepackage{adjustbox}
\usepackage{dashbox}

\newcommand\dboxed[2]{\setlength{\fboxsep}{#2pt} \dbox{\ensuremath{#1}}}

\newcommand{\fidelity}[0]{\overline{\mathcal{F}}}

\def\ketbra#1#2{{\vert#1\rangle\!\langle#2\vert}} 

\begin{document}

\title{Universal qudit gate synthesis for transmons}

\author{Laurin E. Fischer}  \email{aur@zurich.ibm.com}
\affiliation{IBM Quantum, IBM Research Europe – Zurich, S\"aumerstrasse 4, 8803 R\"uschlikon, Switzerland}
\affiliation{Theory and Simulation of Materials (THEOS), {\'E}cole Polytechnique F{\'e}d{\'e}rale de Lausanne, 1015 Lausanne, Switzerland}
\author{Alessandro Chiesa}
\affiliation{Università di Parma, Dipartimento di Scienze Matematiche, Fisiche e Informatiche, I-43124, Parma, Italy}
\affiliation{Gruppo Collegato di Parma, INFN-Sezione Milano-Bicocca, I-43124 Parma, Italy}
\affiliation{UdR Parma, INSTM, I-43124 Parma, Italy}
\author{Francesco Tacchino}
\author{Daniel J. Egger}
\affiliation{IBM Quantum, IBM Research Europe – Zurich, S\"aumerstrasse 4, 8803 R\"uschlikon, Switzerland}
\author{Stefano Carretta}
\affiliation{Università di Parma, Dipartimento di Scienze Matematiche, Fisiche e Informatiche, I-43124, Parma, Italy}
\affiliation{Gruppo Collegato di Parma, INFN-Sezione Milano-Bicocca, I-43124 Parma, Italy}
\affiliation{UdR Parma, INSTM, I-43124 Parma, Italy}
\author{Ivano Tavernelli}  \email{ita@zurich.ibm.com}
\affiliation{IBM Quantum, IBM Research Europe – Zurich, S\"aumerstrasse 4, 8803 R\"uschlikon, Switzerland}

\begin{abstract}

Gate-based quantum computers typically encode and process information in two-dimensional units called qubits. 
Using $d$-dimensional qudits instead may offer intrinsic advantages, including more efficient circuit synthesis, problem-tailored encodings and embedded error correction. 
In this work, we design a superconducting qudit-based quantum processor wherein the logical space of transmon qubits is extended to higher-excited levels. 
We propose a universal gate set featuring a two-qudit cross-resonance entangling gate, for which we predict fidelities beyond $99\%$ in the $d=4$ case of ququarts with realistic experimental parameters. 
Furthermore, we present a decomposition routine that compiles general qudit unitaries into these elementary gates, requiring fewer entangling gates than qubit alternatives.
As proof-of-concept applications, we numerically demonstrate the synthesis of ${\rm SU}(16)$ gates for noisy quantum hardware and an embedded error correction sequence that encodes a qubit memory in a transmon ququart to protect against pure dephasing noise.
We conclude that universal qudit control -- a valuable extension to the operational toolbox of superconducting quantum information processing -- is within reach of current transmon-based architectures and has applications to near-term and long-term hardware. 

\end{abstract}

\maketitle

\section{Introduction}
\label{chap:introduction}
In analogy to their classical counterparts, quantum computers encode information in binary systems -- known as qubits -- that consist of two physically distinct states.
Many experimental implementations, including trapped ions, superconducting qubits, neutral atoms, and spin qubits, embed the logical qubit subspace in a much larger multi-level Hilbert space~\cite{bruzewicz2019trappedion, clarke2008superconducting, shiQuantumLogicEntanglement2021, burkard2021semiconductor}. 
This full Hilbert space allows for a richer set of controls that is forgone by confining the computational space to qubits.
By coherently controlling additional states we enrich the set of available operations and make use of $d$-dimensional \emph{qudits} as the local units of information.

Qudits have several conceptual advantages over their qubit counterparts.
The number of qudits needed to reach the same Hilbert space dimension as a system of qubits is reduced by a factor of $\log_2(d)$.
For instance, ququarts, i.e. $d=4$, cut the number of computational units in half.
Moreover, qudits synthesize arbitrary unitaries more efficiently than qubits with regards to the number of required entangling gates~\cite{di2015optimal},
an advantage that already emerges in the qutrit case of $d=3$~\cite{gokhale2019asymptotic}.
This has led to proposals for efficient implementations of quantum algorithms in the qudit space~\cite{nikolaeva2022efficient, wang2020qudits}.
Applications that are formulated in a product space of multivalued units particularly benefit from a qudit encoding. 
These include the quantum simulation of bosonic modes that arise, e.g., in light-matter interaction processes~\cite{tacchino2021proposal,mazzolag2020,miessen2021spin-boson}, lattice gauge theories~\cite{rico2018nuclear,mathis2020,mazzola2021} and chemical vibrations and reactions~\cite{ollitrault_vib2020,macdonell2021analog}, but also classical problems like multivalued integer optimization~\cite{deller2022quantum}.
Moreover, qudit levels simplify the implementation of qubit gates~\cite{lanyon2009simplifying, galda2021implementing} and POVM measurements~\cite{fischer2022ancillafree, stricker2022experimental}.
Finally, qudits exhibit more complex entanglement than qubits~\cite{kraft2018characterizing}, which can be leveraged to improve protocols such as superdense coding~\cite{hu2018beating} and quantum error correction (QEC) codes~\cite{GKP,scott2004multipartite,Pirandola2008, Cafaro2012,PRXGirvin,Hussain2018,Chiesa2020,Carretta2021,Chizzini2022,Petiziol2021,Chiesa2022}. 
As opposed to block-encoding QEC techniques, which use many physical qubits to encode a single logical one, an error-protected logical qubit can be encoded into the multi-level structure of a single qudit system, as proposed for molecular spins~\cite{Hussain2018, Chiesa2020, Carretta2021}.
This simplifies the implementation of QEC by strongly reducing the number of controlled multi-qubit operations. 

Qudit-based quantum information processing has recently been explored in trapped ions~\cite{ringbauer2021universal}, photonic systems~\cite{chi2022programmable}, Rydberg atoms~\cite{gonzalezcuadra2022hardware}, ultracold atomic mixtures~\cite{kasper2022universal}, and molecular spins~\cite{Hussain2018,Chiesa2020,Gimeno2021,Chicco2021,Carretta2021, Chizzini2022,Chizzini2022physrevres}. 
Here, we conceptualize a superconducting qudit quantum processor, where $d$ qudit levels are encoded into the $d$ energetically lowest states of a transmon~\cite{koch2007chargeinsensitive}.
We propose a concrete scheme to transpile arbitrary unitary circuits into a universal set of hardware-native single- and two-qudit gates, which in principle generalizes to any qudit dimension $d$.

Operating transmons as qutrits has already found many applications including multi-qutrit entanglement studies~\cite{blok2021quantum, cerveralierta2022experimental}, realization of multi-qubit gates~\cite{galda2021implementing, egger2019entanglement, nikolaeva2022decomposing},
excited state promotion readout~\cite{elder2020highfidelity, jurcevic2021demonstration}, quantum metrology~\cite{shlyakhov2018quantum}, fast resets~\cite{egger2018pulsed}, and the realization of two-qutrit quantum algorithms~\cite{roy2022realization}.
The ququart case has been considered in the context of single-qudit applications~\cite{fischer2022ancillafree, kiktenko2015multilevel}, while several experiments have recently reported coherent control of a single ququart~\cite{cao2023emulatinga, seifert2023exploring, liu2023performing}.
However, up to now, it was unclear how to drive general two-qudit unitaries in transmons. 
For qubits, a popular realization of the \textsc{Cnot} gate relies on driving cross-resonance pulses~\cite{rigetti2010fully}. 
Here, we propose a generalization of the echoed cross-resonance (ECR) gate in the qudit space as the fundamental entangling gate between two qudits. 
We study this generalized ECR gate by numerically simulating the time-dynamics of the system, demonstrating that it can reach ququart gate fidelities of $\sim 99\%$ with simple pulse shapes. 

This article is organized as follows. 
In Sec.~\ref{chap:transmon_qudits}, we propose a universal set of qudit operations and numerically benchmark their fidelities with simulations that include leakage, crosstalk and charge-noise errors. 
Sec.~\ref{chap:gate_decompositions} outlines how to decompose general two-qudit gates into the qudit ECR gate and single-qudit gates, enabling the synthesis of arbitrary unitary circuits with fewer entangling gates compared to qubits. 
Finally, in Sec.~\ref{chap:qec_application}, we show that the ECR gate forms the basis of a qudit-based QEC protocol and we demonstrate its basic implementation in the ququart case.
We conclude with a discussion on current and future developments in Sec.~\ref{chap:discussion_conclusion}.

\section{Qudit control in transmons}
\label{chap:transmon_qudits}
Superconducting circuits are a promising architecture to realize large-scale quantum information processors~\cite{bravyi2022future}.
They have fast gates and measurement repetition rates~\cite{wack2021quality} with comparatively long coherence times~\cite{place2021new}. 
The transmon is a particularly popular type of superconducting circuit in which a Josephson junction of energy $E_{\text{J}}$ shunted by a large capacitance with charging energy $E_{\text{C}}$ creates an anharmonic oscillator. 
The eigenenergies $E_n$ of this anharmonic oscillator are characterized by the base excitation frequency $\omega = \left(E_1 - E_0\right)$ between the ground and first excited state, and the anharmonicity $\alpha = (E_2-E_1) - \omega$ (setting $\hbar=1$). 
Transmons are tuned towards large ratios of $E_{\text{J}} / E_{\text{C}} \gg 1$.
This exponentially suppresses charge noise that causes fluctuations in the eigenenergies of the
system, while maintaining sufficient anharmonicities for the selective driving of individual transitions~\cite{koch2007chargeinsensitive}. 
Whereas the two lowest eigenstates of a transmon are typically employed as a qubit, in this work we encode a $d$-dimensional qudit in the $d$ lowest energy states. 
We focus specifically on the ququart case of $d=4$ which represents the minimal unit for embedded QEC.

\subsection{Single qudit control}
\label{sec:single_qudit_control}

\begin{figure}
\includegraphics[width=0.99\columnwidth]{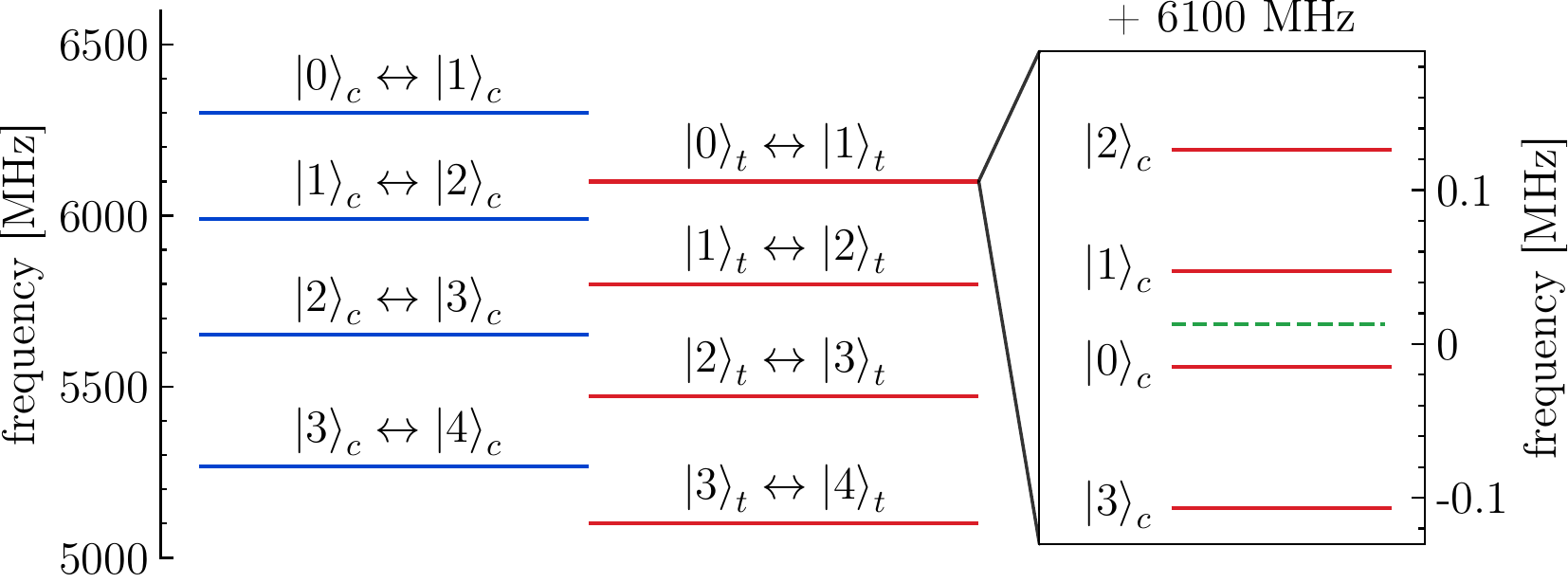}
\caption[]{Transition frequencies of a two-transmon model. Blue and red lines denote the control and target, with bare frequencies of $\omega_c/(2\pi)=6.3\,\text{GHz}$ and $\omega_t/(2\pi)=6.1\,\text{GHz}$, respectively.
The zoomed inset shows the dressing of the bare target frequency due to the capacitive coupling with the average transition frequency of $\overline{\omega}_t/(2\pi)$ in dashed green.} 
\label{fig:two_transmon_level_spectrum}
\end{figure}

We briefly review how single-qudit unitaries are realized in transmons, following Ref.~\cite{fischer2022ancillafree}.
Individual transmons are driven by microwave pulses whose carrier frequency are adjusted to resonantly drive different transitions. 
With base frequencies of $\omega/(2\pi) \sim6\,\text{GHz}$ and anharmonicities of $\alpha/(2\pi) \sim-300\,\text{MHz}$, the transition frequencies of neighboring levels among the first four excited states all lie within $1\,\text{GHz}$ from the base frequency [see Fig.~\ref{fig:two_transmon_level_spectrum}].  
Such frequency shifts are routinely applied in existing microwave control stacks.
We thus assume the ability to drive rotations between neighboring levels, e.g., 
\begin{equation}
\label{eqn:Rx_definition}    
R_x^{n(n+1)}(\varphi) = \mathbb{1}_{n} \oplus \exp(-i\tfrac{\varphi}{2} \sigma_x) \oplus \mathbb{1}_{d-n-2}
\end{equation}
denotes an $x$-rotation between levels $\ket{n}$ and $\ket{n+1}$ and acts as the identity on all other levels. Here, $\sigma_x$ is the Pauli $x$ operator and $\mathbb{1}_n$ is the $n$-dimensional identity operator. 
Single-qudit phase gates can be applied ``virtually'' by adjusting the phases of subsequent drive pulses, which affects the polar angle of their rotation axes in the $xy$-plane. 
These virtual phase gates are near-perfect and come at no additional experimental cost~\cite{mckay2017efficient}.
By keeping track of the relative phase advances between all $d$ levels, the correct frame changes can be implemented on the drives. 
In this setting, any single-qudit operation can be realized with a sequence of at most $d(d-1)/2$ two-level rotations $R_x^{n(n+1)}$ through a decomposition into Givens rotations~\cite{schirmer2002constructive, fischer2022ancillafree}. 

\subsection{Multi-qudit control}
\label{sec:multi_qudit_control}

In addition to single-qudit transformations, one entangling two-qudit gate is required to form a complete set of universal qudit operations~\cite{brennen2005criteria}. 
We therefore investigate extensions of the popular cross-resonance gate to the qubit space.

\subsubsection{The cross-resonance gate}
The cross-resonance (CR) gate is applicable to transmons with a weak interaction mediated by a common resonator, since it is an all-microwave gate~\cite{rigetti2010fully, patterson2019calibration}. 
Here, one transmon, referred to as the \emph{control}, is driven at the $\ket{0}_t \leftrightarrow \ket{1}_t$ transitions frequency $\omega_t$ of the second transmon, referred to as the \emph{target}.
This CR tone entangles the two systems through a complicated interaction dominated by a $Z_c\otimes X_t$ generator in the qubit space~\cite{magesan2020effective}. 
When tuning this rotation to $R_{ZX}(\pi/2) = \exp(-i\tfrac{\pi}{4} \sigma_z \otimes \sigma_x )$, the CR gate is equivalent to a \textsc{Cnot} up to local Clifford gates. 
Analytical studies of CR tones based on perturbation theory show that the effective two-qubit interaction Hamiltonian contains various single-qubit terms ($I\otimes X$, $I\otimes Z$, $Z\otimes I$), as well as a weak $Z\otimes Z$ term~\cite{tripathi2019operation, malekakhlagh2020firstprinciples}. 
A popular approach to largely cancel these unwanted terms employs the echoed pulse sequence shown in Fig.~\ref{fig:ECR_pulse_overview}(a).
In the echoed cross-resonance (ECR) gate the effects of the $Z\otimes I$, $Z\otimes Z$, and $I \otimes X$ destructively interfere, thus isolating the desired $Z\otimes X$ generator~\cite{sheldon2016procedure}.
The echo sequence can further be improved with resonant rotary pulses on the target qubit~\cite{sundaresan2020reducing}.
Previous studies of the CR gate have focused on the qubit subspace~\cite{magesan2020effective, tripathi2019operation, malekakhlagh2020firstprinciples}. 
In Ref.~\cite{galda2021implementing}, an ECR sequence with the control prepared in $\ket{2}_c$ enables a pulse-efficient decomposition of the three-qubit Toffoli gate. 
Going beyond this, we now investigate the action of the CR gate in the full two-ququart subspace through a numerical simulation of the system's dynamics.  

\subsubsection{Numerical model}
\label{sec:numerical_model}
Typical transmon parameters of current IBM Quantum devices are $\omega/(2\pi)\!\sim\! 5\,\text{GHz}$ and $\alpha/(2\pi)\!\sim\!-300\,\text{MHz}$.
With a resulting $E_{\text{J}} / E_{\text{C}}$-ratio of 40--50, the charge noise induced fluctuations of around 20 MHz in the $\ket{2}\leftrightarrow\ket{3}$ transition frequency are intolerable for full ququart operation~\cite{fischer2022ancillafree}.
We therefore choose a two-transmon model with frequencies of $\omega_c/(2\pi) = 6.3\,\text{GHz}$, $\omega_t/(2\pi) = 6.1\,\text{GHz}$, and anharmonicities of $\alpha_c/(2\pi) = -310\,\text{MHz}$, $\alpha_t/(2\pi) = -300\,\text{MHz}$.
This increases the $E_{\text{J}} / E_{\text{C}}$-ratio to $\sim\!70$, pushing the $\ket{2}\leftrightarrow\ket{3}$ frequency fluctuations down to $\sim 180\,\text{kHz}$. 
Moreover, the chosen parameters avoid crosstalk with a gap of at least $100\,\text{MHz}$ between different transitions [see Fig.~\ref{fig:two_transmon_level_spectrum}].
This is smaller than the anharmonicities of each qudit.
Fortunately, any potential leakage is avoidable by shaping the control pulses~\cite{schutjens2013singlequbit, vesterinen2014mitigating}.
The $\sim3\,\text{MHz}$ charge noise on the $\ket{3}\leftrightarrow\ket{4}$ transition renders high-fidelity control of this transition difficult.
We thus focus on the ququart subspace. 

The transmons are coupled by a weak exchange interaction of strength $J/(2\pi)=1.8\,\text{MHz}$ that is routinely achieved in existing systems (see Appendix~\ref{sec:app_model_hamiltonians} for a detailed definition of the model Hamiltonians). 
This always-on coupling $J$ leads to a small shift of the eigenenergies of the joint two-transmon systems.  
From now on, when we denote a basis state as $\ket{n}_c\otimes\ket{m}_t$, we refer to these dressed basis states. 
In this basis, the $\ket{0}_t \leftrightarrow \ket{1}_t$ transition frequency varies by $\sim \pm 100\,\text{kHz}$ depending on the state of the control qudit [see inset of Fig.~\ref{fig:two_transmon_level_spectrum}]. 
We therefore set the drive frequency of the CR tones at $\overline{\omega}_t/(2\pi) = \omega_t/(2\pi) + 13\,\text{kHz}$ obtained by averaging over the lowest four states of the control. This keeps the detuning to each transition as small as possible. 

\subsubsection{Simulation results}
\label{sec:transmon_qudits_simulation_results}

\begin{figure}
\includegraphics[width=0.99\columnwidth]{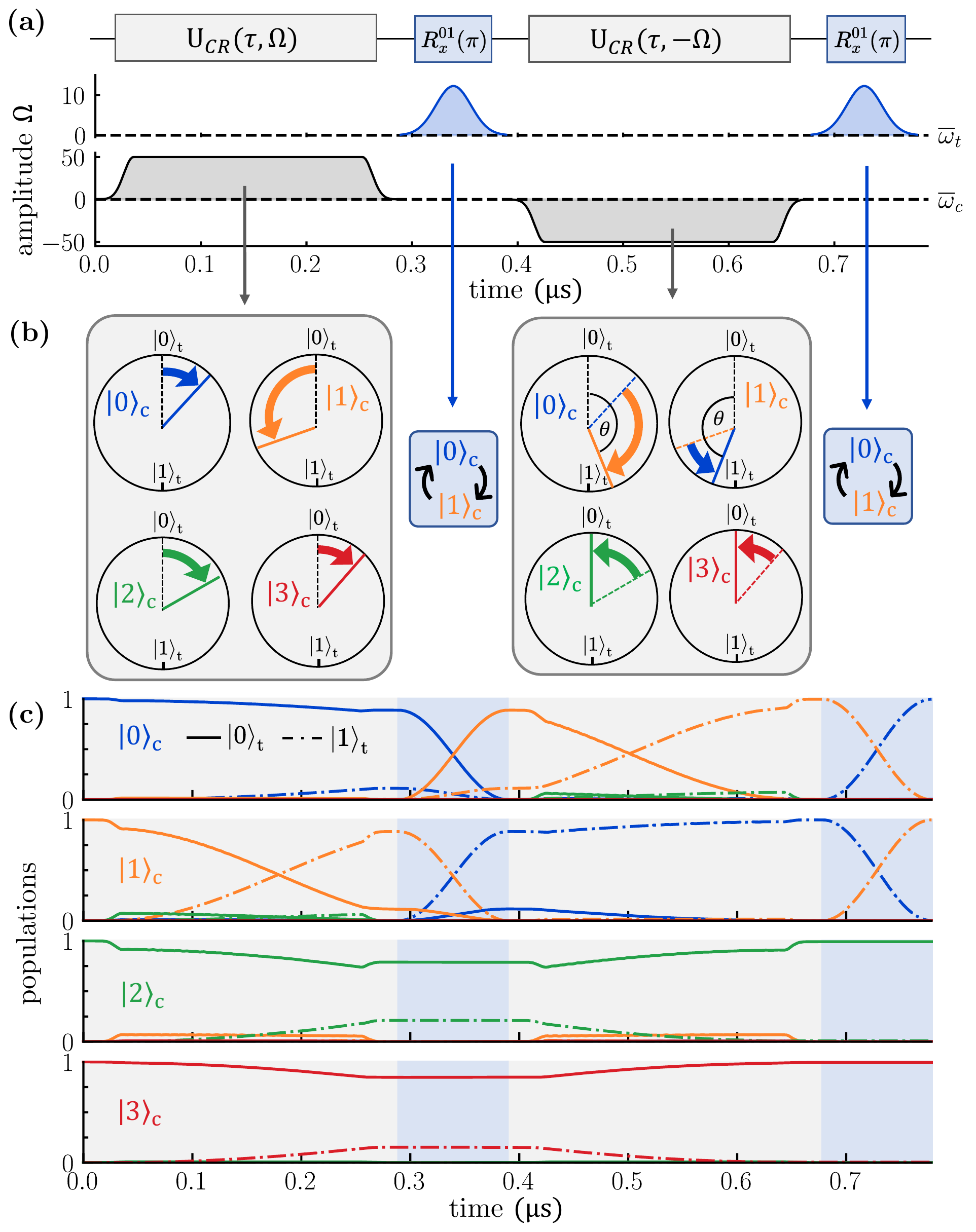}
\caption[]{Qudit space action of the echoed cross-resonance gate. (a) The ECR pulse sequence applied to the control qudit consists of two cross-resonance tones (gray) played at the target qudit frequency $\overline{\omega}_t$  with opposite amplitudes, each followed by a $\pi$-pulse on the control. (b) Action of each pulse on the initial target state $\ket{0}_t$ depending on the control state $\ket{\psi}_c$ on the Bloch sphere spanned by $\{\ket{0}_t, \ket{1}_t\}$. By construction of the echo sequence, a $\theta$ ($-\theta$) rotation is applied in the $\ket{1}_c$ ($\ket{0}_c$) case. (c) Evolution of the populations starting from $\ket{0}_t$ depending on the control state $\ket{\psi}_c$ for the pulse sequence shown in (a). Color denotes the state of the control while line style denotes the state of the target. Pulse durations are calibrated to $\theta = \pi$. 
}
\label{fig:ECR_pulse_overview}
\end{figure}

We now analyze the action of a single CR pulse by numerically integrating the full dynamics of the time-dependent Schr{\"o}dinger equation of the two-transmon system in the $d=4$ subspace. 
We choose a Gaussian-square pulse envelope of amplitude $\Omega/(2\pi)=50\,\text{MHz}$ and duration $\tau$ with a Gaussian rise and fall to suppress leakage out of the $\ket{0}_t - \ket{1}_t$ subspace, as shown in Fig~\ref{fig:ECR_pulse_overview}(a) and detailed in Appendix~\ref{sec:app_single_cr_pulse}.
The chosen model parameters $\omega_c, \omega_t$, $\alpha_c$, $\alpha_t$, $J$ and $\Omega$ yield effective $Z\otimes X$ and $Z\otimes Z$ interaction strengths in the two-qubit subspace of $\omega_{ZX}/(2\pi) = -1.15\,\text{MHz}$ and $\omega_{ZZ}/(2\pi) = 31\,\text{kHz}$, respectively.
Generalizing from the qubit case, we expect the resulting dynamics to create a $R_x^{01}(\varphi)$ rotation on the target, whose rotation angle and direction depend on the state of the control. We write this as  
\begin{align}
\label{eqn:single_Cr_pulse_action}
U_\text{CR}(\vec{\varphi}) = &\ketbra{0}{0}_c \otimes R_x^{01}(-\varphi_0) + \ketbra{1}{1}_c \otimes R_x^{01}(\varphi_1) \\
                            + &\ketbra{2}{2}_c \otimes R_x^{01}(-\varphi_2) + \ketbra{3}{3}_c \otimes R_x^{01}(-\varphi_3) \nonumber
\end{align}
where each rotation angle $\varphi_i$ is proportional to the total area under the pulse envelope.
For the QEC application presented in Sec.~\ref{chap:qec_application}, we require a rotation angle of $\pm \pi$ in the $\ket{0}_c$ case for the echoed sequence. 
We thus aim to calibrate the CR tones such that  $\varphi_0 + \varphi_1 = \pi$. 
For the chosen model parameters, this is achieved with a pulse duration of $\tau = 289\,\text{ns}$ shown in Fig.~\ref{fig:ECR_pulse_overview}(a).
Up to local phases on the control and target, the unitary resulting from our pulse simulation reaches an average gate fidelity of $\fidelity = 99.93 \%$ to Eq.~\eqref{eqn:single_Cr_pulse_action} with angles $\vec{\varphi}=\left(\varphi_0, \dots \varphi_3\right) \approx \left(0.22,  0.78, 0.30, 0.26\right)\pi $. 
This result justifies the intuition behind the schematic illustration in Fig.~\ref{fig:ECR_pulse_overview}(b).
Note the difference in the rotation direction between the states $\ket{0}_c, \ket{2}_c, \ket{3}_c$ and $\ket{1}_c$. 

The echo $\pi$-pulse $R_x^{01}(\pi)$ in the ECR pulse sequence is simulated as a Gaussian at the (average) frequency of the control qudit $\overline{\omega}_c$.
We fix the pulse duration at $100\,\text{ns}$ and calibrate the amplitude to $12.3\,\text{MHz}$, obtaining a fidelity of $\fidelity = 99.99\%$, for details see Appendix~\ref{sec:app_single_qudit_pulse}.
This is slower than current state-of-the-art $X$-gates in transmons~\cite{IBMQuantum, werninghaus2021leakage}, to keep leakage minimal. 
For simplicity, we omit a careful calibration of \textsc{Drag} pulses by which leakage errors and pulse duration could be further reduced~\cite{motzoi2009simple}.
Assuming that the reversed amplitude in the second CR tone of the ECR sequence reverses all rotation angles $\vec{\varphi}$ in Eq.~\eqref{eqn:single_Cr_pulse_action}, we expect that the unitary of the full ECR sequence is
\begin{align}
\label{eqn:ECR_unitary}
U_\text{ECR}(\theta) = &\ketbra{0}{0}_c \otimes R_x^{01}(-\theta) + \ketbra{1}{1}_c \otimes R_x^{01}(\theta) \\
+ &\ketbra{2}{2}_c \otimes \mathbb{1} + \ketbra{3}{3}_c \otimes \mathbb{1} \nonumber
\end{align}
with $\theta = \varphi_0 + \varphi_1$. 
With the CR tones calibrated as described above, our simulation of the entire pulse sequence obtains an average gate fidelity of $99.56 \%$ for the targeted rotation angle $U_\text{ECR}(\theta = \pi)$ (up to local phase gates).
This is the unitary error of the gate. 
To estimate the additional effect of incoherent error channels, we add amplitude damping with a $T_1$-time of $310\,\upmu\text{s}$ and pure dephasing with a $T_2$ time of $170\,\upmu\text{s}$ to the simulation as detailed in Appendix~\ref{appendix:linblad_dynamics}, which correspond to median $T_1$ and $T_2$ times of state-of-the-art IBM Quantum devices~\cite{IBMQuantum}.
This reduces the fidelity to $98.66\%$. 
With values of $474\,\upmu\text{s}$ for $T_1$ and $666\,\upmu\text{s}$ for $T_2$, corresponding to the best available pair of neighboring qubits, the fidelity becomes $99.22\%$.
In this work, we are primarily interested in understanding the limits to the unitary error of this gate. 
We thus leave exploring the trade-off between the unitary gate error -- which is minimal for longer gate durations -- and the incoherent gate error that increases with the gate duration for future work. 

The evolution of the populations in the two-transmon system under the echoed CR sequence is shown in Fig.~\ref{fig:ECR_pulse_overview}(c), with details given in Appendix~\ref{sec:app_echoed_CR_sequence}.
The remaining unitary error of the gate originates mainly from the detuning of the CR tones to the target frequency in the cases where the control is in $\ket{2}_c$ and $\ket{3}_c$ [see Fig.~\ref{fig:two_transmon_level_spectrum}].
This leads to a small $Z$-contribution in the effective rotation axis in each respective subspace, which is not fully reversed by the echoed CR sequence. 
These effects could potentially be resolved by adding rotary tones~\cite{sundaresan2020reducing}, including virtual phase gates into the sequence~\cite{mckay2017efficient} or more advanced qudit-based optimal control techniques~\cite{seifert2022timeefficient, simm2023two}.
We find that leakage out of the $\ket{0}_t - \ket{1}_t$ subspace is not a relevant error source with populations of those levels remaining under $10^{-5}$ after the ECR sequence. 

For qudit-based circuit decomposition, the ECR gate from Eq.~\eqref{eqn:ECR_unitary} is particularly convenient, as it only depends on a single parameter $\theta$ which is tunable through the duration of the CR pulses. 
In Sec.~\ref{chap:gate_decompositions}, we present a decomposition routine that implements general qudit unitaries through the $\textsc{ecr}$ gate and single-qudit gates. 

\subsection{Experimental requirements}
Controlling additional states of a transmon beyond the qubit space comes with an increased complexity. 
We now comment on the requirements to operate the transmon as a ququart in the presence of charge noise, limited lifetimes, and imperfect ququart readout.
Charge noise is exponentially larger in higher-excited states $\ket{n}$~\cite{peterer2015coherence}.
However, charge noise is exponentially suppressed with increasing $E_{\text{J}} / E_{\text{C}}$ at the cost of a polynomial reduction of the anharmonicity $\alpha$~\cite{koch2007chargeinsensitive}. 
Our simulations suggest that reasonable anharmonicities remain with manageable charge noise for ququart operation when choosing $E_{\text{J}} / E_{\text{C}} \sim 70$.
Scaling the system to more than two ququarts may require a careful design of the control pulses to mitigate leakage due to the weak anharmonicity and the frequency crowding introduced by the additional qudit levels.

The lifetime of higher-excited states $\ket{d}$ also generally decreases with $d$.
Fortunately, their decay happens predominantly sequentially, e.g., following $\ket{3} \rightarrow \ket{2} \rightarrow \ket{1} \rightarrow \ket{0}$, 
which leads to workable coherence times for the lowest-lying qudit levels. 
For example, for devices with qubit lifetimes of $\sim 80\,\upmu\text{s}$, which is below current state-of-the art of $\sim 500\,\upmu\text{s}$, experiments have found lifetimes of $\sim30\,\upmu\text{s}$ for state $\ket{3}$~\cite{fischer2022ancillafree, peterer2015coherence}.
In comparison our \texttt{ECR} pulse lasts about $1\,\upmu\text{s}$.

Finally, transmons are measured with a dispersive readout by coupling to a resonator.  
This technique can distinguish between multiple qudit states, as demonstrated for ququarts~\cite{miao2022overcoming}.
In summary, our simulations along with previous experimental demonstrations of coherence times and readout confirm that high-fidelity ququart operation of transmons is possible. 

\section{Universal qudit gate synthesis}
\label{chap:gate_decompositions}
Quantum algorithms are described at the quantum circuit level with abstract gate instructions that typically do not match those of the hardware.
A vast amount of work is dedicated to producing practical and efficient hardware-executable gate decompositions for qubits~\cite{li2019tackling, earnest2021pulse, miller2022hardwaretailored, galda2021implementing}.
However, little attention has been given to the qudit case aside for general algorithms.
Here, we show how to transpile an arbitrary $d^n \times d^n$ unitary on a system of $n$ $d$-dimensional transmon qudits into the hardware-native gate set of single-qudit rotations and the $\texttt{ECR}$ gate. 
This constitutes the first practical blueprint for qudit unitary gate synthesis on superconducting qudits, as detailed in Appendices~\ref{app:transpiler} and~\ref{app:general_synthesis}.  

\subsection{Qudit transpilation}

Any set of arbitrary single-qudit gates combined with a single entangling two-qudit gate is in principle exact-universal~\cite{brylinski2002universal, brennen2005criteria}.
Several constructive decomposition routines exist which rely on different choices of two-qudit gates~\cite{wang2020qudits, brennen2005efficient}. 
For qubits, the quantum Shannon decomposition (QSD) is a powerful tool to synthesize arbitrary unitaries~\cite{shende2006synthesis}. 
We build on the multivalued QSD that generalizes the QSD to the qudit setting~\cite{di2013synthesis}. 
Within this framework, the circuit complexity, quantified by the number of two-qudit gates required to achieve arbitrary $N$-qudit unitaries, is reduced by a factor of $d-1$ compared to the qubit setting of $d=2$, highlighting the comparative efficiency of qudit circuits~\cite{di2015optimal}. 
The multivalued QSD iteratively reduces the desired unitary to block matrices that contain only one-qudit and two-qudit unitaries. The remaining two-qudit blocks consist of singly-controlled gates 
\begin{align}
\label{eq:def_singly_controlled_gate}
C^m[U] = \ketbra{m}{m} \otimes U + \sum_{i\neq m} \ketbra{i}{i} \otimes \mathbb{1}
\end{align}
that apply a unitary $U\in{\rm SU}(d)$ on the target qudit if and only if the control qudit is in the basis state $\ket{m}$~\cite{di2013synthesis}.

\begin{figure}
\includegraphics[width=0.99\columnwidth]{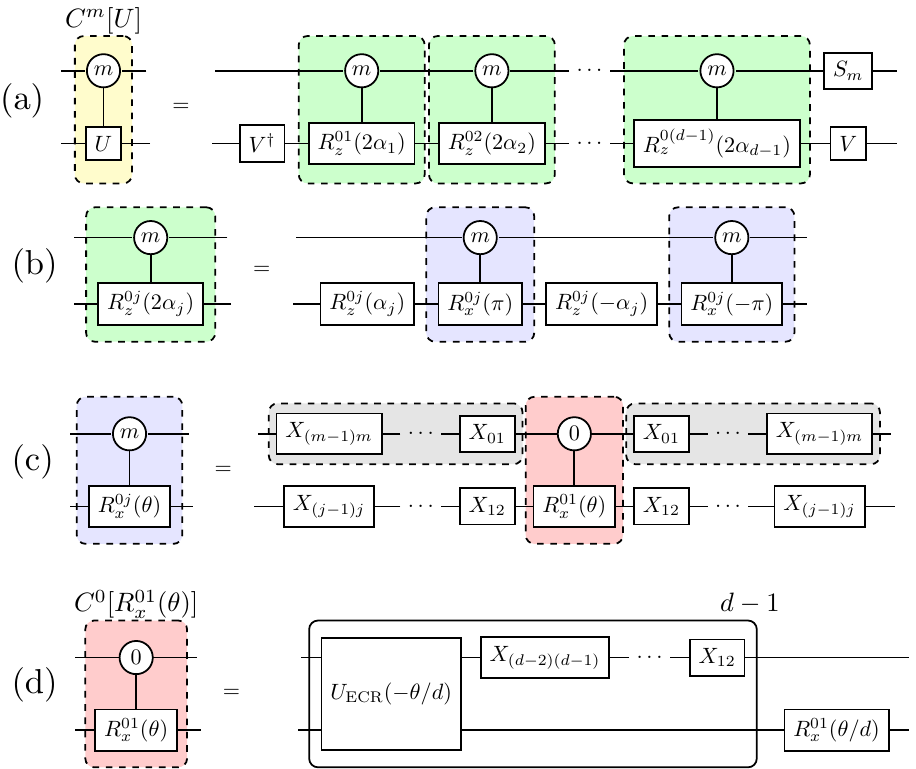}
\caption[]{Gate decompositions to implement a general singly-controlled two-qudit gate $C^m[U]$. (a) Diagonalization of $C^m[U]$ leads to a sequence of controlled $z$-rotations $R_z^{0j}$ between the $0^{\text{th}}$ and $j^{\text{th}}$ level. (b) Each $C^m[R_z^{0j}]$ is implemented through two $C^m[R_x^{0j}]$ rotations and local phase gates. (c) Decomposition of $C^m[R_x^{0j}]$ into local permutations and a single $C^0[R_x^{01}]$ gate. Gray shaded gates cancel between consecutive $C^m[R_x^{0j}]$ gates as they arise in (b). (d) Realization of the $C^0[R_x^{01}(\theta)]$ gate through echoed cross-resonance gates.}
\label{fig:decomposition_circuits}
\end{figure}

We now present a decomposition routine to realize $C^m[U]$ through single-qudit gates and the $\texttt{ECR}(\theta)$ gate.
This decomposition holds for any $d$, as long as the $\texttt{ECR}$ gate in Eq.~\eqref{eqn:ECR_unitary} acts as $\ketbra{n}{n}\otimes \mathbb{1}$ for all states $n\geq 2$. 
We diagonalize $U = V D V^\dagger$ such that the diagonal matrix $D = e^{i\gamma} \text{diag}(e^{-i(\sum_{j=0}^{d-1} \alpha_j)}, e^{i\alpha_1}, \dots, e^{i\alpha_{d-1}})$ can be decomposed into $R_z^{0j}$ rotations as $ D = e^{i\gamma} \prod_{j=1}^{d-1}  R_z^{0j}(2 \alpha_j)$.
$C^m[U]$ can then be implemented as a product of controlled phase gates between the target's $\ket{0}$ and $\ket{j}$ states
\begin{align}
\label{eqn:D_decomposition_RZ}
C^m[U] = \left(S_m \otimes V \right) \prod_{j=1}^{d-1}  C^m[R_z^{0j}(2 \alpha_j)] \left(\mathbb{1} \otimes V^\dagger \right)
\end{align}
with a phase gate $S_m = \sum_{j = 0}^{d-1} e^{i\gamma \delta_{jm}}\ket{j}\bra{j}$ [see Fig.~\ref{fig:decomposition_circuits}(a)].
Next, each $C^m[R_z^{0j}(2\alpha_j)]$ gate is expressed through two controlled $x$-rotations, yielding [see Fig.~\ref{fig:decomposition_circuits}(b)]
\begin{align}
\label{eqn:CZ_rot_through_CX_rot}
C^m[R_z^{0j}(2\alpha_j)] = C^m[&R_x^{0j}(-\pi)] \left(\mathbb{1}\otimes R_z^{0j}(-\alpha_j)\right) \\
\times &C^m[R_x^{0j}(\pi)] \left(\mathbb{1}\otimes R_z^{0j}(\alpha_j)\right). \nonumber
\end{align}
This shifts all angular dependence into local phase gates we implement virtually. 
The general $m$-controlled $R_x^{0j}$ gates from Eq.~\eqref{eqn:CZ_rot_through_CX_rot} can be realized through $0$-controlled $R_x^{01}$ rotations by applying single-qudit permutation gates on the control and the target, as shown in Fig.~\ref{fig:decomposition_circuits}(c). 
Here, $X_{n(n+1)}$ denotes a swap of the levels $i$ and $i+1$ which is equivalent to $R_x^{n(n+1)}(\pi)$ up to virtual local phases. 
Note that the permutation gates on the control qudit to shift the control from $m$ to 0, highlighted in gray in Fig.~\ref{fig:decomposition_circuits}(c), need to be applied only once at the beginning and the end of the decomposition sequence.

\begin{figure*}
\includegraphics[width=0.99\textwidth]{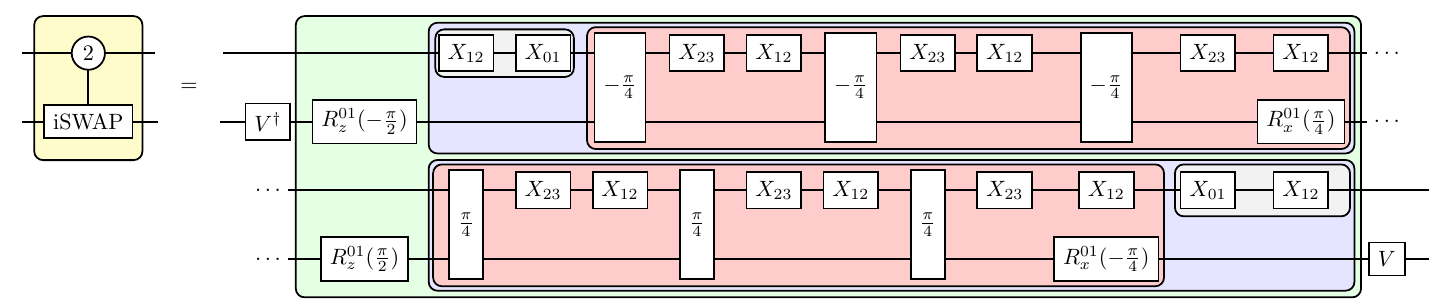}
\caption[]{
Example application of the general qudit transpiler decomposing a $\ket{2}$-controlled $\texttt{iSWAP}$ gate between two ququarts. 
Colors indicate the different steps of the transpilation shown in Fig.~\ref{fig:decomposition_circuits}, while two-qudit boxes denote $\texttt{ECR}(\theta)$ gates. 
}
\label{fig:Iswap_example}
\end{figure*}

The final step is now to realize the remaining $C^0[R_x^{01}(\theta)]$ gates with the $\texttt{ECR}$ gate defined in Eq.~\eqref{eqn:ECR_unitary}. 
Since the $\texttt{ECR}$ gate acts non-trivially on the target for both control states $\ket{0}_c$ and $\ket{1}_c$, $C^0[R_x^{01}(\theta)]$ can not be implemented with a single $U_\text{ECR}(\theta)$ gate.
Instead, we split the rotation into $d-1$ steps of $U_\text{ECR}(-\theta/d)$ and permute the levels of the control between each step, as shown in Fig.~\ref{fig:decomposition_circuits}(d). 
This way, the action on the target is $R_x^{01}(\theta(d-1)/d)$ when the control is in $\ket{0}_c$ and $R_x^{01}(-\theta/d)$ when the control is in any other state.  
Applying an $R_x^{01}(\theta / d)$ gate on the target finally recovers the desired $C^0[R_x^{01}(\theta)]$ rotation, since
\begin{align}
\label{eqn:C0RX_to_ECR}
C^0[R_x^{01}(\theta)] = & \left(\mathbb{1} \otimes  R_x^{01}\left(\tfrac{\theta}{d}\right) \right) \\
    & \times \left(  \left( \prod_{j=2}^{d-1} X_{(j-1)j} \otimes \mathbb{1} \right) U_\text{ECR}(-\tfrac{\theta}{d})  \right)^{d-1} . \nonumber
\end{align}

Fig.~\ref{fig:decomposition_circuits} shows the decomposition in the most general case.
In practice, the complexity of a desired two-qudit gate can be much simpler, as illustrated by the decomposition of a $\ket{2}$-controlled \texttt{iSWAP} gate $C^2[R_x^{12}(-\pi)]$, shown in Fig.~\ref{fig:Iswap_example}. 
Following the notation from above, $\alpha_2, \alpha_3$ and $\gamma$ of the diagonal $D$ are zero, and $\alpha_1 = -\pi/2$.
Therefore, only one of the green blocks from Fig.~\ref{fig:decomposition_circuits}(a) appears in the decomposition.
The single-qudit diagonalization gates $V$ can be realized with two $R_x^{01}$ gates, one $R_x^{12}$ gate and virtual phase gates. 

\subsection{Comparison to qubits}
\label{sec:compiler_comparison_qubits}

In general, the presented decomposition routine requires $O(d^2)$ two-qudit entangling gates and $O(d^3)$ single-qudit gates.
In the $d=4$ case, any $C^m[U]$ unitary can be synthesized with 18 $U_\text{ECR}(\pm \pi/4)$ gates and $56 + 2m$ single-qudit gates (not counting virtually implemented phase gates). 
This construction adds an overhead of single-qudit gates, but makes an efficient use of the entanglement generation rate of the CR effect as the total duration of the \texttt{ECR} pulses is proportional to $\theta (d-1)/d$. 
In comparison, a qubit \textsc{Cnot} gate is locally equivalent to a $U_\text{ECR}(\pm \pi/2)$ pulse sequence, and thus takes roughly twice as long as a $U_\text{ECR}(\pm \pi/4)$ qudit gate.

To benchmark the efficacy of our qudit transpiler, we compare its performance against a state-of-the-art qubit transpiler available in the software package Qiskit~\cite{Qiskit}.  
Qiskit's transpiler uses a column-by-column decomposition developed in Ref.~\cite{iten2016quantum}.
We focus on the task of unitary synthesis of general $16 \times 16$ matrices $U_\text{target}$. 
For the qudit case, we use two ququarts with a basis gate set made of bidirectional $U_\text{ECR}(\pm \pi/4)$ gates, single-qudit gates and virtual phase gates. 
We make use of an iterative cosine-sine decomposition~\cite{chen2013qcompiler} to synthesize $U_\text{target}$ from a sequence of block-diagonal $C^m[U]$ gates, see Appendix~\ref{app:general_synthesis} for details. 
For the qubit case, we consider four linearly connected qubits and a basis gate set of bidirectional \textsc{Cnot} gates as well as single-qubit $\sqrt{X}$ gates and parametric virtual phase gates.

To compare the gate cost between the qubit and the qudit case, we equate two $U_\text{ECR}(\pi/4)$ gates with one \textsc{Cnot} gate, and decompose all single-qudit gates into $R_x^{n, n+1}(\pi/2)$ gates which we equate to a single-qubit $\sqrt{X}$ gates.
We do not count virtual phase gates since they are implemented by a phase change of subsequent gates.
We find that the qubit transpiler requires 286 \textsc{Cnot} and 244 $\sqrt{X}$ gates to synthesize random SU$(16)$ gates.
The ququarts transpiler achieves the same unitary with an equivalent of 170 \textsc{Cnot} and 2776 $\sqrt{X}$ gates.
This highlights the central tradeoff between qudit and qubit transpilation: The two ququarts reduce the required number of entangling gates by $\approx 40\%$, at the cost of increasing the number of single-qudit gates by a factor of~$\sim 11$.
Since single-qudit gates are typically an order of magnitude faster with at least an order of magnitude higher fidelity, there might be a regime where the qudit transpilation is favorable despite the larger total gate count.

\begin{figure}
\includegraphics[width=0.99\columnwidth]{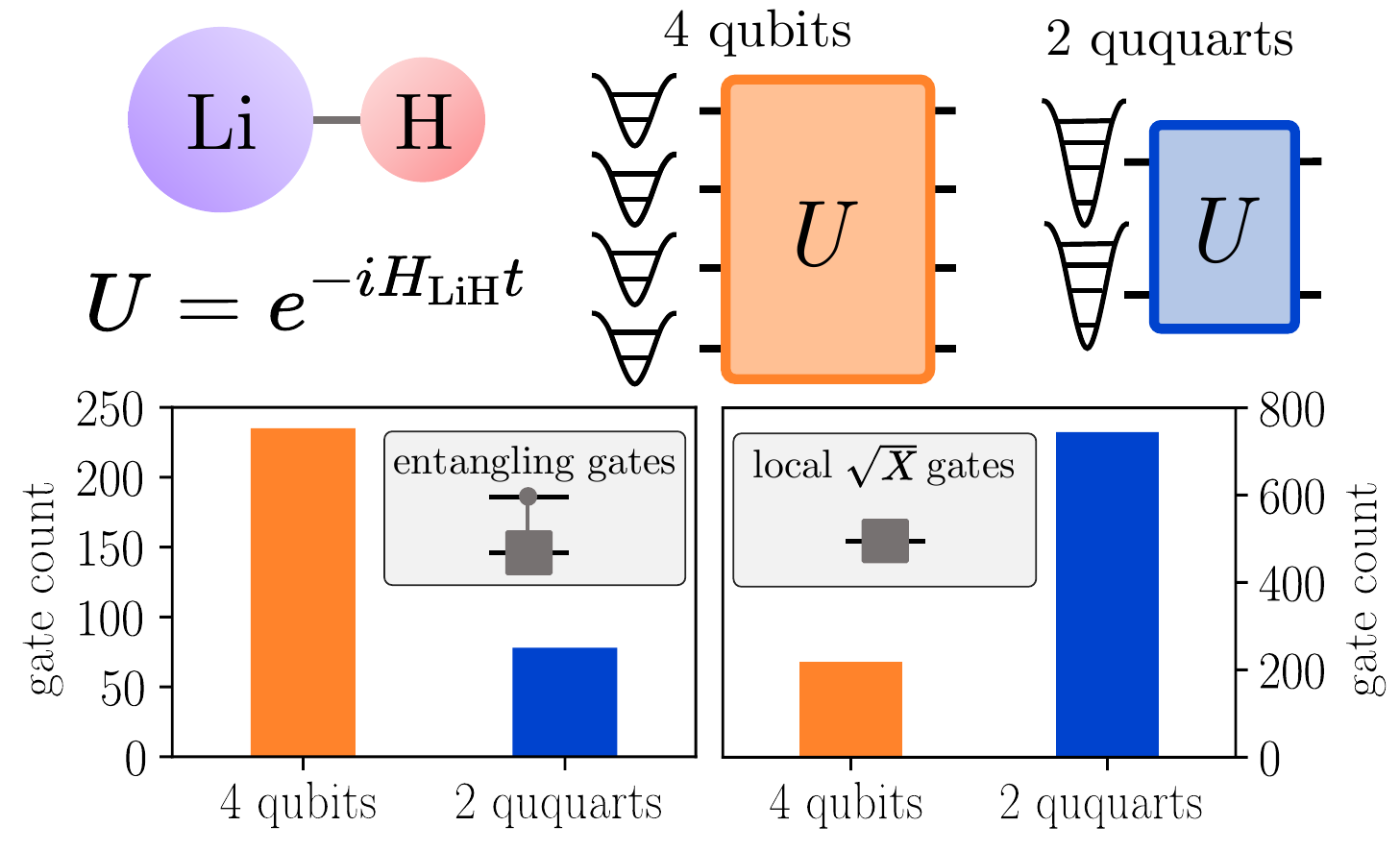}
\caption[]{Benchmark study of the qudit transpiler on the Hamiltonian exponentiation of a lithium hydride molecule. 
Two ququarts require only a third of the entangling gates that four linearly connected qubits require, at the expense of 3.4 times as many single-qudit $\sqrt{X}$ gates.}
\label{fig:gatecount_comparison}
\end{figure}

Besides dense, random unitaries, we consider the physically motivated example of the exponentiation of a Hamiltonian associated to a small molecule. 
We choose the second-quantized Hamiltonian corresponding to the LiH molecule in a minimal Sto-3g basis set, mapped to a four-qubit Hamiltonian $H_\text{LiH}$ with a Jordan-Wigner mapping. 
Next, we synthesize the target unitary $U_\text{target} = e^{-iH_\text{LiH} t} $ for a time $t=10$ and find that qubit transpilation requires 235 \textsc{Cnot}s and 218 $\sqrt{X}$ gates.
The ququart case takes the equivalent of 78 \textsc{Cnot}s and 744 $\sqrt{X}$ gates, as shown in Fig.~\ref{fig:gatecount_comparison}.
This amounts to a reduction by a factor of $3.01$ in entangling gate cost while increasing the single-qudit gate count only by a factor of $3.41$. 
This example illustrates that our qudit transpiler can already rival a heavily optimized qubit transpiler.

\subsection{Qudit gate extensions}

As exemplified above, the qudit-based unitary synthesis creates a substantial overhead in single-qudit gates. 
This is because the entangling gate we use only acts non-trivially when the control is in $\ket{0}_c$ or $\ket{1}_c$ and only affects the target states $\ket{0}_t$ and $\ket{1}_t$, see Eq.~\eqref{eqn:ECR_unitary}. 
This creates the need for single-qudit permutation gates on both the target and the control, such as the single-qudit gates in Fig.~\ref{fig:decomposition_circuits}(c). 
We now outline two strategies to overcome this bottleneck. 
(i) The \texttt{ECR} gate acts trivially when the control is in $\ket{2}_c$ and $\ket{3}_c$ due to the single-qudit $R_x^{01}$ gates applied in the echoed sequence, see Fig.~\ref{fig:ECR_pulse_overview}. 
Replacing these gates with $R_x^{12}$ or $R_x^{23}$ rotations creates a gate that instead acts non-trivially on the target when the control is in states $\ket{1}_c / \ket{2}_c$ or  $\ket{2}_c / \ket{3}_c$, respectively.
(ii) We envision that changing the frequency of the ECR drive tones, such that it is resonant with the $\ket{1}_t - \ket{2}_t$ or $\ket{2}_t - \ket{3}_t$ transitions of the target would enable a gate that directly addresses the subspaces spanned by $\{ \ket{1}_t, \ket{2}_t\}$ or $\{ \ket{2}_t, \ket{3}_t\}$, respectively. 
The added flexibility of either approach would strongly increase the effective ``connectivity'' of the different levels of two coupled qudits; the permutation gates would no longer be needed to move states into the $\{\ket{0}, \ket{1}\}$ interaction subspace.

Moreover, the number of gates in our decomposition could potentially be improved by working with a direct (non-echoed) CR gate~\cite{jurcevic2021demonstration}, which, however, increases the complexity as this depends on multiple parameters $\vec{\varphi}$, see Eq.~\eqref{eqn:single_Cr_pulse_action}.
More broadly, our decomposition could be adapted to different interactions as those provided by, e.g., tunable couplers, or frequency-tunable transmons~\cite{goss2022highfidelity, roy2022realization}.
Another approach to improve the single-qudit gate overhead is given by optimal control pulse shaping techniques~\cite{simm2023two}.
Finally, note also that the step in Eq.~\eqref{eqn:CZ_rot_through_CX_rot} introduces controlled $R_x$ gates with a maximal rotation angle of $\pm \pi$, irrespective of the original rotation angle $\alpha$. Thus, the resulting pulse schedules could be shortened by pulse-efficient circuit transpilation techniques as developed in~\cite{earnest2021pulse}, resulting in better gate fidelities.

\section{Correction of dephasing errors}
\label{chap:qec_application}
The multi-level transmon structure and the ECR gate can implement qudit-based QEC protocols~\cite{Pirandola2008,PRXGirvin},
which were recently proposed in the context of molecular spin qudits with embedded error correction~\cite{Hussain2018,Chiesa2020,Carretta2021,Petiziol2021,Chizzini2022}. 
We now first translate the same ideas to the here-proposed transmon setup and then present a detailed simulation of a QEC routine. 
This highlights how the universal qudit logic toolbox could be used to engineer lower error rates for qubit computations.

\subsection{Embedding error correction into qudits}
\label{sec:error_correction_in_qudits}
A 4-level qudit is the minimal unit needed to embed QEC against a single kind of noise source~\cite{Chiesa2020}, such as amplitude damping or pure dephasing~\cite{Cafaro2012}. 
Indeed, for an error to be identified and corrected, its action on a superposition of logical states ({\it code words}) must bring them to distinguishable ones, thus requiring a Hilbert space of dimension $\ge 4$.
Specifically, we consider a correction of pure dephasing errors, which for a bosonic system are well approximated by an error operator $\sqrt{\Gamma} n$, where $n=a^\dagger a$ and $a^\dagger$ ($a$) is the creation (annihilation) operator of the bosonic mode~\cite{PRXGirvin}. 
We model dephasing as Markovian noise described by the Lindblad equation
\begin{equation}
    \dot{\rho}(t) = -\frac{i}{\hbar} \left[H(t), \rho(t) \right]  + \Gamma \left(  2 n \rho(t)  n - \{ n^2, \rho(t) \} \right)
    \label{eq:lindblad}
\end{equation}
where $\rho$ is the system single-qudit density matrix subject to an external driving Hamiltonian $H(t)$ and $\Gamma = 1/T_2$ is the dephasing rate.
Expanding the solution to this equation in series for small $\Gamma t$ yields a leading error term proportional to $n$ in the Kraus representation~\cite{Chiesa2020}.
An analogous result, with error operators represented by powers of $n$, can be derived for non-Markovian noise \cite{PRXGirvin}.
The same form of noise also describes pure dephasing in a slightly anharmonic transmon system as considered here, by simply making the replacement $n \rightarrow \sum_m m \ketbra{m}{m}$ in the truncated Hilbert space $m=0,1,2,3$~\cite{blais2021circuit}.
We require a pair of code words $\ket{0_L}$ and $\ket{1_L}$ protected against the set of errors $E_k \in \{ I, n \}$, i.e., satisfying the Knill-Laflamme conditions \cite{KnillLaflamme}:
\begin{align} \label{eq:KL1}
\begin{split}
    \bra{0_L} E_k E_j^\dagger \ket{0_L} &= \bra{1_L} E_k E_j^\dagger \ket{1_L}  \\
    \bra{0_L} E_k E_j^\dagger \ket{1_L} &= 0 . 
\end{split}
\end{align}
A possible choice is
\begin{equation} 
\label{eq:logical_state}
    \ket{0_L} = \frac{\ket{0} + \sqrt{3} \ket{2} }{2} \quad\text{and} \quad
    \ket{1_L} = \frac{\sqrt{3} \ket{1} + \ket{3} }{2} .
\end{equation}
These code words fulfill Eq.~\eqref{eq:KL1} by construction and it can be easily checked that $\bra{0_L} n \ket{0_L} = \bra{1_L} n \ket{1_L} = 3/2$ and $\bra{0_L} n^2 \ket{0_L} = \bra{1_L} n^2 \ket{1_L} = 3$. Hence, the effect of a $n$ error is to bring a generic encoded logical state $\ket{\psi_L} = \alpha \ket{0_L} + \beta \ket{1_L}$ into the error state
\begin{equation}
    \frac{n \ket{\psi_L}}{\sqrt{\bra{\psi_L} n^2 \ket{\psi_L}}} = \frac{\sqrt{3}}{2} \ket{\psi_L}
    - \frac{1}{2} \underbrace{\left( \alpha \ket{e_0} + \beta \ket{e_1} \right)  }_{ = \ket{\psi_e}}.
    \label{eq:ew}
\end{equation}
This is a superposition of the code words and of the error words $\ket{e_0} = (\sqrt{3}\ket{0} -\ket{2})/2$ and $\ket{e_1} = (\ket{1} -\sqrt{3}\ket{3})/2$. Crucially, $\ket{\psi_e}$ preserves $\alpha$ and $\beta$.
Note that the Lindblad dynamics of Eq.~\eqref{eq:lindblad} yield the same time evolution as for a spin $S$ subject to pure dephasing, see, e.g., Ref.~\cite{Chiesa2020}. 
This leads to an independent decay of each coherence $\rho_{m m^\prime}$ with an exponential rate $(m-m^\prime)^2/T_2$. 
Indeed, the number operator appearing in Eq.~\eqref{eq:lindblad} is equivalent to the spin operator $S_z$ apart from an irrelevant shift, i.e. $S_z = n - S$. 

\begin{figure}
\includegraphics[width=0.99\columnwidth]{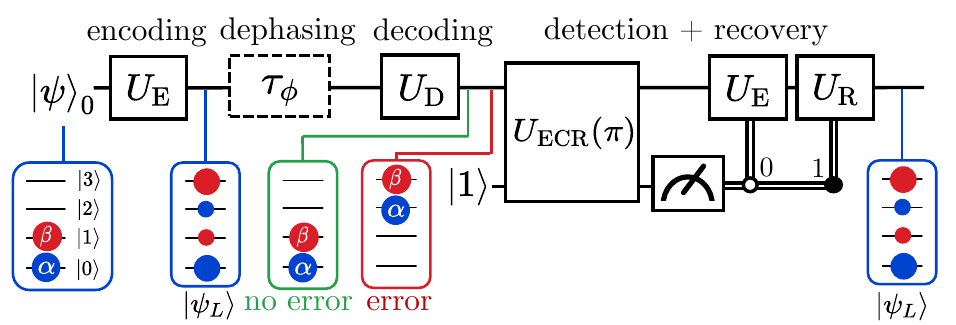}
\caption[]{Sequence to correct pure dephasing errors in a transmon. 
The initial qubit state $\ket{\psi}_0$ is encoded into a ququart logical state $\ket{\psi_L}$.
After dephasing, a decoding unitary $U_{\text{D}}$ maps the ideal and erroneous cases to orthogonal states.
The error is detected by applying a $U_\text{ECR}(\pi)$ gate where the target is an ancilla qubit prepared in $\ket{1}$.
In the ideal case where the ancilla is de-excited to 0, the state is re-encoded, while in the error case, where the ancilla is measured in 1, a recovery operation $U_{\text{R}}$ restores $\ket{\psi_L}$.}
\label{fig:error_correction_sequence}
\end{figure}

To test the performance of this qudit code in protecting a transmon memory from dephasing, we consider a pair of transmons. 
The first one, used as a ququart, encodes the logical state defined by Eq.~\eqref{eq:logical_state}. 
The second one acts as an ancillary qubit to detect errors. 
The protocol is summarized in Fig.~\ref{fig:error_correction_sequence} and consists of the following steps:
An initial qubit state $\ket{\psi}_0 = \alpha \ket{0} + \beta \ket{1}$ is encoded into the logical state $\ket{\psi_L} = \alpha \ket{0_L} + \beta \ket{1_L}$ by an encoding unitary realized with pulses on neighboring levels $U_{\text{E}} = R_y^{12}(-\pi) R_y^{01}(-2\pi/3) R_y^{23}(\pi/3) R_y^{12}(\pi)$.
The ququart then evolves freely for a memory time $\tau_\phi$ subject to pure dephasing according to Eq.~\eqref{eq:lindblad}.
Next, the decoding sequence $U_{\text{D}} = U_{\text{E}}^\dagger$ is applied to map the basis of the code and error words back into the basis of qudit eigenstates such that $\ket{\psi_L}$ is mapped back to $\ket{\psi}_0$ and $\ket{\psi_e}$ is mapped to $\alpha \ket{2} + \beta \ket{3}$. 

To detect $n$ errors, i.e., the qudit in $\ket{2}$ or $\ket{3}$, the qudit is coupled to a flag ancilla qubit prepared in $\ket{1}$. 
The $U_\text{ECR}(\pi)$ gate with the data qudit as the control and the ancilla as the target de-excites the ancilla when the control is in $\ket{0}$ or $\ket{1}$.
The ancilla is subsequently measured. 
This projects the qudit either to the $01$-subspace or the $23$-subspace while preserving the encoded superposition. 
If the ancilla is found in $\ket{0}$, i.e., no error has occurred, the encoding sequence $U_{\text{E}}$ can be reapplied, whereas, if the ancilla is found in $\ket{1}$, a recovery sequence $U_{\text{R}}$ is applied to restore $\ket{\psi_L}$ from $U_{\text{D}} \ket{\psi_e}$. 
This is accomplished with two-level rotations as 
$U_{\text{R}} = R_y^{12} (-\pi) R_y^{01} ( \pi/3) R_y^{23} (-2\pi/3) R_y^{12} (\pi)$.

\subsection{Implementation on transmons}

\begin{figure}
\includegraphics[width=0.99\columnwidth]{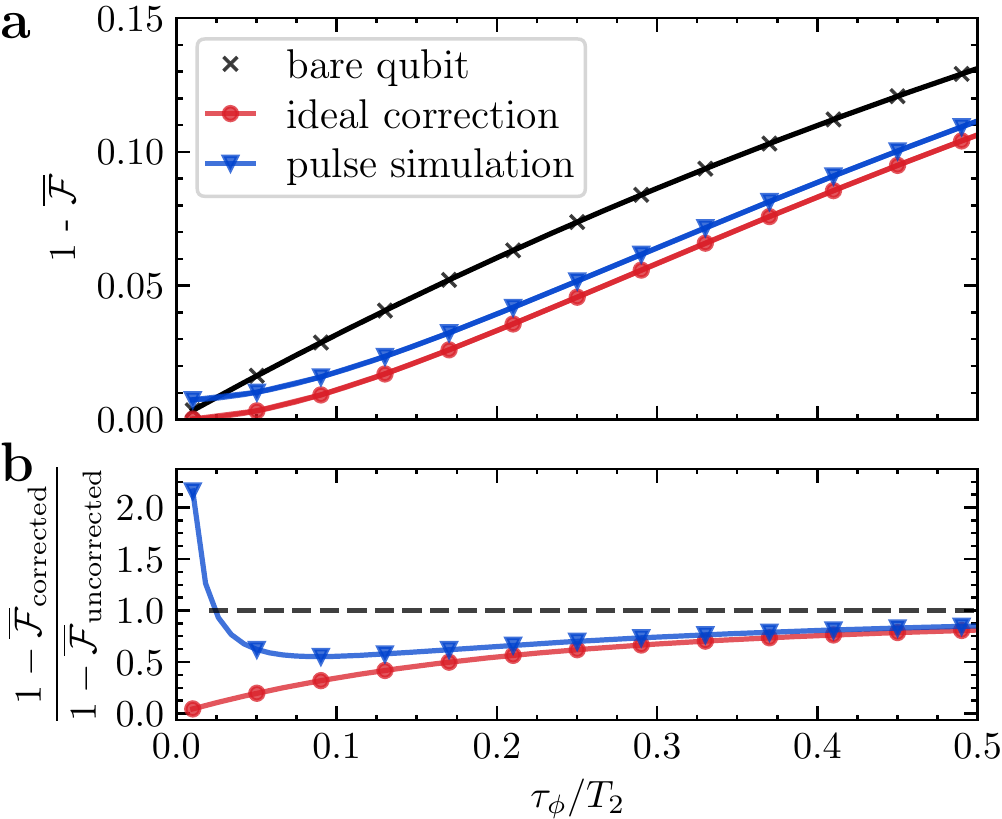}
\caption[]{Numerical simulations of a single correction cycle applied after dephasing for a duration of $\tau_\phi$. 
(a) Qubit error $1-\overline{\mathcal{F}}$.
(b) Error reduction compared to the bare qubit case.
}
\label{fig:single_correction}
\end{figure}

\begin{figure}
\includegraphics[width=0.99\columnwidth]{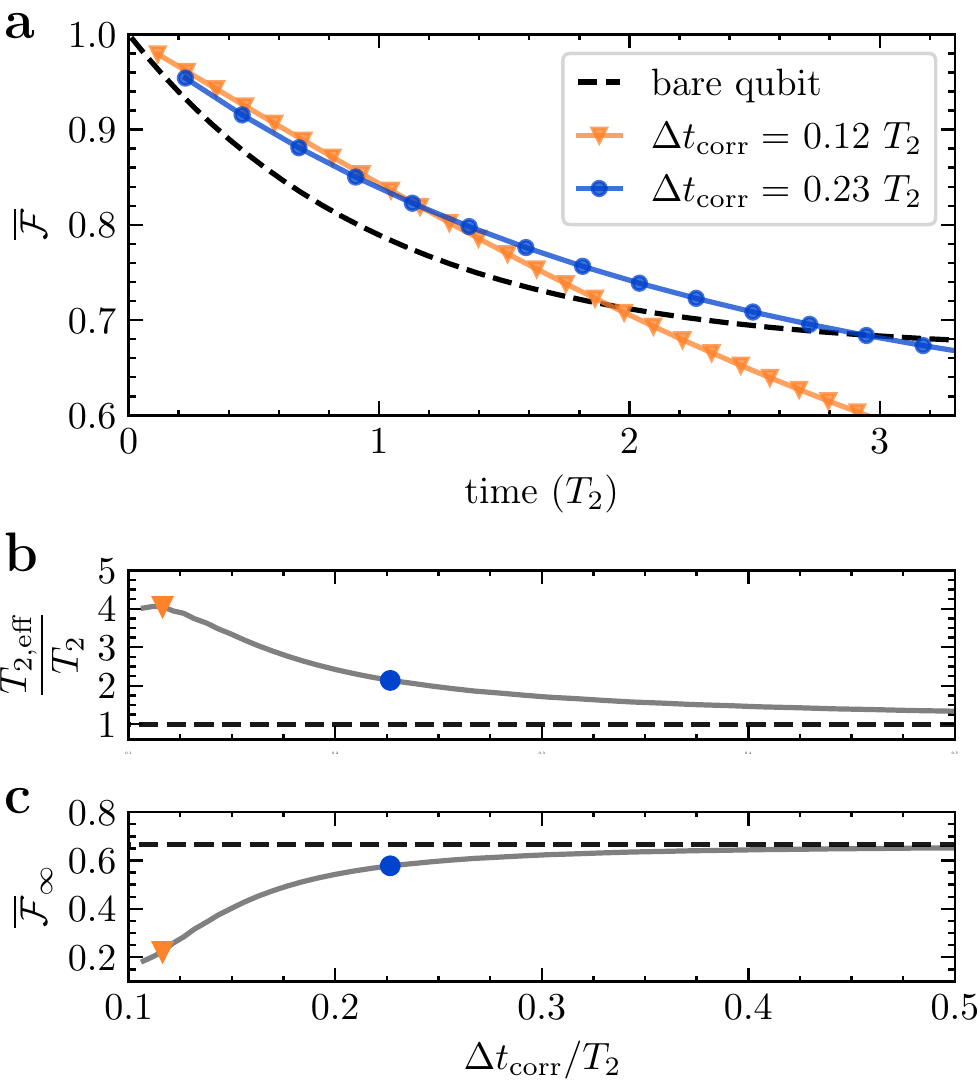}
\caption[]{Numerical simulations of the error correction pulse sequence with repeated cycles where $\Delta t_{\text{corr}}$ is the time between each cycle. 
(a) Exponential decay of the fidelity $\overline{\mathcal{F}}$ for the bare qubit case and two different correction cycle times.
(b) Time constant of the exponential decay $T_{2, \text{eff}}$ as a function of the cycle time $\Delta t_{\text{corr}}$.
(c) Asymptotic fidelity for $t\rightarrow \infty$.
}
\label{fig:multiple_corrections}
\end{figure}

We now benchmark the ququart error correction sequence in a transmon qudit through numerical simulations. 
We consider three cases:
(1) The bare qubit case where the qubit state $\ket{\psi}_0$ dephases for a duration of $\tau_\phi$ according to Eq.~\eqref{eq:lindblad} with no external drives and no encoding, 
(2) the ideal correction sequence, where all gates in Fig.~\ref{fig:error_correction_sequence} are applied as instantaneous and perfect unitaries, and
(3) the correction sequence with simulations of the pulses that implement each gate by integrating Eq.~\eqref{eq:lindblad} with the appropriate drive Hamiltonians.
We choose a dephasing time of $200\,\upmu\text{s}$, which is comparable to state-of-the-art $T_2$ values for IBM Quantum devices~\cite{IBMQuantum}.  
In cases (2) and (3), the encoded qudit dephases for a time $\tau_\phi$ before the correction cycle, after which a final decoding operation $U_{\text{D}}$ and tracing out of the ancilla obtains the resulting qubit state. 
For case (3), we use the two-transmon model and calibration of the $U_\text{ECR}(\pi)$ as presented in Sec.~\ref{sec:multi_qudit_control} where the control qudit becomes the data qudit and the target qubit becomes the ancilla.
For details on pulse shapes and numerics see Appendix~\ref{app:numerical_model}.

For all cases, we compute the channel $\mathcal{E}$ that maps $\ket{\psi}_0$ to the final qubit state of the sequence by constructing its 4x4 Liouvillian superoperator matrix.
The performance of each qubit channel in preserving an input qubit state is benchmarked by the average gate fidelity to the identity $\overline{\mathcal{F}}(\mathcal{E}, \mathbb{1}_{2})$. 
In the bare qubit case, $\overline{\mathcal{F}}$ decreases exponentially with a time constant given by $T_2$. 
The ideal correction sequence always improves on this for all dephasing times $\tau_\phi$, see Fig.~\ref{fig:single_correction}(a). 
The reduction in the error $1-\overline{\mathcal{F}}$ is most pronounced after only short dephasing times $ \tau_\phi< 0.1 \,T_2$ and becomes less and less significant for larger $\tau_\phi$, see Fig.~\ref{fig:single_correction}(b). 
This is because higher-order powers of $n$ become important for longer times, while the code only protects for first-order $n$-errors.

The pulse-level simulation introduces imperfections into the correction cycle that arise from finite unitary gate errors, leakage, non-zero durations of the gate and measurement pulses, during which the data qubit is unprotected and subject to dephasing.  
Therefore, after very short dephasing times, the correction sequence increases the qubit error compared to the bare qubit case [see Fig.~\ref{fig:single_correction}(b)]. 
However, for $\tau_\phi > 0.025\, T_2$, the code breaks even. 
With our choice of parameters, the best achievable error reduction of $ 45\%$ is found after a dephasing time $\tau_\phi = 0.09\, T_2$.

The error reduction can be extended to longer dephasing times by repeatedly applying the correction cycle, see Fig.~\ref{fig:multiple_corrections}(a). 
We denote the time between individual correction cycles as $\Delta t_{\text{corr}}$.
Ideally, the fidelity is kept arbitrarily high by making $\Delta t_{\text{corr}}$ correspondingly small. 
In reality, each correction cycle introduces a small error.
For our pulse simulations, we empirically find that the fidelity after each correction cycle is well-described by an exponential decay of the form
\begin{equation}
\label{eq:exp_decay}
\fidelity(t) = (1-\fidelity_\infty) \exp(- t/ T_{2,\text{eff}}) + \fidelity_\infty . 
\end{equation}
Here, $T_{2,\text{eff}}$ is an effective $T_2$ time and $\fidelity_\infty$ is the fidelity that is reached asymptotically for long times. 
We extract $T_{2,\text{eff}}$ and $\fidelity_\infty$ from fits to the data. 
Both quantities are plotted as a function of $\Delta t_{\text{corr}}$ in Fig.~\ref{fig:multiple_corrections}(b) and Fig.~\ref{fig:multiple_corrections}(c), respectively.  
We find that for a short cycle time, the effective $T_2$ time reaches up to four times the bare qubit $T_2$ time. 
However, this comes at the expense of a much worse asymptotic fidelity at longer times, which, in the bare qubit case, is given by $\fidelity_\infty = 2/3$.
The optimal correction frequency thus comes with a trade-off of short-term gain vs. long term infidelity. 
Nonetheless, a significant reduction in the error can be upheld for a duration of $\approx T_2$ without drastically worsening the asymptotic fidelity as shown by the blue curve in Fig.~\ref{fig:multiple_corrections}. 

\begin{figure}
\includegraphics[width=0.99\columnwidth]{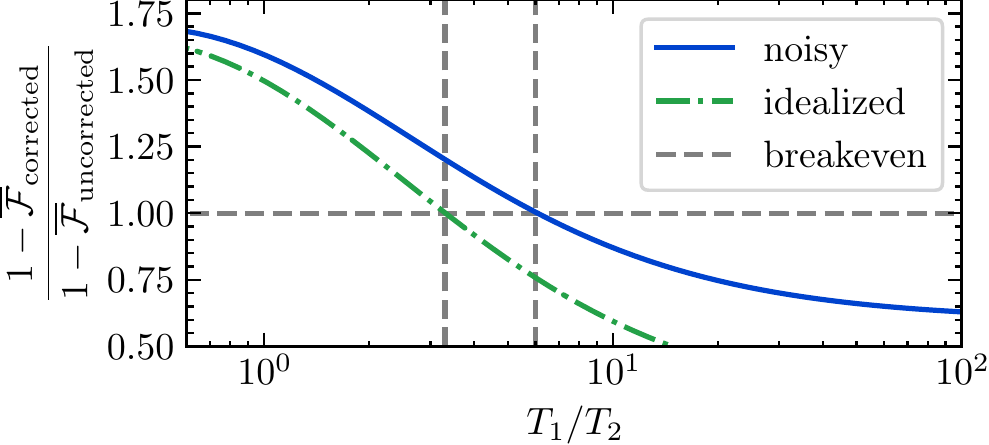}
\caption[]{Effect of amplitude damping on the error reduction of one correction cycle. 
$\tau_\phi =0.12 \, T_2$ and $T_2 = 200\,\upmu s$ are fixed while the strength of $T_1$ is varied. 
For the noisy simulation, the code breaks even for $T_1 > 6\, T_2$. 
With idealized gates (no unitary error), this decreases to $T_1 > 3.3\, T_2$}.
\label{fig:T1_effect}
\end{figure}

In a more realistic setting, additional error terms such as amplitude damping are present, which is not handled by this code.
We investigate how amplitude damping affects the protocol by adding jump operators of the form $\sqrt{n/T_1}\ketbra{n-1}{n}$ for $n\in \{1, 2, 3\}$ to the Lindbladian evolution of Eq.~\eqref{eq:lindblad}, see Appendix.~\ref{appendix:linblad_dynamics} for details.
Here, $T_1$ is the time scale of the jumps which is chosen to decrease inverse proportionally with $n$.
This model describes, in accordance with experiments~\cite{peterer2015coherence, fischer2022ancillafree}, a cascading decay from higher to lower-excited states.
With additional errors present the correction sequence fails to reduce the error compared to the bare qubit case [see Fig.~\ref{fig:T1_effect}]. 
In our model, the code reaches a break even point for $T_1/T_2 > 6$, when pure dephasing is strongly dominant over amplitude damping. 
Improving the fidelities of the gates in the correction cycle lowers the break-even threshold.
Fully eliminating the unitary errors of all gates reduces it to $T_1/T_2  > 3.3$ for the chosen parameters of $T_2$ and $\tau_\phi$.
In this idealized case, we find that reducing $\tau_\phi$ further improves the threshold to $T_1/T_2  > 2$.
These conditions can arise in practice for transmons.
For example, of the 127 qubits on \textit{ibm{\_}sherbrooke}, $9\%$, $18 \%$, and $36 \%$ of the qubits have $T_1/T_2$ ratios greater then $6$, $3.3$, and $2$, respectively~\cite{IBMQuantum}.
Finally, using more than four levels could enable a full-fledged error correction sequence. 
This is pursued for molecular nanomagnets, where the system size can be increased without a significant impact on decoherence~\cite{Hussain2018,Chiesa2020,Carretta2021,Chiesa2022}. 
For example, using $d=7$ levels of a bosonic mode with code words $\ket{0_L} = \left(\ket{0} + \sqrt{3} \ket{4}\right)/2$ and $\ket{1_L} = \left(\sqrt{3} \ket{2} + \ket{6}\right)/2$, it is possible to correct for both dephasing errors ($n$) and relaxation errors ($a$) simultaneously~\cite{PRXGirvin, Cafaro2012}.

\section{Discussion}
\label{chap:discussion_conclusion}
We have proposed a method to perform universal quantum computation in superconducting transmon qudits.
Our universal basis gate set consists of general single-qudit unitaries and an entangling cross-resonance gate between two qudits. 
With this gate set, we have developed a decomposition routine to realize arbitrary $m$-controlled two-qudit gates $C^m[U]$ as building blocks to synthesize general qudit unitaries. 
This decomposition is the center piece of a general purpose qudit transpiler to map application-level qudit-based quantum circuits to hardware-native instructions.
Compared to a state-of-the-art qubit transpiler, this qudit decomposition reduces the number of entangling gates required to synthesize general unitaries.
We have further proposed several strategies to overcome the single-qudit gate overhead opening future research directions for optimizing qudit unitary synthesis with cross-resonance gates.

The proposed entangling gate is suitable for dispersively coupled transmons. 
It represents a higher-dimensional extension of the echoed cross-resonance gate previously employed for qubits.
Our numerical model that includes charge noise and leakage errors predicts average gate fidelities of up to $99.6 \%$ with simple Gaussian pulse profiles.
The main contribution to the unitary error is the frequency dependence of the dressed target states on the control states.
These findings would benefit from systematic complementary analytical studies of the qudit-space interactions. 
For example, we chose qudit frequencies and anharmonicities such that transition frequencies between neighboring states are well separated.
Further research could therefore seek more optimal parameter regions to maximize the desired $Z\otimes X$ cross-resonance rate, akin to analytical studies of the qubit case~\cite{malekakhlagh2020firstprinciples}.
Other open questions are whether a direct cross-resonance driving without echoes benefits qudit operation or whether rotary tones reduce the unitary error, as demonstrated for qubits~\cite{sundaresan2020reducing}.
Finally, we note that cross-Kerr interactions, tunable couplers, or frequency-tunable transmons may offer different qudit gates~\cite{goss2022highfidelity, miao2022overcoming, roy2022realization}.

Qudit operation of transmons is attractive from a theoretical standpoint, as it makes full use of the available quantum resources of the system.
Compared to current standard transmon setups, the proposed gates require no additional microwave drive lines.
Moreover, the fact that higher-excited states suffer increasingly from charge noise can be mitigated by moving the transmon parameters towards higher $E_{\text{J}}/E_{\text{C}}$-ratios than those typically employed for qubit operation.
Our numerical simulations suggest that high-fidelity multi-ququart operations are possible under realistic experimental conditions. 
The maximum number of usable levels $d_\text{max}$ in a transmon qudit is limited by frequency crowding, charge noise and decoherence. 
Identifying $d_\text{max}$ is left to future work.
Crucially, our work shows how to extend the recent experimental demonstrations of coherent single-ququart operations~\cite{fischer2022ancillafree, cao2023emulatinga, seifert2023exploring, liu2023performing} into fully-fledged quantum information processors capable of executing a quantum algorithm.

As an example application, the additional levels of a qudit space can encode logical qubit states for quantum error correction. 
By embedding logical states within a single object, this scheme significantly reduces the complexity of QEC.
To date, its implementation has been proposed for molecular spin systems~\cite{Petiziol2021, Chiesa2020, Chicco2021, Chizzini2022, Carretta2021}, leveraging the large number of addressable levels and their intrinsic coherence.
Our work shows that transmons can also embed self-corrected logical units. 
As a proof-of-principle, we used four qudit levels and an ancilla flag qubit to reduce pure dephasing errors for an encoded qubit memory.
As a tool to engineer lower noise rates, these techniques could complement some of the most advanced
error mitigation strategies known in the literature, such as probabilistic error cancellation~\cite{temme2017error}, to reduce the associated overheads.
A fully-fledged quantum error correction code requires more complicated code words that cover more than four qudit states.
Although going beyond $d=4$ represents a non-trivial challenge for transmons, one may nevertheless envision optimal hybrid encoding schemes, where -- making use of a two-qudit universal gate set as proposed here -- the code space is spread over multiple qudits.

In conclusion, our results provide a blueprint for the implementation of multivalued quantum logic in transmons and has applications for both noisy as well as error-corrected hardware.
By enabling a richer set of operation modes and embedded functionalities, this approach could open up new design possibilities for the next generation of quantum processors. 

\section{Acknowledgments}
We thank David Sutter for fruitful discussions and Daniel Miller for helpful comments and a careful proofread of the manuscript. 
This research has received funding from the European Union’s Horizon 2020 research and innovation program under the Marie Sk\l{}odowska-Curie grant agreement No.~955479 (MOQS – Molecular Quantum Simulations).
This work received financial support from the European
Union’s Horizon 2020 program under Grant Agreement No.
862893 (FET-OPEN project FATMOLS) and from European Union – NextGenerationEU, PNRR MUR project PE0000023-NQSTI.
IBM, the IBM logo, and ibm.com are trademarks of International Business Machines Corp., registered in many jurisdictions worldwide. Other product and service names might be trademarks of IBM or other companies. 
The current list of IBM trademarks is available at \url{https://www.ibm.com/legal/copytrade}.

\newpage
\appendix

\section{Details on the numerical model}
\label{app:numerical_model}
Here, we summarize the technical details of the numerical simulations presented in Sec.~\ref{chap:transmon_qudits} and Sec.~\ref{chap:qec_application}.
Our analysis of the two-transmon system is largely based on Ref.~\cite{malekakhlagh2020firstprinciples}.
We start form the standard transmon Hamiltonian
\begin{align}
\label{eqn:transmon_hamiltonian}
\hat H = 4E_\text{C}\left( \hat{n} - n_g \right)^2 - E_\text{J} \cos(\hat{\phi}), 
\end{align}
where $\hat{n}$ and $\hat{\phi}$ are dimensionless conjugate variables for the charge and phase and $n_g$ represents the offset charge.

\subsection{Treatment of charge noise}
The eigenenergies $E_n$ of $\hat{H}$ are subject to fluctuations with $n_g$, with a maximal value given by the charge dispersion $\Delta E_n = E_n(n_g=0) - E_n(n_g = 1/2)$. 
We compute $\Delta E_n$ by diagonalizing $\hat{H}$ in the charge representation truncating to 20 Fourier modes in $\hat{\phi}$~\cite{gambetta2013quantum}.
We can then estimate the deviation $\Delta \omega^{n,n+1}$ of the transition frequencies $\omega^{n,n+1} = E_{n+1}(n_g) - E_{n}(n_g)$ from their mean as a uniform average over $n_g$. 
For the chosen transmon model parameters with $E_{\text{J}}/E_{\text{C}}\sim70$ as presented in Sec.~\ref{sec:numerical_model}, we obtain 
$\Delta \omega^{01}/(2\pi) = 0.2\,\text{kHz}$, 
$\Delta \omega^{12}/(2\pi) = 7.7\,\text{kHz}$, and
$\Delta \omega^{23}/(2\pi) = 185\,\text{kHz}$.
To account for this uncertainty caused by charge dispersion, we apply $\Delta \omega^{n,n+1}$ as a detuning to each pulse played on the $n\leftrightarrow n+1$ transition. 

\subsection{Model Hamiltonians}
\label{sec:app_model_hamiltonians}

\begin{figure*}
\includegraphics[width=0.99\textwidth]{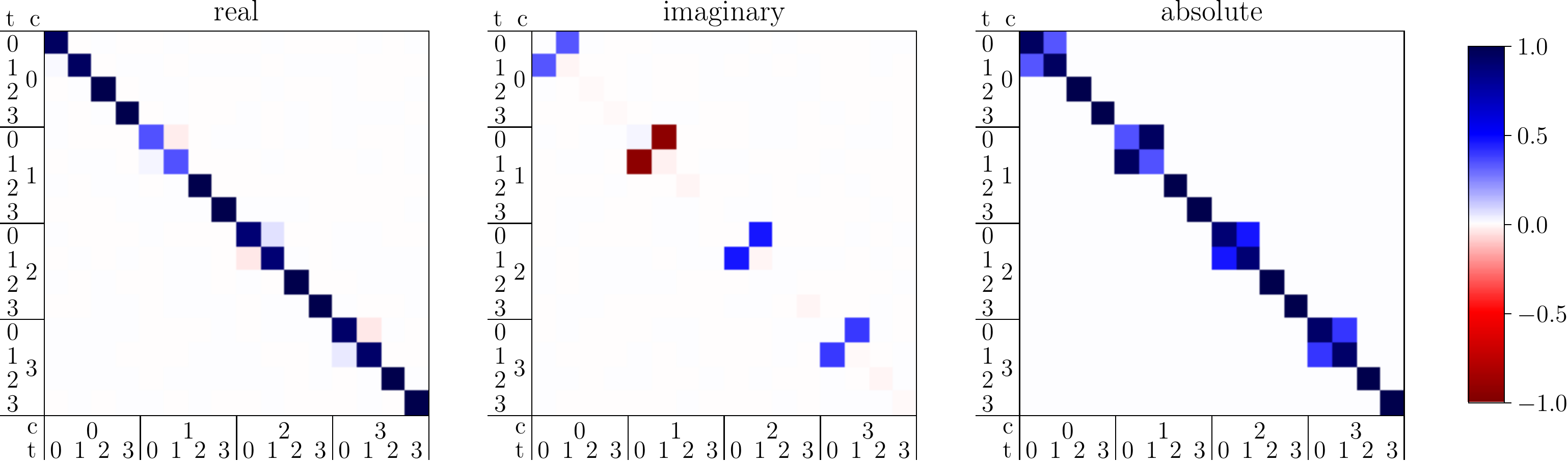}
\caption[]{Numerical simulation of the unitary matrix $\tilde{R}_{\text{CR}}$ of a single CR pulse in the two-transmon dressed basis. The resulting operation is well described by an $R_x^{01}$ rotation on the target with a direction and angle that depend on the state of the control.}
\label{fig:single_cr_pulse_unitary}
\end{figure*}

The transmon Hamiltonian from Eq.~\eqref{eqn:transmon_hamiltonian} is commonly simplified in the Kerr approximation by expanding the cosine term to fourth order. 
Taking also the next-highest order into account, the Hamiltonian in its eigenbasis truncated to the $d=4$ subspace becomes~\cite{malekakhlagh2020firstprinciples}
\begin{align}
\label{eq:single_qudit_approx_hamiltonian}
H^\prime & = \omega \ketbra{1}{1} + \left(2\omega + \alpha\right) \ketbra{2}{2} \\ 
+ &3 \left( \omega + \alpha - \frac{E_{\text{C}}}{8E_{\text{J}}} \sqrt{2E_{\text{C}}E_{\text{J}}} \right) \ketbra{3}{3}
\nonumber
\end{align}
with the qubit frequency $\omega$ and the anharmonicity $\alpha$. 

The capacitive coupling Hamiltonian between control (c) and target (t) is given by $H_J = J y_c \otimes y_t$. 
It is expressed up to $O(\epsilon^3)$ in the unitless anharmonicity measure $\epsilon = \sqrt{2 E_{\text{C}} / E_{\text{J}}} \sim 0.168$ through the unitless charge operator $y = -i(b - b^\dagger)$ with 
\begin{equation}
\label{eqn:lowering_operator}
b = \begin{pmatrix}
0 & b_{01} & 0      & b_{03} \\
0 & 0      & b_{12} & 0      \\
0 & 0      & 0      & b_{23} \\
0 & 0      & 0      & 0
\end{pmatrix}
\end{equation}
and
\begin{subequations}
\begin{align}
b_{01} &= 1 - \frac{\epsilon}{8} - \frac{11 \epsilon^2}{256} \\
b_{12} &= \left(1 - \frac{\epsilon}{4} - \frac{73 \epsilon^2}{512}\right)\sqrt{2} \\
b_{23} &= \left(1 - \frac{3\epsilon}{8} - \frac{79 \epsilon^2}{256}\right)\sqrt{3} \\
b_{03} &= - \frac{\sqrt{6}\epsilon}{16} - \frac{5\sqrt{6} \epsilon^2}{128}.
\end{align}
\end{subequations}
This always-on coupling leads to a dressing of the basis states of the two-transmon system. 
The states $\ket{n}_c \otimes \ket{m}_t$ referred to in the context of the coupled system, denote the dressed eigenstates of the time-independent Hamiltonian
\begin{equation}
\label{eqn:H0}
H_0 = H^\prime_c \otimes \mathbb{1} + \mathbb{1} \otimes H^\prime_t + H_J.
\end{equation}
An external microwave drive with a carrier frequency of $\omega_{\text{d}}$ and pulse envelope $\Omega(t)$ leads to an interaction Hamiltonian (in the rotating wave approximation) of
\begin{equation}
\label{eqn:interaction_hamiltonian}
H_{\text{int}}(t) = \frac{\Omega(t)}{2} \left( b e^{i \omega_{\text{d}} t} + b^\dagger e^{-i\omega_{\text{d}} t} \right).
\end{equation}
We simulate the action of a pulse applied to the control by numerically solving the time-dependent Schr{\"o}dinger equation for the full Hamiltonian $H_{\text{tot}} = H_0 + H_{\text{int}}(t)\otimes \mathbb{1}$ with the master equation solver provided by QuTip~\cite{johansson2012qutip}. 

\subsection{Single CR pulse}
\label{sec:app_single_cr_pulse}

\begin{figure*}
\includegraphics[width=0.99\textwidth]{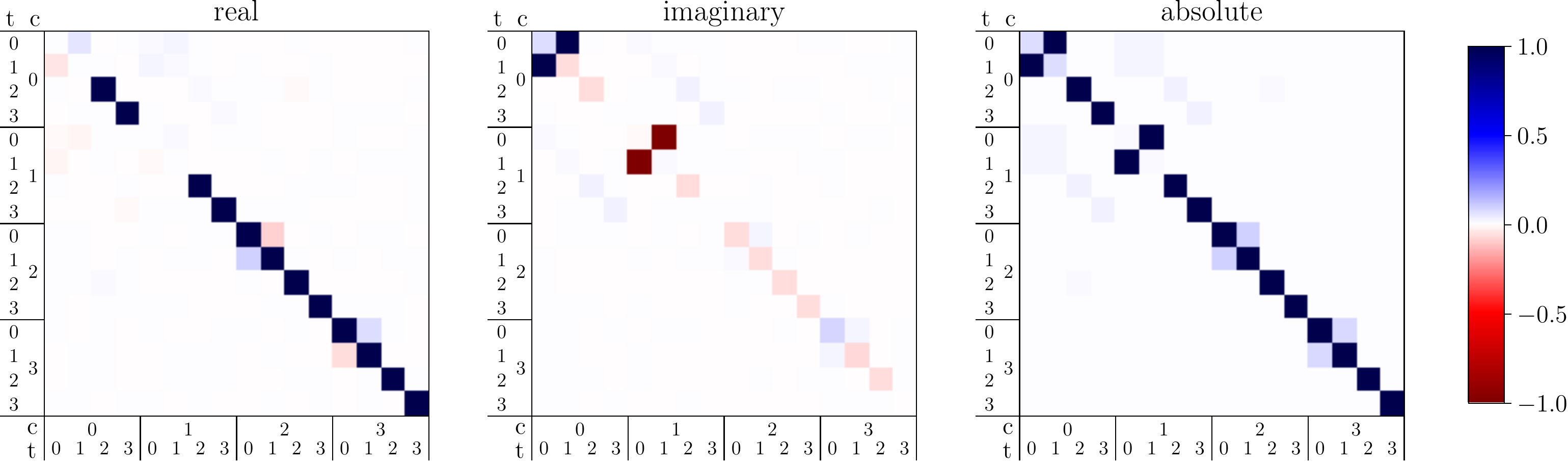}
\caption[]{Numerical simulation of the unitary matrix of the full echoed CR sequence in the two-transmon dressed basis. The resulting operation is well described by an $U_\text{ECR}(\pi)$ gate.}
\label{fig:ecr_pi_unitary}
\end{figure*}

\begin{figure*}
\includegraphics[width=0.99\textwidth]{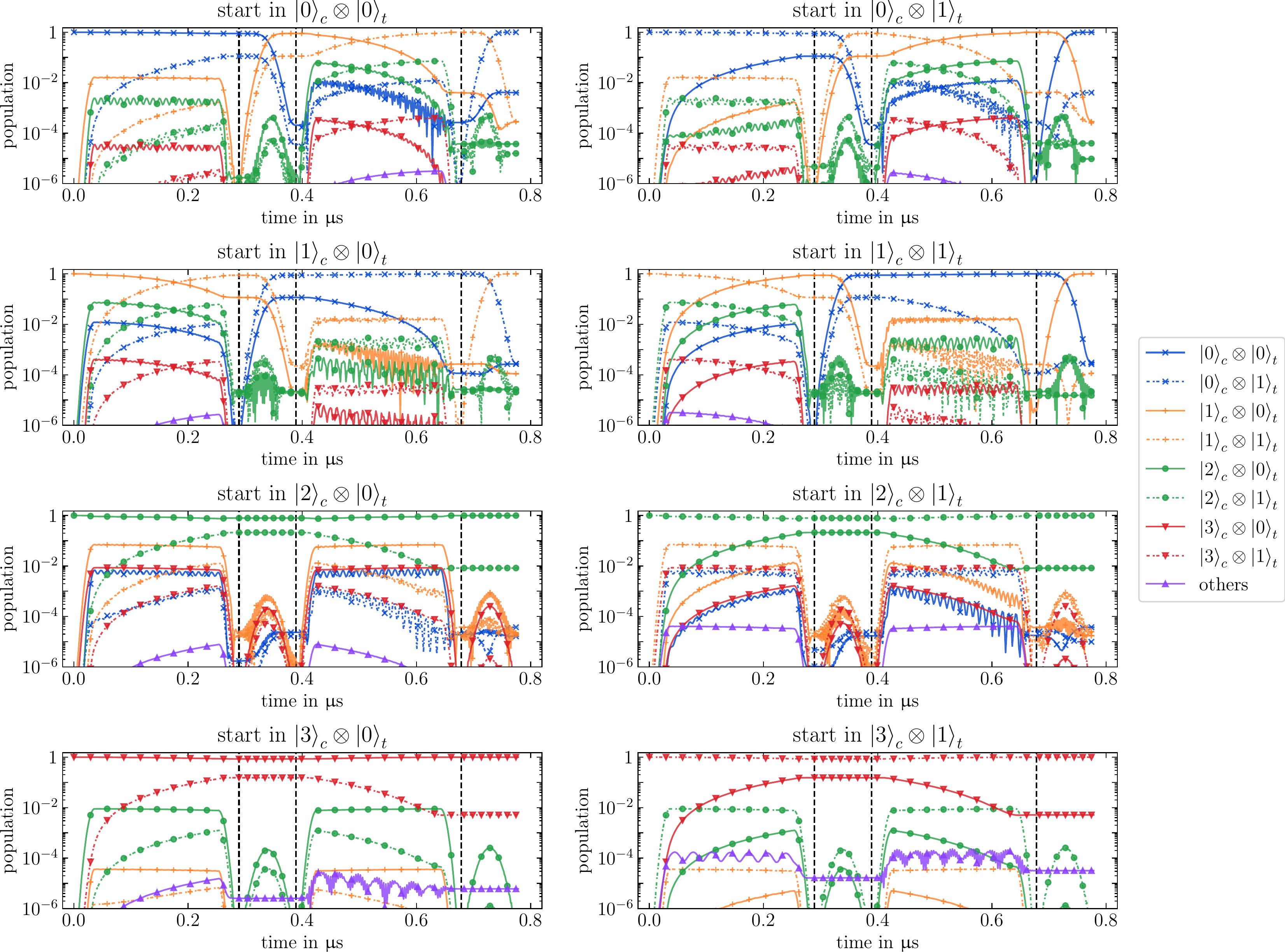}
\caption[]{Evolution of the populations in the two-transmon dressed basis during the echoed CR sequence for different initial states. 
Color indicates the state of the control, while the linestyle indicates the state of the target.
Vertical dashed black lines form four regions that correspond to the four total pulses of the ECR sequence. }
\label{fig:ecr_dynamics_log}
\end{figure*}

All CR pulses we apply to the control are played at a frequency of $\omega_{\text{d}} = \overline{\omega}_{\text{t}}$, i.e., the target qudit frequency in the dressed basis averaged over the four lowest levels of the control qudit. 
We choose a Gaussian-square pulse shape which consists of a plateau of duration $\tau_s$ between a Gaussian rise and fall with duration $\tau_g$ and standard deviation $\sigma$. 
The pulse envelope is thus
\begin{equation}
\label{eqn:gaussian_square_pulse}
\Omega(t) = \tilde\Omega \cdot
\begin{cases}
     \dfrac{e^{-\frac{1}{2}\frac{(t - \tau_g)^2}{\sigma^2}} - \chi}{1  - \chi} &,\,0 < t \leq \tau_g \\
    1 &,\,\tau_g  < t < \tau_g + \tau_s \\
     \dfrac{e^{-\frac{1}{2}\frac{(t - \tau_g - \tau_s)^2}{\sigma^2}} - \chi}{1  - \chi} &,\,\tau_g + \tau_s \leq t < \tau \\    
\end{cases}
\end{equation}
where $\tilde\Omega$ is the maximal amplitude, $\chi = e^{-\frac{1}{2}\frac{(1+\tau_g)^2}{\sigma^2}}$ is a rescaling constant, and $\tau = \tau_s + 2\tau_g$ is the total duration. 
We choose parameter values of $\tilde\Omega/(2\pi) = 50\,\text{MHz}$, $\tau_g=36\,\text{ns}$, and $\sigma = \tau_g/4$ for the simulated CR pulses, as we empirically find that this leads to a tolerable amount of leakage [see Figs.~\labelcref{fig:single_cr_pulse_unitary,fig:ecr_pi_unitary,fig:ecr_dynamics_log}]. 
To calibrate the rotation angles, we fix $ \tilde\Omega $, $\tau_g$, and $\sigma$, adjusting only the width $\tau_s$, and denote the obtained unitary as $R_{\text{CR}}(\tau)$.

We benchmark how well the action of a CR pulse is described by the conditional $R_x^{01}$ rotations from Eq.~\eqref{eqn:single_Cr_pulse_action} (denoted $U_\text{CR}(\vec{\varphi})$) through the average gate fidelity $\fidelity$ as defined in Ref.~\cite{nielsen2002simple}. 
Under the time evolution of the CR pulse, each basis state $\ket{n}_c \otimes \ket{m}_t$ acquires a phase $e^{-i\alpha_{nm}}$, which we assume to be uncorrelated, i.e., $\alpha_{nm} = \alpha_n \alpha_m \,\forall n, m \in \{0, 1, 2, 3\} $.
We apply local phases after the action of the pulse to obtain the phase-corrected unitaries
\begin{align}
\label{eqn:local_phase_correction}
\tilde{R}_{\text{CR}}(\tau) = [&\text{diag}(e^{i \alpha_{c_0}}, e^{i \alpha_{c_1}}, e^{i \alpha_{c_2}}, e^{i \alpha_{c_3}})\,\, \otimes \\
&\text{diag}(e^{i \alpha_{t_0}}, e^{i \alpha_{t_1}}, e^{i \alpha_{t_2}}, e^{i \alpha_{t_3}}) ] R_{\text{CR}}(\tau). \nonumber
\end{align}
In practice, these can be applied virtually on each qudit, as discussed in Sec.~\ref{sec:single_qudit_control}.
For a duration of $\tau_\pi = 289\,\text{ns}$, we numerically optimize the angles $\vec{\varphi}$ and phases $\vec{\alpha_c}$, $\vec{\alpha_t}$, obtaining an optimal fidelity $\fidelity(\tilde{R}_{\text{CR}}(\tau), U_\text{CR}(\vec{\varphi}))  = 99.93\%$.
Such a fidelity is possible since the optimized unitary $\tilde{R}_{\text{CR}}(\tau)$ has a uniform phase structure with vanishing imaginary parts on the diagonal and little leakage, see Fig.~\ref{fig:single_cr_pulse_unitary}, . 
The duration $\tau_\pi$ is calibrated such that $\varphi_0 + \varphi_1 = \pi$, which leads to an echoed CR sequence with a rotation angle of $\theta = \pi$.

\subsection{Single qudit pulses}
\label{sec:app_single_qudit_pulse}

Here, we summarize details on the simulation of single-qudit pulses.
The echoed CR sequence reported in Sec.~\ref{sec:transmon_qudits_simulation_results} requires $R_x^{01}(\pm \pi)$ gates, whereas the single-qudit encoding, decoding, and recovery operations as defined in Sec.~\ref{sec:error_correction_in_qudits} are built from $R_y^{01}(\pm 2\pi/3)$, $R_y^{01}(\pi/3)$, $R_y^{12}(\pm \pi)$, $R_y^{23}(\pm \pi/3)$, and $R_y^{23}(-2\pi/3)$ rotations.
For simplicity, we simulate the single-qudit system of the control with the Hamiltonian $H^\prime_c + H_{\text{int}}(t)$ and set the drive frequency to $\omega_d = \omega_c^{i, i+1}$ to drive $R^{i,i+1}$ rotations. 
We set the phase of the pulse envelope $\Omega(t)$ to define the axis of rotation. 
While a real and positive (negative) $\Omega(t)$ drives $x$ ($-x$) rotations, an imaginary positive (negative) $\Omega(t)$ drives $y$ ($-y$) rotations. 

Each single qudit pulse is played with a Gaussian envelope given by Eq.~\eqref{eqn:gaussian_square_pulse} when setting $\tau_s=0$ and $\sigma = \tau_g/4$ (up to a global phase setting the rotation direction).
We fix the duration of the pulses $\tau_g$ to $ 100\,\text{ns}$, $66.6\,\text{ns}$, and $33.3\,\text{ns}$ for rotation angles $\pm \pi$,  $\pm 2\pi/3$, and $\pm \pi/3$, respectively. 
The amplitude $\tilde{\Omega}$ of the pulses is then tuned such that the area under the envelope matches the desired rotation angle. 
Within this model, we achieve average gate fidelities of $>99.99 \%$ (unitary error) in the ququart subspace for all single-qudit pulses. 

\subsection{Echoed CR sequence}
\label{sec:app_echoed_CR_sequence}

Here, we discuss observations from the simulation of the echoed CR sequence in further detail. 
We simulate the CR pulses individually with the parameters given in Sec.~\ref{sec:app_single_cr_pulse}, applying the second CR pulse of the sequence with a negative amplitude. 
The durations of the CR tones $\tau_s$ are optimized such that the total rotation angle in Eq.~\eqref{eqn:ECR_unitary} is $\theta = \pi$.
The unitaries of the $R_x^{01}(\pi)$ rotations on the control are simulated as discussed in Sec.~\ref{sec:app_single_qudit_pulse}.
We apply local phase corrections for the final unitary to maximize the fidelity with $U_\text{ECR}(\pi)$, similar to Eq.~\eqref{eqn:local_phase_correction}.
The final unitary, which achieves an average gate fidelity of $99.6\%$, is shown in Fig.~\ref{fig:ecr_pi_unitary}. 
To further investigate the dominant contributions to the unitary error, we plot the evolution of all populations throughout the pulse sequence in Fig.~\ref{fig:ecr_dynamics_log}. This is the same data as shown in Fig.~\ref{fig:ECR_pulse_overview}(c), plotted on a logarithmic ordinate to highlight the small contributions in the undesired levels. 

We can identify small non-zero off-diagonal entries in the $\ket{2}_c$ and $\ket{3}_c$ subspaces, which manifest in remaining populations of $0.005$ -- $0.01$ for $\ket{1}_t$ and $\ket{0}_t$ in the bottom two panels of Fig.~\ref{fig:ecr_dynamics_log}.
This means that the second CR pulse does not fully reverse the rotations that were applied to these states by the first CR pulse. 
We attribute this to the fact that the driving frequency $\overline{\omega}_t$ is detuned by $120\,\text{kHz}$ to the $\ket{2}_c \ket{0}_t \leftrightarrow \ket{2}_c \ket{1}_t$ transition and by $-110\,\text{kHz}$
to the $\ket{3}_c \ket{0}_t \leftrightarrow \ket{3}_c \ket{1}_t$ transition~[see Fig.~\ref{fig:two_transmon_level_spectrum}].
Therefore, both the $x$-rotation of the first CR pulse and the $-x$-rotation of the second CR pulse carry a small $Z$-component, which results in a misalignment between the rotation axis of the two CR pulses. 
We also attribute the non-trivial phase structure on the diagonal that is discernible in Fig.~\ref{fig:ecr_pi_unitary} to this effect.
The populations of the higher-excited states of the target, shown in purple in Fig.~\ref{fig:ecr_dynamics_log}, remain below $10^{-4}$ from which we conclude that the chosen pulse shape successfully limits leakage. 

Finally, we summarize qualitatively the effect that the different parameters of our model have on the fidelity of the $\texttt{ECR}$ gate.
The transmon frequencies $\omega_c$, $\omega_t$ and $\alpha_c$, $\alpha_t$ define the level structure of the system. Through the $E_{\text{J}}/E_{\text{C}}$-ratio, this defines the amount of charge dispersion in each level. The closeness of frequencies in the system determine the susceptibility to leakage and crosstalk. 
The maximum drive amplitude $\tilde \Omega$ determines the entangling speed of the gate and has a strong influence on leakage and crosstalk.
With increasing coupling strength $J$ the entangling speed of the gate also increases. 
However, it also increases the dressing of the bare qudit eigenstates, which leads to a dependency of the $\ket{0}_t\leftrightarrow \ket{1}_t$ frequency on the state of the control and limits the unitary of the $\texttt{ECR}$ gate in the idle levels $\ket{2}_c$ and $\ket{3}_c$ as detailed above. 

\subsection{Lindbladian dynamics}
\label{appendix:linblad_dynamics}
We model non-unitary dynamics under a Markovian noise approximation with a Lindblad master equation 
\begin{align}
    \label{eq:lindblad_general}
    \dot{\rho}(t) = -\frac{i}{\hbar}  &\left[H(t),\rho(t) \right]  + \\ &\frac{1}{2} \sum_i  2 L_i \rho(t)  L_i^\dagger - \{  L_i^\dagger L_i, \rho(t) \} \nonumber
\end{align}
where $\{L_i\}$ are a set of Lindblad jump operators. 
In Sec.~\ref{chap:qec_application}, we consider two types of errors for ququarts, pure dephasing and amplitude damping. 
Firstly, we model pure dephasing with a jump operator $L_0 = \sqrt{2/T_2} \sum_{m=0}^3 \ketbra{m}{m}$.
The choice of $T_2$ is motivated by the fact that Eq.~\eqref{eq:lindblad_general} with only pure dephasing leads to a quantum channel $\mathcal{E}(t)$ whose fidelity $\fidelity(\mathcal{E}(t), \mathbb{1})$ in the qubit subspace decays exponentially in time with a time constant of $T_2$. 
Secondly, our model of amplitude damping contains three individual jump operators of the form $L_n = \sqrt{n/T_1}\ketbra{n-1}{n}$ for $n\in \{1, 2, 3\}$.
These lead to an exponential decay of the population in the $n$-th level, with a time constant given by $T_1/n$.
Note that these definitions of $T_1$ and $T_2$ might differ from experimental usage in the context of measuring $T_1$, $T_2$, or $T_2^{*}$ times for a qubit. 
Rather, in this work, $T_1$ and $T_2$ simply characterize the time-scale of dephasing and amplitude damping errors in our model. 
To simulate the action of the two-transmon \texttt{ECR} gate under noise, we include the jump operators on each qudit individually, i.e., adding operators $L_j \otimes \mathbb{1}$ and $\mathbb{1}\otimes L_j$ to Eq.~\eqref{eq:lindblad_general}.

The shortest current measurement pulses on IBM Quantum devices have a duration of $t_{\text{meas}} = 675\,\text{ns}$~\cite{IBMQuantum}.
This affects the error correction sequence of Fig.~\ref{fig:error_correction_sequence} as the data qudit is in a decoded and thus unprotected state during the measurement of the ancilla. 
We account for the measurement duration in our simulations by adding an idle time of $t_{\text{meas}}$ before processing the measurement. 

We compute the quantum channel $\mathcal{E}$ of a pulse sequence by simulating its action on a complete set of (not necessarily physical) input states to obtain the full Liouvillian superoperator, leveraging the quantum information package of Qiskit~\cite{Qiskit}.
The error correction sequence from Fig.~\ref{fig:error_correction_sequence} features a measurement of the ancilla and subsequent classically controlled operations on the data qudit.
Simulating this measurement of the ancilla is problematic for non-physical input states as it can result in invalid measurement probabilities. 
We thus make use of the ``delayed measurement principle''~\cite{nielsen_chuang_2010} and simulate the classically-controlled operations as quantum-controlled operations which are well-defined also for non-physical input states. 

\section{Details on the decomposition of $C^m\left[U\right]$}
\label{app:transpiler}
Here, we point out technical details relating to  the decomposition of a general unitary into the \texttt{ECR} and single-qudit gates and distinguish our work from previous state-of-the-art.  
Our transpilation routine presented in Sec.~\ref{chap:gate_decompositions} builds on the multivalued quantum Shannon decomposition as developed in Ref.~\cite{di2013synthesis}. 
The authors of Ref.~\cite{di2013synthesis} show how to synthesize arbitrary unitaries with an $m$-controlled $X$ gate
\begin{equation}
\label{eq:def_gcx}
\textsc{gcx} = C^m[\ketbra{0}{1} + \ketbra{1}{0} + \sum_{k=2}^d \ketbra{k}{k}]
\end{equation}
as the fundamental entangling gate.   
One might assume that existing implementations of the \textsc{Cnot} gate naturally generalize to the \textsc{gcx} gate. 
However, this is not the case for both direct and echoed cross-resonance \textsc{Cnot} gates, as discussed below.
This necessitates the modification we propose in Sec.~\ref{chap:gate_decompositions} to the transpilation provided in Ref.~\cite{di2013synthesis}.

As discussed in the main text, the cross-resonance effect applies an $R_x^{01}(\varphi_j)$ rotation to the target qudit, where $\varphi_j$ depends on the control state $\ket{j}_c$ and the gate duration, see Eq.~\eqref{eqn:single_Cr_pulse_action}.
To implement a \emph{direct} \textsc{Cnot}, the angles are tuned such that $\varphi_0 - \varphi_1 = \pi$. 
In the qubit subspace, the resulting unitary is equivalent to the \textsc{Cnot} up to local operations. 
However, in contrast to Eq.~\eqref{eq:def_gcx}, the gate acts non-trivially on higher-excited states of the control; an effect that can not be reversed by single-qudit gates. 
For qutrits, this can be compensated for by tuning the drive strength such that $\varphi_0 = \varphi_2$, as demonstrated in Ref.~\cite{blok2021quantum}.
This approach does not scale to higher qudit dimension. 

The action of higher-excited control states can be cancelled by an echoed CR sequence, leading to the $U_\text{ECR}(\theta)$ gate defined in Eq.~\eqref{eqn:ECR_unitary}. 
While $U_\text{ECR}(\pi/2)$ is local-Clifford equivalent to the \textsc{Cnot} for qubits, it is again not possible to transform $U_\text{ECR}(\pi/2)$ to the \textsc{gcx} gate with local operations. 
As shown in Sec.~\ref{chap:gate_decompositions}, $d-1$ $U_\text{ECR}(-\pi/d)$ gates can be combined to synthesize a $C^m[R_x^{01}(\pi)]$ gate.
However, there is still a subtle difference to the \textsc{gcx} gate. 
A $C^m[R_x^{01}(\pi)]$ gate introduces a relative phase factor of $i$ between the qubit subspace and higher-excited states, as
\begin{equation}
\label{eq:CRX_GCX_comparison}
R_x^{01}(\pm \pi) =\mp i \ketbra{0}{1} \mp i \ketbra{1}{0} + \sum_{k=2}^d \ketbra{k}{k}.
\end{equation}
Cancelling this relative phase on the target requires a controlled phase gate.
The unitary synthesis proposed in Ref.~\cite{di2013synthesis} is thus not suitable for architectures that implement $m$-controlled $x$ rotations, such as the \texttt{ECR} gate. 
We circumvent this issue by using pairs of $C^m[R_x^{01}(\pi)]$ and $C^m[R_x^{01}(-\pi)]$ gates instead of the \textsc{gcx} gate. 

\section{General unitary synthesis with $C^m\left[U\right]$}
\label{app:general_synthesis}

In Sec.~\ref{chap:gate_decompositions} we present a decomposition of $C^m[U]$ gates into ECR pulses.
However, making a quantum computer based on ququarts requires realizing general ${\rm SU}(16)$ gates.
We now show how to do this with bidirectional $C^m[U]$ gates.
We decompose a general unitary $U$ into a product of $C^m[U]$ gates with an iterative cosine-sine decomposition (CSD)~\cite{chen2013qcompiler}.
In general, a cosine-sine decomposition of a $2^n\times 2^n$ matrix $U$ is
\begin{align}
    U=
    \begin{pmatrix}
    u  & \\
    & v
    \end{pmatrix}
    \begin{pmatrix}
    C  & -S \\
    S & C
    \end{pmatrix}
    \begin{pmatrix}
    x  & \\
    & y
    \end{pmatrix},
\end{align}
with $C={\rm diag}(\cos\theta_1, ..., \cos\theta_{2^{n-1}})$, $S={\rm diag}(\sin\theta_1,$ $ ..., \sin\theta_{2^n-1})$.
Here, $u$, $v$, $x$, and $y$ are $2^{n-1}\times 2^{n-1}$ unitary matrices that can be further decomposed.
We label two quqaurts according to $\textit{control}\,\otimes\,\textit{target}$. 
Applying the CSD five times to a dense two-ququart unitary $U$ yields
\begin{widetext}
\begin{align}
\label{eq:cosinesine_decomp}
U =  
\left( \arraycolsep=0pt \begin{array}{cccc} 
    \setlength{\fboxsep}{1.5pt} \fbox{$U_0$}  &  &  & \\ 
       & \setlength{\fboxsep}{1.5pt} \fbox{$U_1$} &  & \\
       &  &  \setlength{\fboxsep}{1.5pt} \fbox{$U_2$} & \\ 
       &  &  & \setlength{\fboxsep}{1.5pt} \fbox{$U_3$}  
    \end{array} \right) 
\left( \arraycolsep=0pt \begin{array}{cc} 
    \dboxed{
    \arraycolsep=2pt \begin{array}{|c|c|} 
    \hline
    C_0 & -S_0  \\ \hline 
    S_0 & C_0  \\ \hline
     \end{array}
    }{2} & \\
     & \dboxed{
    \arraycolsep=2pt \begin{array}{|c|c|} 
    \hline
    C_1 & -S_1  \\ \hline 
    S_1 & C_1  \\ \hline
     \end{array}
    }{2} \end{array} \right) 
&\left( \arraycolsep=0pt \begin{array}{cccc} 
    \setlength{\fboxsep}{1.5pt} \fbox{$U_4$}  &  &  & \\ 
       & \setlength{\fboxsep}{1.5pt} \fbox{$U_5$} &  & \\
       &  &  \setlength{\fboxsep}{1.5pt} \fbox{$U_6$} & \\ 
       &  &  & \setlength{\fboxsep}{1.5pt} \fbox{$U_7$}  
    \end{array} \right) 
\left(  \vspace{-15pt}\arraycolsep=0pt \begin{array}{cc} 
    \dbox{\begin{minipage}[t][20pt][t]{22pt}\vspace{5pt}\large{$\,\,C_2$}\end{minipage}}  & \dbox{\begin{minipage}[t][20pt][t]{22pt}\vspace{5pt}\large{$-S_2$}\end{minipage}}     
    \vspace{-5pt} \\ 
    \dbox{\begin{minipage}[t][20pt][t]{22pt}\vspace{5pt}\large{$\,\,S_2$}\end{minipage}}  & \dbox{\begin{minipage}[t][20pt][t]{22pt}\vspace{5pt}\large{$\,\,C_2$}\end{minipage}}  
    \end{array} \right) \\ \nonumber
    \times
&\left( \arraycolsep=0pt \begin{array}{cccc} 
    \setlength{\fboxsep}{1.5pt} \fbox{$U_8$}  &  &  & \\ 
    & \setlength{\fboxsep}{1.5pt} \fbox{$U_{9}$} &  & \\
    &  &  \setlength{\fboxsep}{1.5pt} \fbox{$U_{10}$} & \\ 
    &  &  & \setlength{\fboxsep}{1.5pt} \fbox{$U_{11}$}  
    \end{array} \right) 
\left( \arraycolsep=0pt \begin{array}{cc} 
    \dboxed{
    \arraycolsep=2pt \begin{array}{|c|c|} 
    \hline
    C_3 & -S_3  \\ \hline 
    S_3 & C_3  \\ \hline
     \end{array}
    }{2} & \\
     & \dboxed{
    \arraycolsep=2pt \begin{array}{|c|c|} 
    \hline
    C_4 & -S_4  \\ \hline 
    S_4 & C_4  \\ \hline
     \end{array}
    }{2} \end{array} \right) 
\left( \arraycolsep=0pt \begin{array}{cccc} 
    \setlength{\fboxsep}{1.5pt} \fbox{$U_{12}$}  &  &  & \\ 
    & \setlength{\fboxsep}{1.5pt} \fbox{$U_{13}$} &  & \\
    &  &  \setlength{\fboxsep}{1.5pt} \fbox{$U_{14}$} & \\ 
    &  &  & \setlength{\fboxsep}{1.5pt} \fbox{$U_{15}$}  
\end{array} \right). 
\end{align} 
\end{widetext}
Here, solid and dashed boxes indicate $4\times4$ and $8\times8$ subspaces, respectively.  
$C_i$ and $S_i$ are diagonal cosine and sine matrices defined by  $C_i = \text{diag}(\cos \theta_{i_0}, \dots, \cos \theta_{i_{N-1}})$ and $S_i = \text{diag}(\sin\theta_{i_0}, \dots, \sin\theta_{i_{N-1}})$ for some angles $\theta_j$ with $N=4$ for $i\in\{0, 1, 3, 4\}$ and $N=8$ for $i=2$.
Crucially, the $U_i$ are arbitrary four-dimensional unitaries and the block-diagonal matrices of which they are blocks are directly realized by a sequence of $C^m[U]$ gates as derived in the main text.
For example, the first block-diagonal matrix in Eq.~\eqref{eq:cosinesine_decomp} is implemented by $\prod_{i=0}^3 C^i[U_i]$.

Next, we implement the cosine and sine matrices with $R_y^{ij}$ rotations applied to the control qubit depending on the state of the target.
This requires a bidirectional \texttt{ECR} gate such that the role of control and target can be reversed. 
For example, the first block of the second matrix in Eq.~\eqref{eq:cosinesine_decomp} is 
\begin{equation}
\label{eq:example_cosinesine}
    \dboxed{
        \arraycolsep=2pt \begin{array}{|c|c|} 
        \hline
        C_0 & -S_0  \\ \hline 
        S_0 & C_0  \\ \hline
         \end{array}
        }{2}
    = \sum_{n=0}^3 R_y^{01}(\theta_{0_{n}}) \otimes \ketbra{n}{n}_t
\end{equation}
while the center matrix is
\begin{align}
\vspace{-15pt}\arraycolsep=0pt \begin{array}{cc} 
    \dbox{\begin{minipage}[t][15pt][t]{17pt}\vspace{5pt}{$\,\,C_2$}\end{minipage}}  & \dbox{\begin{minipage}[t][15pt][t]{17pt}\vspace{5pt}{$-S_2$}\end{minipage}}     
    \vspace{-5pt} \\ 
    \dbox{\begin{minipage}[t][15pt][t]{17pt}\vspace{5pt}{$\,\,S_2$}\end{minipage}}  & \dbox{\begin{minipage}[t][15pt][t]{17pt}\vspace{5pt}{$\,\,C_2$}\end{minipage}}  
    \end{array} 
    = \sum_{n=0}^3 R_y^{02}(\theta_{2_{n}}) R_y^{13}(\theta_{2_{4+n}}) \otimes \ketbra{n}{n}_t .
\end{align}
Up to a phase on the \texttt{ECR} drive pulses, these controlled $R_y$ rotations are equivalent to the $C^m[R_x^{0j}]$ rotations derived in Sec.~\ref{sec:compiler_comparison_qubits}, and can be accomplished through the decomposition shown in Fig.~\ref{fig:decomposition_circuits}(c).
Since our \texttt{ECR} only operates in the $\ket{0}/\ket{1}$-subspaces of the control and the target, this necessitates single-qudit permutation gates on both the control and target. 
These additional permutation gates needed to shift the control gates are accounted for in the gate counts presented in Sec.~\ref{sec:compiler_comparison_qubits}.

The CSD of an arbitrary ${\rm SU}(16)$ gate shown here serves as the building block for an arbitrary quantum circuit that is executed on a ququart-based hardware.
It requires a total of 340 \texttt{ECR} pulses and, as discussed in Sec.~\ref{sec:compiler_comparison_qubits}, is equivalent to 170 CNOT gates in a qubit-based architecture.


\begin{thebibliography}{95}%
\makeatletter
\providecommand \@ifxundefined [1]{%
 \@ifx{#1\undefined}
}%
\providecommand \@ifnum [1]{%
 \ifnum #1\expandafter \@firstoftwo
 \else \expandafter \@secondoftwo
 \fi
}%
\providecommand \@ifx [1]{%
 \ifx #1\expandafter \@firstoftwo
 \else \expandafter \@secondoftwo
 \fi
}%
\providecommand \natexlab [1]{#1}%
\providecommand \enquote  [1]{``#1''}%
\providecommand \bibnamefont  [1]{#1}%
\providecommand \bibfnamefont [1]{#1}%
\providecommand \citenamefont [1]{#1}%
\providecommand \href@noop [0]{\@secondoftwo}%
\providecommand \href [0]{\begingroup \@sanitize@url \@href}%
\providecommand \@href[1]{\@@startlink{#1}\@@href}%
\providecommand \@@href[1]{\endgroup#1\@@endlink}%
\providecommand \@sanitize@url [0]{\catcode `\\12\catcode `\$12\catcode
  `\&12\catcode `\#12\catcode `\^12\catcode `\_12\catcode `\%12\relax}%
\providecommand \@@startlink[1]{}%
\providecommand \@@endlink[0]{}%
\providecommand \url  [0]{\begingroup\@sanitize@url \@url }%
\providecommand \@url [1]{\endgroup\@href {#1}{\urlprefix }}%
\providecommand \urlprefix  [0]{URL }%
\providecommand \Eprint [0]{\href }%
\providecommand \doibase [0]{http://dx.doi.org/}%
\providecommand \selectlanguage [0]{\@gobble}%
\providecommand \bibinfo  [0]{\@secondoftwo}%
\providecommand \bibfield  [0]{\@secondoftwo}%
\providecommand \translation [1]{[#1]}%
\providecommand \BibitemOpen [0]{}%
\providecommand \bibitemStop [0]{}%
\providecommand \bibitemNoStop [0]{.\EOS\space}%
\providecommand \EOS [0]{\spacefactor3000\relax}%
\providecommand \BibitemShut  [1]{\csname bibitem#1\endcsname}%
\let\auto@bib@innerbib\@empty
\bibitem [{\citenamefont {Bruzewicz}\ \emph {et~al.}(2019)\citenamefont
  {Bruzewicz}, \citenamefont {Chiaverini}, \citenamefont {McConnell},\ and\
  \citenamefont {Sage}}]{bruzewicz2019trappedion}%
  \BibitemOpen
  \bibfield  {author} {\bibinfo {author} {\bibfnamefont {Colin~D.}\
  \bibnamefont {Bruzewicz}}, \bibinfo {author} {\bibfnamefont {John}\
  \bibnamefont {Chiaverini}}, \bibinfo {author} {\bibfnamefont {Robert}\
  \bibnamefont {McConnell}}, \ and\ \bibinfo {author} {\bibfnamefont
  {Jeremy~M.}\ \bibnamefont {Sage}},\ }\bibfield  {title} {\enquote {\bibinfo
  {title} {Trapped-ion quantum computing: {{Pprogress}} and challenges},}\
  }\href {\doibase 10.1063/1.5088164} {\bibfield  {journal} {\bibinfo
  {journal} {Applied Physics Reviews}\ }\textbf {\bibinfo {volume} {6}},\
  \bibinfo {pages} {021314} (\bibinfo {year} {2019})}\BibitemShut {NoStop}%
\bibitem [{\citenamefont {Clarke}\ and\ \citenamefont
  {Wilhelm}(2008)}]{clarke2008superconducting}%
  \BibitemOpen
  \bibfield  {author} {\bibinfo {author} {\bibfnamefont {John}\ \bibnamefont
  {Clarke}}\ and\ \bibinfo {author} {\bibfnamefont {Frank~K.}\ \bibnamefont
  {Wilhelm}},\ }\bibfield  {title} {\enquote {\bibinfo {title} {Superconducting
  quantum bits},}\ }\href {\doibase 10.1038/nature07128} {\bibfield  {journal}
  {\bibinfo  {journal} {Nature}\ }\textbf {\bibinfo {volume} {453}},\ \bibinfo
  {pages} {1031--1042} (\bibinfo {year} {2008})}\BibitemShut {NoStop}%
\bibitem [{\citenamefont {Shi}(2021)}]{shiQuantumLogicEntanglement2021}%
  \BibitemOpen
  \bibfield  {author} {\bibinfo {author} {\bibfnamefont {Xiaofeng}\
  \bibnamefont {Shi}},\ }\bibfield  {title} {\enquote {\bibinfo {title}
  {Quantum logic and entanglement by neutral {{Rydberg}} atoms: Methods and
  fidelity},}\ }\href {\doibase 10.1088/2058-9565/ac18b8} {\bibfield  {journal}
  {\bibinfo  {journal} {Quantum Science and Technology}\ }\textbf {\bibinfo
  {volume} {7}},\ \bibinfo {pages} {023002} (\bibinfo {year}
  {2021})}\BibitemShut {NoStop}%
\bibitem [{\citenamefont {Burkard}\ \emph {et~al.}(2021)\citenamefont
  {Burkard}, \citenamefont {Ladd}, \citenamefont {Nichol}, \citenamefont
  {Pan},\ and\ \citenamefont {Petta}}]{burkard2021semiconductor}%
  \BibitemOpen
  \bibfield  {author} {\bibinfo {author} {\bibfnamefont {Guido}\ \bibnamefont
  {Burkard}}, \bibinfo {author} {\bibfnamefont {Thaddeus~D.}\ \bibnamefont
  {Ladd}}, \bibinfo {author} {\bibfnamefont {John~M.}\ \bibnamefont {Nichol}},
  \bibinfo {author} {\bibfnamefont {Andrew}\ \bibnamefont {Pan}}, \ and\
  \bibinfo {author} {\bibfnamefont {Jason~R.}\ \bibnamefont {Petta}},\
  }\bibfield  {title} {\enquote {\bibinfo {title} {Semiconductor {{spin
  qubits}}},}\ }\href {http://arxiv.org/abs/2112.08863} {\bibfield  {journal}
  {\bibinfo  {journal} {arXiv:2112.08863}\ } (\bibinfo {year}
  {2021})}\BibitemShut {NoStop}%
\bibitem [{\citenamefont {Di}\ and\ \citenamefont {Wei}(2015)}]{di2015optimal}%
  \BibitemOpen
  \bibfield  {author} {\bibinfo {author} {\bibfnamefont {Yao-Min}\ \bibnamefont
  {Di}}\ and\ \bibinfo {author} {\bibfnamefont {Hai-Rui}\ \bibnamefont {Wei}},\
  }\bibfield  {title} {\enquote {\bibinfo {title} {Optimal synthesis of
  multivalued quantum circuits},}\ }\href {\doibase 10.1103/PhysRevA.92.062317}
  {\bibfield  {journal} {\bibinfo  {journal} {Physical Review A}\ }\textbf
  {\bibinfo {volume} {92}},\ \bibinfo {pages} {062317} (\bibinfo {year}
  {2015})}\BibitemShut {NoStop}%
\bibitem [{\citenamefont {Gokhale}\ \emph {et~al.}(2019)\citenamefont
  {Gokhale}, \citenamefont {Baker}, \citenamefont {Duckering}, \citenamefont
  {Brown}, \citenamefont {Brown},\ and\ \citenamefont
  {Chong}}]{gokhale2019asymptotic}%
  \BibitemOpen
  \bibfield  {author} {\bibinfo {author} {\bibfnamefont {Pranav}\ \bibnamefont
  {Gokhale}}, \bibinfo {author} {\bibfnamefont {Jonathan~M.}\ \bibnamefont
  {Baker}}, \bibinfo {author} {\bibfnamefont {Casey}\ \bibnamefont
  {Duckering}}, \bibinfo {author} {\bibfnamefont {Natalie~C.}\ \bibnamefont
  {Brown}}, \bibinfo {author} {\bibfnamefont {Kenneth~R.}\ \bibnamefont
  {Brown}}, \ and\ \bibinfo {author} {\bibfnamefont {Frederic~T.}\ \bibnamefont
  {Chong}},\ }\bibfield  {title} {\enquote {\bibinfo {title} {Asymptotic
  improvements to quantum circuits via qutrits},}\ }in\ \href {\doibase
  10.1145/3307650.3322253} {\emph {\bibinfo {booktitle} {Proceedings of the
  46th {International} {Symposium} on {Computer} {Architecture}}}}\ (\bibinfo
  {publisher} {ACM},\ \bibinfo {address} {Phoenix Arizona},\ \bibinfo {year}
  {2019})\ pp.\ \bibinfo {pages} {554--566}\BibitemShut {NoStop}%
\bibitem [{\citenamefont {Nikolaeva}\ \emph
  {et~al.}(2022{\natexlab{a}})\citenamefont {Nikolaeva}, \citenamefont
  {Kiktenko},\ and\ \citenamefont {Fedorov}}]{nikolaeva2022efficient}%
  \BibitemOpen
  \bibfield  {author} {\bibinfo {author} {\bibfnamefont {Anastasiia~S.}\
  \bibnamefont {Nikolaeva}}, \bibinfo {author} {\bibfnamefont {Evgeniy~O.}\
  \bibnamefont {Kiktenko}}, \ and\ \bibinfo {author} {\bibfnamefont
  {Aleksey~K.}\ \bibnamefont {Fedorov}},\ }\bibfield  {title} {\enquote
  {\bibinfo {title} {Efficient realization of quantum algorithms with
  qudits},}\ }\href {http://arxiv.org/abs/2111.04384} {\bibfield  {journal}
  {\bibinfo  {journal} {arXiv:2111.04384}\ } (\bibinfo {year}
  {2022}{\natexlab{a}})}\BibitemShut {NoStop}%
\bibitem [{\citenamefont {Wang}\ \emph {et~al.}(2020)\citenamefont {Wang},
  \citenamefont {Hu}, \citenamefont {Sanders},\ and\ \citenamefont
  {Kais}}]{wang2020qudits}%
  \BibitemOpen
  \bibfield  {author} {\bibinfo {author} {\bibfnamefont {Yuchen}\ \bibnamefont
  {Wang}}, \bibinfo {author} {\bibfnamefont {Zixuan}\ \bibnamefont {Hu}},
  \bibinfo {author} {\bibfnamefont {Barry~C.}\ \bibnamefont {Sanders}}, \ and\
  \bibinfo {author} {\bibfnamefont {Sabre}\ \bibnamefont {Kais}},\ }\bibfield
  {title} {\enquote {\bibinfo {title} {Qudits and {high}-{dimensional}
  {quantum} {computing}},}\ }\href {\doibase 10.3389/fphy.2020.589504}
  {\bibfield  {journal} {\bibinfo  {journal} {Frontiers in Physics}\ }\textbf
  {\bibinfo {volume} {8}},\ \bibinfo {pages} {589504} (\bibinfo {year}
  {2020})}\BibitemShut {NoStop}%
\bibitem [{\citenamefont {Tacchino}\ \emph {et~al.}(2021)\citenamefont
  {Tacchino}, \citenamefont {Chiesa}, \citenamefont {Sessoli}, \citenamefont
  {Tavernelli},\ and\ \citenamefont {Carretta}}]{tacchino2021proposal}%
  \BibitemOpen
  \bibfield  {author} {\bibinfo {author} {\bibfnamefont {Fracesco}\
  \bibnamefont {Tacchino}}, \bibinfo {author} {\bibfnamefont {Alessandro}\
  \bibnamefont {Chiesa}}, \bibinfo {author} {\bibfnamefont {Roberta}\
  \bibnamefont {Sessoli}}, \bibinfo {author} {\bibfnamefont {Ivano}\
  \bibnamefont {Tavernelli}}, \ and\ \bibinfo {author} {\bibfnamefont
  {Stefano}\ \bibnamefont {Carretta}},\ }\bibfield  {title} {\enquote {\bibinfo
  {title} {A proposal for using molecular spin qudits as quantum simulators of
  light\textendash matter interactions},}\ }\href {\doibase 10.1039/D1TC00851J}
  {\bibfield  {journal} {\bibinfo  {journal} {Journal of Materials Chemistry
  C}\ }\textbf {\bibinfo {volume} {9}},\ \bibinfo {pages} {10266--10275}
  (\bibinfo {year} {2021})}\BibitemShut {NoStop}%
\bibitem [{\citenamefont {Ollitrault}\ \emph
  {et~al.}(2020{\natexlab{a}})\citenamefont {Ollitrault}, \citenamefont
  {Mazzola},\ and\ \citenamefont {Tavernelli}}]{mazzolag2020}%
  \BibitemOpen
  \bibfield  {author} {\bibinfo {author} {\bibfnamefont {Pauline~J.}\
  \bibnamefont {Ollitrault}}, \bibinfo {author} {\bibfnamefont {Guglielmo}\
  \bibnamefont {Mazzola}}, \ and\ \bibinfo {author} {\bibfnamefont {Ivano}\
  \bibnamefont {Tavernelli}},\ }\bibfield  {title} {\enquote {\bibinfo {title}
  {Nonadiabatic molecular quantum dynamics with quantum computers},}\ }\href
  {https://doi.org/10.1103/PhysRevLett.125.260511} {\bibfield  {journal}
  {\bibinfo  {journal} {Phys. Rev. Lett.}\ }\textbf {\bibinfo {volume} {125}},\
  \bibinfo {pages} {260511} (\bibinfo {year} {2020}{\natexlab{a}})}\BibitemShut
  {NoStop}%
\bibitem [{\citenamefont {Miessen}\ \emph {et~al.}(2021)\citenamefont
  {Miessen}, \citenamefont {Ollitrault},\ and\ \citenamefont
  {Tavernelli}}]{miessen2021spin-boson}%
  \BibitemOpen
  \bibfield  {author} {\bibinfo {author} {\bibfnamefont {Alexander}\
  \bibnamefont {Miessen}}, \bibinfo {author} {\bibfnamefont {Pauline~J.}\
  \bibnamefont {Ollitrault}}, \ and\ \bibinfo {author} {\bibfnamefont {Ivano}\
  \bibnamefont {Tavernelli}},\ }\bibfield  {title} {\enquote {\bibinfo {title}
  {Quantum algorithms for quantum dynamics: A performance study on the
  spin-boson model},}\ }\href
  {https://doi.org/10.1103/PhysRevResearch.3.043212} {\bibfield  {journal}
  {\bibinfo  {journal} {Phys. Rev. Research}\ }\textbf {\bibinfo {volume}
  {3}},\ \bibinfo {pages} {043212} (\bibinfo {year} {2021})}\BibitemShut
  {NoStop}%
\bibitem [{\citenamefont {Rico}\ \emph {et~al.}(2018)\citenamefont {Rico},
  \citenamefont {Dalmonte}, \citenamefont {Zoller}, \citenamefont {Banerjee},
  \citenamefont {B{\"o}gli}, \citenamefont {Stebler},\ and\ \citenamefont
  {Wiese}}]{rico2018nuclear}%
  \BibitemOpen
  \bibfield  {author} {\bibinfo {author} {\bibfnamefont {Enrique}\ \bibnamefont
  {Rico}}, \bibinfo {author} {\bibfnamefont {Marcello}\ \bibnamefont
  {Dalmonte}}, \bibinfo {author} {\bibfnamefont {Peter}\ \bibnamefont
  {Zoller}}, \bibinfo {author} {\bibfnamefont {Debarghya}\ \bibnamefont
  {Banerjee}}, \bibinfo {author} {\bibfnamefont {Michael}\ \bibnamefont
  {B{\"o}gli}}, \bibinfo {author} {\bibfnamefont {Pascal}\ \bibnamefont
  {Stebler}}, \ and\ \bibinfo {author} {\bibfnamefont {Uwe-Jens}\ \bibnamefont
  {Wiese}},\ }\bibfield  {title} {\enquote {\bibinfo {title} {S{O}(3)
  “{nuclear} {physics}” with ultracold {gases}},}\ }\href {\doibase
  10.1016/j.aop.2018.03.020} {\bibfield  {journal} {\bibinfo  {journal} {Annals
  of Physics}\ }\textbf {\bibinfo {volume} {393}},\ \bibinfo {pages} {466--483}
  (\bibinfo {year} {2018})}\BibitemShut {NoStop}%
\bibitem [{\citenamefont {Mathis}\ \emph {et~al.}(2020)\citenamefont {Mathis},
  \citenamefont {Mazzola},\ and\ \citenamefont {Tavernelli}}]{mathis2020}%
  \BibitemOpen
  \bibfield  {author} {\bibinfo {author} {\bibfnamefont {Simon~V.}\
  \bibnamefont {Mathis}}, \bibinfo {author} {\bibfnamefont {Guglielmo}\
  \bibnamefont {Mazzola}}, \ and\ \bibinfo {author} {\bibfnamefont {Ivano}\
  \bibnamefont {Tavernelli}},\ }\bibfield  {title} {\enquote {\bibinfo {title}
  {Toward scalable simulations of lattice gauge theories on quantum
  computers},}\ }\href {https://doi.org/10.1103/PhysRevD.102.094501} {\bibfield
   {journal} {\bibinfo  {journal} {Phys. Rev. D}\ }\textbf {\bibinfo {volume}
  {102}},\ \bibinfo {pages} {094501} (\bibinfo {year} {2020})}\BibitemShut
  {NoStop}%
\bibitem [{\citenamefont {Mazzola}\ \emph {et~al.}(2021)\citenamefont
  {Mazzola}, \citenamefont {Mathis}, \citenamefont {Mazzola},\ and\
  \citenamefont {Tavernelli}}]{mazzola2021}%
  \BibitemOpen
  \bibfield  {author} {\bibinfo {author} {\bibfnamefont {Giulia}\ \bibnamefont
  {Mazzola}}, \bibinfo {author} {\bibfnamefont {Simon~V.}\ \bibnamefont
  {Mathis}}, \bibinfo {author} {\bibfnamefont {Guglielmo}\ \bibnamefont
  {Mazzola}}, \ and\ \bibinfo {author} {\bibfnamefont {Ivano}\ \bibnamefont
  {Tavernelli}},\ }\bibfield  {title} {\enquote {\bibinfo {title}
  {Gauge-invariant quantum circuits for $u$(1) and yang-mills lattice gauge
  theories},}\ }\href {https://doi.org/10.1103/PhysRevResearch.3.043209}
  {\bibfield  {journal} {\bibinfo  {journal} {Phys. Rev. Research}\ }\textbf
  {\bibinfo {volume} {3}},\ \bibinfo {pages} {043209} (\bibinfo {year}
  {2021})}\BibitemShut {NoStop}%
\bibitem [{\citenamefont {Ollitrault}\ \emph
  {et~al.}(2020{\natexlab{b}})\citenamefont {Ollitrault}, \citenamefont
  {Baiardi}, \citenamefont {Reiher},\ and\ \citenamefont
  {Tavernelli}}]{ollitrault_vib2020}%
  \BibitemOpen
  \bibfield  {author} {\bibinfo {author} {\bibfnamefont {Pauline~J.}\
  \bibnamefont {Ollitrault}}, \bibinfo {author} {\bibfnamefont {Alberto}\
  \bibnamefont {Baiardi}}, \bibinfo {author} {\bibfnamefont {Markus}\
  \bibnamefont {Reiher}}, \ and\ \bibinfo {author} {\bibfnamefont {Ivano}\
  \bibnamefont {Tavernelli}},\ }\bibfield  {title} {\enquote {\bibinfo {title}
  {Hardware efficient quantum algorithms for vibrational structure
  calculations},}\ }\href {https://doi.org/10.1039/D0SC01908A} {\bibfield
  {journal} {\bibinfo  {journal} {Chem. Sci.}\ }\textbf {\bibinfo {volume}
  {11}},\ \bibinfo {pages} {6842--6855} (\bibinfo {year}
  {2020}{\natexlab{b}})}\BibitemShut {NoStop}%
\bibitem [{\citenamefont {MacDonell}\ \emph {et~al.}(2021)\citenamefont
  {MacDonell}, \citenamefont {Dickerson}, \citenamefont {Birch}, \citenamefont
  {Kumar}, \citenamefont {Edmunds}, \citenamefont {Biercuk}, \citenamefont
  {Hempel},\ and\ \citenamefont {Kassal}}]{macdonell2021analog}%
  \BibitemOpen
  \bibfield  {author} {\bibinfo {author} {\bibfnamefont {Ryan~J.}\ \bibnamefont
  {MacDonell}}, \bibinfo {author} {\bibfnamefont {Claire~E.}\ \bibnamefont
  {Dickerson}}, \bibinfo {author} {\bibfnamefont {Clare J.~T.}\ \bibnamefont
  {Birch}}, \bibinfo {author} {\bibfnamefont {Alok}\ \bibnamefont {Kumar}},
  \bibinfo {author} {\bibfnamefont {Claire~L.}\ \bibnamefont {Edmunds}},
  \bibinfo {author} {\bibfnamefont {Michael~J.}\ \bibnamefont {Biercuk}},
  \bibinfo {author} {\bibfnamefont {Cornelius}\ \bibnamefont {Hempel}}, \ and\
  \bibinfo {author} {\bibfnamefont {Ivan}\ \bibnamefont {Kassal}},\ }\bibfield
  {title} {\enquote {\bibinfo {title} {Analog quantum simulation of chemical
  dynamics},}\ }\href {\doibase 10.1039/D1SC02142G} {\bibfield  {journal}
  {\bibinfo  {journal} {Chemical Science}\ }\textbf {\bibinfo {volume} {12}},\
  \bibinfo {pages} {9794--9805} (\bibinfo {year} {2021})}\BibitemShut {NoStop}%
\bibitem [{\citenamefont {Deller}\ \emph {et~al.}(2022)\citenamefont {Deller},
  \citenamefont {Schmitt}, \citenamefont {Lewenstein}, \citenamefont {Lenk},
  \citenamefont {Federer}, \citenamefont {Jendrzejewski}, \citenamefont
  {Hauke},\ and\ \citenamefont {Kasper}}]{deller2022quantum}%
  \BibitemOpen
  \bibfield  {author} {\bibinfo {author} {\bibfnamefont {Yannick}\ \bibnamefont
  {Deller}}, \bibinfo {author} {\bibfnamefont {Sebastian}\ \bibnamefont
  {Schmitt}}, \bibinfo {author} {\bibfnamefont {Maciej}\ \bibnamefont
  {Lewenstein}}, \bibinfo {author} {\bibfnamefont {Steve}\ \bibnamefont
  {Lenk}}, \bibinfo {author} {\bibfnamefont {Marika}\ \bibnamefont {Federer}},
  \bibinfo {author} {\bibfnamefont {Fred}\ \bibnamefont {Jendrzejewski}},
  \bibinfo {author} {\bibfnamefont {Philipp}\ \bibnamefont {Hauke}}, \ and\
  \bibinfo {author} {\bibfnamefont {Valentin}\ \bibnamefont {Kasper}},\
  }\bibfield  {title} {\enquote {\bibinfo {title} {Quantum approximate
  optimization algorithm for qudit systems with long-range interactions},}\
  }\href {http://arxiv.org/abs/2204.00340} {\bibfield  {journal} {\bibinfo
  {journal} {arXiv:2204.00340}\ } (\bibinfo {year} {2022})}\BibitemShut
  {NoStop}%
\bibitem [{\citenamefont {Lanyon}\ \emph {et~al.}(2009)\citenamefont {Lanyon},
  \citenamefont {Barbieri}, \citenamefont {Almeida}, \citenamefont {Jennewein},
  \citenamefont {Ralph}, \citenamefont {Resch}, \citenamefont {Pryde},
  \citenamefont {O'Brien}, \citenamefont {Gilchrist},\ and\ \citenamefont
  {White}}]{lanyon2009simplifying}%
  \BibitemOpen
  \bibfield  {author} {\bibinfo {author} {\bibfnamefont {Benjamin~P.}\
  \bibnamefont {Lanyon}}, \bibinfo {author} {\bibfnamefont {Marco}\
  \bibnamefont {Barbieri}}, \bibinfo {author} {\bibfnamefont {Marcelo~P.}\
  \bibnamefont {Almeida}}, \bibinfo {author} {\bibfnamefont {Thomas}\
  \bibnamefont {Jennewein}}, \bibinfo {author} {\bibfnamefont {Timothy~C.}\
  \bibnamefont {Ralph}}, \bibinfo {author} {\bibfnamefont {Kevin~J.}\
  \bibnamefont {Resch}}, \bibinfo {author} {\bibfnamefont {Geoff~J.}\
  \bibnamefont {Pryde}}, \bibinfo {author} {\bibfnamefont {Jeremy~L.}\
  \bibnamefont {O'Brien}}, \bibinfo {author} {\bibfnamefont {Alexei}\
  \bibnamefont {Gilchrist}}, \ and\ \bibinfo {author} {\bibfnamefont
  {Andrew~G.}\ \bibnamefont {White}},\ }\bibfield  {title} {\enquote {\bibinfo
  {title} {Simplifying quantum logic using higher-dimensional {{Hilbert}}
  spaces},}\ }\href {\doibase 10.1038/nphys1150} {\bibfield  {journal}
  {\bibinfo  {journal} {Nature Physics}\ }\textbf {\bibinfo {volume} {5}},\
  \bibinfo {pages} {134--140} (\bibinfo {year} {2009})}\BibitemShut {NoStop}%
\bibitem [{\citenamefont {Galda}\ \emph {et~al.}(2021)\citenamefont {Galda},
  \citenamefont {Cubeddu}, \citenamefont {Kanazawa}, \citenamefont {Narang},\
  and\ \citenamefont {Earnest-Noble}}]{galda2021implementing}%
  \BibitemOpen
  \bibfield  {author} {\bibinfo {author} {\bibfnamefont {Alexey}\ \bibnamefont
  {Galda}}, \bibinfo {author} {\bibfnamefont {Michael}\ \bibnamefont
  {Cubeddu}}, \bibinfo {author} {\bibfnamefont {Naoki}\ \bibnamefont
  {Kanazawa}}, \bibinfo {author} {\bibfnamefont {Prineha}\ \bibnamefont
  {Narang}}, \ and\ \bibinfo {author} {\bibfnamefont {Nathan}\ \bibnamefont
  {Earnest-Noble}},\ }\bibfield  {title} {\enquote {\bibinfo {title}
  {Implementing a {ternary} {decomposition} of the {Toffoli} {gate} on
  {fixed}-{frequency transmon} {qutrits}},}\ }\href
  {http://arxiv.org/abs/2109.00558} {\bibfield  {journal} {\bibinfo  {journal}
  {arXiv:2109.00558}\ } (\bibinfo {year} {2021})}\BibitemShut {NoStop}%
\bibitem [{\citenamefont {Fischer}\ \emph {et~al.}(2022)\citenamefont
  {Fischer}, \citenamefont {Miller}, \citenamefont {Tacchino}, \citenamefont
  {Barkoutsos}, \citenamefont {Egger},\ and\ \citenamefont
  {Tavernelli}}]{fischer2022ancillafree}%
  \BibitemOpen
  \bibfield  {author} {\bibinfo {author} {\bibfnamefont {Laurin~E.}\
  \bibnamefont {Fischer}}, \bibinfo {author} {\bibfnamefont {Daniel}\
  \bibnamefont {Miller}}, \bibinfo {author} {\bibfnamefont {Francesco}\
  \bibnamefont {Tacchino}}, \bibinfo {author} {\bibfnamefont {Panagiotis~Kl.}\
  \bibnamefont {Barkoutsos}}, \bibinfo {author} {\bibfnamefont {Daniel~J.}\
  \bibnamefont {Egger}}, \ and\ \bibinfo {author} {\bibfnamefont {Ivano}\
  \bibnamefont {Tavernelli}},\ }\bibfield  {title} {\enquote {\bibinfo {title}
  {Ancilla-free implementation of generalized measurements for qubits embedded
  in a qudit space},}\ }\href {\doibase 10.1103/PhysRevResearch.4.033027}
  {\bibfield  {journal} {\bibinfo  {journal} {Physical Review Research}\
  }\textbf {\bibinfo {volume} {4}},\ \bibinfo {pages} {033027} (\bibinfo {year}
  {2022})}\BibitemShut {NoStop}%
\bibitem [{\citenamefont {Stricker}\ \emph {et~al.}(2022)\citenamefont
  {Stricker}, \citenamefont {Meth}, \citenamefont {Postler}, \citenamefont
  {Edmunds}, \citenamefont {Ferrie}, \citenamefont {Blatt}, \citenamefont
  {Schindler}, \citenamefont {Monz}, \citenamefont {Kueng},\ and\ \citenamefont
  {Ringbauer}}]{stricker2022experimental}%
  \BibitemOpen
  \bibfield  {author} {\bibinfo {author} {\bibfnamefont {Roman}\ \bibnamefont
  {Stricker}}, \bibinfo {author} {\bibfnamefont {Michael}\ \bibnamefont
  {Meth}}, \bibinfo {author} {\bibfnamefont {Lukas}\ \bibnamefont {Postler}},
  \bibinfo {author} {\bibfnamefont {Claire}\ \bibnamefont {Edmunds}}, \bibinfo
  {author} {\bibfnamefont {Chris}\ \bibnamefont {Ferrie}}, \bibinfo {author}
  {\bibfnamefont {Rainer}\ \bibnamefont {Blatt}}, \bibinfo {author}
  {\bibfnamefont {Philipp}\ \bibnamefont {Schindler}}, \bibinfo {author}
  {\bibfnamefont {Thomas}\ \bibnamefont {Monz}}, \bibinfo {author}
  {\bibfnamefont {Richard}\ \bibnamefont {Kueng}}, \ and\ \bibinfo {author}
  {\bibfnamefont {Martin}\ \bibnamefont {Ringbauer}},\ }\bibfield  {title}
  {\enquote {\bibinfo {title} {Experimental {{single-setting quantum state
  tomography}}},}\ }\href {\doibase 10.1103/PRXQuantum.3.040310} {\bibfield
  {journal} {\bibinfo  {journal} {PRX Quantum}\ }\textbf {\bibinfo {volume}
  {3}},\ \bibinfo {pages} {040310} (\bibinfo {year} {2022})}\BibitemShut
  {NoStop}%
\bibitem [{\citenamefont {Kraft}\ \emph {et~al.}(2018)\citenamefont {Kraft},
  \citenamefont {Ritz}, \citenamefont {Brunner}, \citenamefont {Huber},\ and\
  \citenamefont {G{\"u}hne}}]{kraft2018characterizing}%
  \BibitemOpen
  \bibfield  {author} {\bibinfo {author} {\bibfnamefont {Tristan}\ \bibnamefont
  {Kraft}}, \bibinfo {author} {\bibfnamefont {Christina}\ \bibnamefont {Ritz}},
  \bibinfo {author} {\bibfnamefont {Nicolas}\ \bibnamefont {Brunner}}, \bibinfo
  {author} {\bibfnamefont {Marcus}\ \bibnamefont {Huber}}, \ and\ \bibinfo
  {author} {\bibfnamefont {Otfried}\ \bibnamefont {G{\"u}hne}},\ }\bibfield
  {title} {\enquote {\bibinfo {title} {Characterizing {{genuine multilevel
  1ntanglement}}},}\ }\href {\doibase 10.1103/PhysRevLett.120.060502}
  {\bibfield  {journal} {\bibinfo  {journal} {Physical Review Letters}\
  }\textbf {\bibinfo {volume} {120}},\ \bibinfo {pages} {060502} (\bibinfo
  {year} {2018})}\BibitemShut {NoStop}%
\bibitem [{\citenamefont {Hu}\ \emph {et~al.}(2018)\citenamefont {Hu},
  \citenamefont {Guo}, \citenamefont {Liu}, \citenamefont {Huang},
  \citenamefont {Li},\ and\ \citenamefont {Guo}}]{hu2018beating}%
  \BibitemOpen
  \bibfield  {author} {\bibinfo {author} {\bibfnamefont {Xiao-Min}\
  \bibnamefont {Hu}}, \bibinfo {author} {\bibfnamefont {Yu}~\bibnamefont
  {Guo}}, \bibinfo {author} {\bibfnamefont {Bi-Heng}\ \bibnamefont {Liu}},
  \bibinfo {author} {\bibfnamefont {Yun-Feng}\ \bibnamefont {Huang}}, \bibinfo
  {author} {\bibfnamefont {Chuan-Feng}\ \bibnamefont {Li}}, \ and\ \bibinfo
  {author} {\bibfnamefont {Guang-Can}\ \bibnamefont {Guo}},\ }\bibfield
  {title} {\enquote {\bibinfo {title} {Beating the channel capacity limit for
  superdense coding with entangled ququarts},}\ }\href {\doibase
  10.1126/sciadv.aat9304} {\bibfield  {journal} {\bibinfo  {journal} {Science
  Advances}\ }\textbf {\bibinfo {volume} {4}},\ \bibinfo {pages} {eaat9304}
  (\bibinfo {year} {2018})}\BibitemShut {NoStop}%
\bibitem [{\citenamefont {Gottesman}\ \emph {et~al.}(2001)\citenamefont
  {Gottesman}, \citenamefont {Kitaev},\ and\ \citenamefont {Preskill}}]{GKP}%
  \BibitemOpen
  \bibfield  {author} {\bibinfo {author} {\bibfnamefont {Daniel}\ \bibnamefont
  {Gottesman}}, \bibinfo {author} {\bibfnamefont {Alexei}\ \bibnamefont
  {Kitaev}}, \ and\ \bibinfo {author} {\bibfnamefont {John}\ \bibnamefont
  {Preskill}},\ }\bibfield  {title} {\enquote {\bibinfo {title} {Encoding a
  qubit in an oscillator},}\ }\href {\doibase 10.1103/PhysRevA.64.012310}
  {\bibfield  {journal} {\bibinfo  {journal} {Physical Review A}\ }\textbf
  {\bibinfo {volume} {64}},\ \bibinfo {pages} {012310} (\bibinfo {year}
  {2001})}\BibitemShut {NoStop}%
\bibitem [{\citenamefont {Scott}(2004)}]{scott2004multipartite}%
  \BibitemOpen
  \bibfield  {author} {\bibinfo {author} {\bibfnamefont {Andrew~J.}\
  \bibnamefont {Scott}},\ }\bibfield  {title} {\enquote {\bibinfo {title}
  {Multipartite entanglement, quantum-error-correcting codes, and entangling
  power of quantum evolutions},}\ }\href {\doibase 10.1103/PhysRevA.69.052330}
  {\bibfield  {journal} {\bibinfo  {journal} {Physical Review A}\ }\textbf
  {\bibinfo {volume} {69}},\ \bibinfo {pages} {052330} (\bibinfo {year}
  {2004})}\BibitemShut {NoStop}%
\bibitem [{\citenamefont {Pirandola}\ \emph {et~al.}(2008)\citenamefont
  {Pirandola}, \citenamefont {Mancini}, \citenamefont {Braunstein},\ and\
  \citenamefont {Vitali}}]{Pirandola2008}%
  \BibitemOpen
  \bibfield  {author} {\bibinfo {author} {\bibfnamefont {Stefano}\ \bibnamefont
  {Pirandola}}, \bibinfo {author} {\bibfnamefont {Stefano}\ \bibnamefont
  {Mancini}}, \bibinfo {author} {\bibfnamefont {Samuel~L.}\ \bibnamefont
  {Braunstein}}, \ and\ \bibinfo {author} {\bibfnamefont {David}\ \bibnamefont
  {Vitali}},\ }\bibfield  {title} {\enquote {\bibinfo {title} {Minimal qudit
  code for a qubit in the phase-damping channel},}\ }\href {\doibase
  10.1103/PhysRevA.77.032309} {\bibfield  {journal} {\bibinfo  {journal}
  {Physical Review A}\ }\textbf {\bibinfo {volume} {77}},\ \bibinfo {pages}
  {032309} (\bibinfo {year} {2008})}\BibitemShut {NoStop}%
\bibitem [{\citenamefont {Cafaro}\ \emph {et~al.}(2012)\citenamefont {Cafaro},
  \citenamefont {Maiolini},\ and\ \citenamefont {Mancini}}]{Cafaro2012}%
  \BibitemOpen
  \bibfield  {author} {\bibinfo {author} {\bibfnamefont {Carlo}\ \bibnamefont
  {Cafaro}}, \bibinfo {author} {\bibfnamefont {Federico}\ \bibnamefont
  {Maiolini}}, \ and\ \bibinfo {author} {\bibfnamefont {Stefano}\ \bibnamefont
  {Mancini}},\ }\bibfield  {title} {\enquote {\bibinfo {title} {Quantum
  stabilizer codes embedding qubits into qudits},}\ }\href {\doibase
  10.1103/PhysRevA.86.022308} {\bibfield  {journal} {\bibinfo  {journal}
  {Physical Review A}\ }\textbf {\bibinfo {volume} {86}},\ \bibinfo {pages}
  {022308} (\bibinfo {year} {2012})}\BibitemShut {NoStop}%
\bibitem [{\citenamefont {Michael}\ \emph {et~al.}(2016)\citenamefont
  {Michael}, \citenamefont {Silveri}, \citenamefont {Brierley}, \citenamefont
  {Albert}, \citenamefont {Salmilehto}, \citenamefont {Jiang},\ and\
  \citenamefont {Girvin}}]{PRXGirvin}%
  \BibitemOpen
  \bibfield  {author} {\bibinfo {author} {\bibfnamefont {Marios~H.}\
  \bibnamefont {Michael}}, \bibinfo {author} {\bibfnamefont {Matti}\
  \bibnamefont {Silveri}}, \bibinfo {author} {\bibfnamefont {Richard}\
  \bibnamefont {Brierley}}, \bibinfo {author} {\bibfnamefont {Victor~V.}\
  \bibnamefont {Albert}}, \bibinfo {author} {\bibfnamefont {Juha}\ \bibnamefont
  {Salmilehto}}, \bibinfo {author} {\bibfnamefont {Liang}\ \bibnamefont
  {Jiang}}, \ and\ \bibinfo {author} {\bibfnamefont {Steven~M.}\ \bibnamefont
  {Girvin}},\ }\bibfield  {title} {\enquote {\bibinfo {title} {New class of
  quantum error-correcting codes for a bosonic mode},}\ }\href {\doibase
  https://doi.org/10.1103/PhysRevX.6.031006} {\bibfield  {journal} {\bibinfo
  {journal} {Physical Review X}\ }\textbf {\bibinfo {volume} {6}},\ \bibinfo
  {pages} {031006} (\bibinfo {year} {2016})}\BibitemShut {NoStop}%
\bibitem [{\citenamefont {Hussain}\ \emph {et~al.}(2018)\citenamefont
  {Hussain}, \citenamefont {Allodi}, \citenamefont {Chiesa}, \citenamefont
  {Garlatti}, \citenamefont {Mitcov}, \citenamefont {Konstantatos},
  \citenamefont {Pedersen}, \citenamefont {De~Renzi}, \citenamefont
  {Piligkos},\ and\ \citenamefont {Carretta}}]{Hussain2018}%
  \BibitemOpen
  \bibfield  {author} {\bibinfo {author} {\bibfnamefont {Riaz}\ \bibnamefont
  {Hussain}}, \bibinfo {author} {\bibfnamefont {Giuseppe}\ \bibnamefont
  {Allodi}}, \bibinfo {author} {\bibfnamefont {Alessandro}\ \bibnamefont
  {Chiesa}}, \bibinfo {author} {\bibfnamefont {Elena}\ \bibnamefont
  {Garlatti}}, \bibinfo {author} {\bibfnamefont {Dmitri}\ \bibnamefont
  {Mitcov}}, \bibinfo {author} {\bibfnamefont {Andreas}\ \bibnamefont
  {Konstantatos}}, \bibinfo {author} {\bibfnamefont {Kasper~S.}\ \bibnamefont
  {Pedersen}}, \bibinfo {author} {\bibfnamefont {Roberto}\ \bibnamefont
  {De~Renzi}}, \bibinfo {author} {\bibfnamefont {Stergios}\ \bibnamefont
  {Piligkos}}, \ and\ \bibinfo {author} {\bibfnamefont {Stefano}\ \bibnamefont
  {Carretta}},\ }\bibfield  {title} {\enquote {\bibinfo {title} {Coherent
  {{manipulation}} of a {{molecular Ln-Based nuclear qudit coupled}} to an
  {{electron qubit}}},}\ }\href {\doibase 10.1021/jacs.8b05934} {\bibfield
  {journal} {\bibinfo  {journal} {Journal of the American Chemical Society}\
  }\textbf {\bibinfo {volume} {140}},\ \bibinfo {pages} {9814--9818} (\bibinfo
  {year} {2018})}\BibitemShut {NoStop}%
\bibitem [{\citenamefont {Chiesa}\ \emph {et~al.}(2020)\citenamefont {Chiesa},
  \citenamefont {Macaluso}, \citenamefont {Petiziol}, \citenamefont
  {Wimberger}, \citenamefont {Santini},\ and\ \citenamefont
  {Carretta}}]{Chiesa2020}%
  \BibitemOpen
  \bibfield  {author} {\bibinfo {author} {\bibfnamefont {Alessandro}\
  \bibnamefont {Chiesa}}, \bibinfo {author} {\bibfnamefont {Emilio}\
  \bibnamefont {Macaluso}}, \bibinfo {author} {\bibfnamefont {Francesco}\
  \bibnamefont {Petiziol}}, \bibinfo {author} {\bibfnamefont {Sandro}\
  \bibnamefont {Wimberger}}, \bibinfo {author} {\bibfnamefont {Paolo}\
  \bibnamefont {Santini}}, \ and\ \bibinfo {author} {\bibfnamefont {Stefano}\
  \bibnamefont {Carretta}},\ }\bibfield  {title} {\enquote {\bibinfo {title}
  {Molecular nanomagnets as qubits with embedded quantum-error correction},}\
  }\href {\doibase https://doi.org/10.1021/acs.jpclett.0c02213} {\bibfield
  {journal} {\bibinfo  {journal} {Journal of Physical Chemistry Letters}\
  }\textbf {\bibinfo {volume} {11}},\ \bibinfo {pages} {8610--8615} (\bibinfo
  {year} {2020})}\BibitemShut {NoStop}%
\bibitem [{\citenamefont {Carretta}\ \emph {et~al.}(2021)\citenamefont
  {Carretta}, \citenamefont {Zueco}, \citenamefont {Chiesa}, \citenamefont
  {{G{\'o}mez-Le{\'o}n}},\ and\ \citenamefont {Luis}}]{Carretta2021}%
  \BibitemOpen
  \bibfield  {author} {\bibinfo {author} {\bibfnamefont {Stefano}\ \bibnamefont
  {Carretta}}, \bibinfo {author} {\bibfnamefont {David}\ \bibnamefont {Zueco}},
  \bibinfo {author} {\bibfnamefont {Alessandro}\ \bibnamefont {Chiesa}},
  \bibinfo {author} {\bibfnamefont {{\'A}lvaro}\ \bibnamefont
  {{G{\'o}mez-Le{\'o}n}}}, \ and\ \bibinfo {author} {\bibfnamefont {Fernando}\
  \bibnamefont {Luis}},\ }\bibfield  {title} {\enquote {\bibinfo {title} {A
  perspective on scaling up quantum computation with molecular spins},}\ }\href
  {\doibase 10.1063/5.0053378} {\bibfield  {journal} {\bibinfo  {journal}
  {Applied Physics Letters}\ }\textbf {\bibinfo {volume} {118}},\ \bibinfo
  {pages} {240501} (\bibinfo {year} {2021})}\BibitemShut {NoStop}%
\bibitem [{\citenamefont {Chizzini}\ \emph
  {et~al.}(2022{\natexlab{a}})\citenamefont {Chizzini}, \citenamefont {Crippa},
  \citenamefont {Zaccardi}, \citenamefont {Macaluso}, \citenamefont {Carretta},
  \citenamefont {Chiesa},\ and\ \citenamefont {Santini}}]{Chizzini2022}%
  \BibitemOpen
  \bibfield  {author} {\bibinfo {author} {\bibfnamefont {Mario}\ \bibnamefont
  {Chizzini}}, \bibinfo {author} {\bibfnamefont {Luca}\ \bibnamefont {Crippa}},
  \bibinfo {author} {\bibfnamefont {Luca}\ \bibnamefont {Zaccardi}}, \bibinfo
  {author} {\bibfnamefont {Emilio}\ \bibnamefont {Macaluso}}, \bibinfo {author}
  {\bibfnamefont {Stefano}\ \bibnamefont {Carretta}}, \bibinfo {author}
  {\bibfnamefont {Alessandro}\ \bibnamefont {Chiesa}}, \ and\ \bibinfo {author}
  {\bibfnamefont {Paolo}\ \bibnamefont {Santini}},\ }\bibfield  {title}
  {\enquote {\bibinfo {title} {Quantum error correction with molecular spin
  qudits},}\ }\href {\doibase 10.1039/D2CP01228F} {\bibfield  {journal}
  {\bibinfo  {journal} {Physical Chemistry Chemical Physics}\ }\textbf
  {\bibinfo {volume} {24}},\ \bibinfo {pages} {20030--20039} (\bibinfo {year}
  {2022}{\natexlab{a}})}\BibitemShut {NoStop}%
\bibitem [{\citenamefont {Petiziol}\ \emph {et~al.}(2021)\citenamefont
  {Petiziol}, \citenamefont {Chiesa}, \citenamefont {Wimberger}, \citenamefont
  {Santini},\ and\ \citenamefont {Carretta}}]{Petiziol2021}%
  \BibitemOpen
  \bibfield  {author} {\bibinfo {author} {\bibfnamefont {Francesco}\
  \bibnamefont {Petiziol}}, \bibinfo {author} {\bibfnamefont {Alessandro}\
  \bibnamefont {Chiesa}}, \bibinfo {author} {\bibfnamefont {Sandro}\
  \bibnamefont {Wimberger}}, \bibinfo {author} {\bibfnamefont {Paolo}\
  \bibnamefont {Santini}}, \ and\ \bibinfo {author} {\bibfnamefont {Stefano}\
  \bibnamefont {Carretta}},\ }\bibfield  {title} {\enquote {\bibinfo {title}
  {Counteracting dephasing in molecular nanomagnets by optimized qudit
  encodings},}\ }\href {\doibase 10.1038/s41534-021-00466-3} {\bibfield
  {journal} {\bibinfo  {journal} {npj Quantum Information}\ }\textbf {\bibinfo
  {volume} {7}},\ \bibinfo {pages} {133} (\bibinfo {year} {2021})}\BibitemShut
  {NoStop}%
\bibitem [{\citenamefont {Chiesa}\ \emph {et~al.}(2022)\citenamefont {Chiesa},
  \citenamefont {Petiziol}, \citenamefont {Chizzini}, \citenamefont {Santini},\
  and\ \citenamefont {Carretta}}]{Chiesa2022}%
  \BibitemOpen
  \bibfield  {author} {\bibinfo {author} {\bibfnamefont {Alessandro}\
  \bibnamefont {Chiesa}}, \bibinfo {author} {\bibfnamefont {Francesco}\
  \bibnamefont {Petiziol}}, \bibinfo {author} {\bibfnamefont {Mario}\
  \bibnamefont {Chizzini}}, \bibinfo {author} {\bibfnamefont {Paolo}\
  \bibnamefont {Santini}}, \ and\ \bibinfo {author} {\bibfnamefont {Stefano}\
  \bibnamefont {Carretta}},\ }\bibfield  {title} {\enquote {\bibinfo {title}
  {Theoretical design of optimal molecular qudits for quantum error
  correction},}\ }\href {https://doi.org/10.1021/acs.jpclett.2c01602}
  {\bibfield  {journal} {\bibinfo  {journal} {Journal of Physical Chemistry
  Letters}\ }\textbf {\bibinfo {volume} {13}},\ \bibinfo {pages} {6468--6474}
  (\bibinfo {year} {2022})}\BibitemShut {NoStop}%
\bibitem [{\citenamefont {Ringbauer}\ \emph {et~al.}(2022)\citenamefont
  {Ringbauer}, \citenamefont {Meth}, \citenamefont {Postler}, \citenamefont
  {Stricker}, \citenamefont {Blatt}, \citenamefont {Schindler},\ and\
  \citenamefont {Monz}}]{ringbauer2021universal}%
  \BibitemOpen
  \bibfield  {author} {\bibinfo {author} {\bibfnamefont {Martin}\ \bibnamefont
  {Ringbauer}}, \bibinfo {author} {\bibfnamefont {Michael}\ \bibnamefont
  {Meth}}, \bibinfo {author} {\bibfnamefont {Lukas}\ \bibnamefont {Postler}},
  \bibinfo {author} {\bibfnamefont {Roman}\ \bibnamefont {Stricker}}, \bibinfo
  {author} {\bibfnamefont {Rainer}\ \bibnamefont {Blatt}}, \bibinfo {author}
  {\bibfnamefont {Philipp}\ \bibnamefont {Schindler}}, \ and\ \bibinfo {author}
  {\bibfnamefont {Thomas}\ \bibnamefont {Monz}},\ }\bibfield  {title} {\enquote
  {\bibinfo {title} {A universal qudit quantum processor with trapped ions},}\
  }\href {\doibase 10.1038/s41567-022-01658-0} {\bibfield  {journal} {\bibinfo
  {journal} {Nature Physics}\ }\textbf {\bibinfo {volume} {18}},\ \bibinfo
  {pages} {1053--1057} (\bibinfo {year} {2022})}\BibitemShut {NoStop}%
\bibitem [{\citenamefont {Chi}\ \emph {et~al.}(2022)\citenamefont {Chi},
  \citenamefont {Huang}, \citenamefont {Zhang}, \citenamefont {Mao},
  \citenamefont {Zhou}, \citenamefont {Chen}, \citenamefont {Zhai},
  \citenamefont {Bao}, \citenamefont {Dai}, \citenamefont {Yuan}, \citenamefont
  {Zhang}, \citenamefont {Dai}, \citenamefont {Tang}, \citenamefont {Yang},
  \citenamefont {Li}, \citenamefont {Ding}, \citenamefont {Oxenløwe},
  \citenamefont {Thompson}, \citenamefont {O’Brien}, \citenamefont {Li},
  \citenamefont {Gong},\ and\ \citenamefont {Wang}}]{chi2022programmable}%
  \BibitemOpen
  \bibfield  {author} {\bibinfo {author} {\bibfnamefont {Yulin}\ \bibnamefont
  {Chi}}, \bibinfo {author} {\bibfnamefont {Jieshan}\ \bibnamefont {Huang}},
  \bibinfo {author} {\bibfnamefont {Zhanchuan}\ \bibnamefont {Zhang}}, \bibinfo
  {author} {\bibfnamefont {Jun}\ \bibnamefont {Mao}}, \bibinfo {author}
  {\bibfnamefont {Zinan}\ \bibnamefont {Zhou}}, \bibinfo {author}
  {\bibfnamefont {Xiaojiong}\ \bibnamefont {Chen}}, \bibinfo {author}
  {\bibfnamefont {Chonghao}\ \bibnamefont {Zhai}}, \bibinfo {author}
  {\bibfnamefont {Jueming}\ \bibnamefont {Bao}}, \bibinfo {author}
  {\bibfnamefont {Tianxiang}\ \bibnamefont {Dai}}, \bibinfo {author}
  {\bibfnamefont {Huihong}\ \bibnamefont {Yuan}}, \bibinfo {author}
  {\bibfnamefont {Ming}\ \bibnamefont {Zhang}}, \bibinfo {author}
  {\bibfnamefont {Daoxin}\ \bibnamefont {Dai}}, \bibinfo {author}
  {\bibfnamefont {Bo}~\bibnamefont {Tang}}, \bibinfo {author} {\bibfnamefont
  {Yan}\ \bibnamefont {Yang}}, \bibinfo {author} {\bibfnamefont {Zhihua}\
  \bibnamefont {Li}}, \bibinfo {author} {\bibfnamefont {Yunhong}\ \bibnamefont
  {Ding}}, \bibinfo {author} {\bibfnamefont {Leif~K.}\ \bibnamefont
  {Oxenløwe}}, \bibinfo {author} {\bibfnamefont {Mark~G.}\ \bibnamefont
  {Thompson}}, \bibinfo {author} {\bibfnamefont {Jeremy~L.}\ \bibnamefont
  {O’Brien}}, \bibinfo {author} {\bibfnamefont {Yan}\ \bibnamefont {Li}},
  \bibinfo {author} {\bibfnamefont {Qihuang}\ \bibnamefont {Gong}}, \ and\
  \bibinfo {author} {\bibfnamefont {Jianwei}\ \bibnamefont {Wang}},\ }\bibfield
   {title} {\enquote {\bibinfo {title} {A programmable qudit-based quantum
  processor},}\ }\href {\doibase 10.1038/s41467-022-28767-x} {\bibfield
  {journal} {\bibinfo  {journal} {Nature Communications}\ }\textbf {\bibinfo
  {volume} {13}},\ \bibinfo {pages} {1166} (\bibinfo {year}
  {2022})}\BibitemShut {NoStop}%
\bibitem [{\citenamefont {Gonz\'alez-Cuadra}\ \emph {et~al.}(2022)\citenamefont
  {Gonz\'alez-Cuadra}, \citenamefont {Zache}, \citenamefont {Carrasco},
  \citenamefont {Kraus},\ and\ \citenamefont
  {Zoller}}]{gonzalezcuadra2022hardware}%
  \BibitemOpen
  \bibfield  {author} {\bibinfo {author} {\bibfnamefont {Daniel}\ \bibnamefont
  {Gonz\'alez-Cuadra}}, \bibinfo {author} {\bibfnamefont {Torsten~V.}\
  \bibnamefont {Zache}}, \bibinfo {author} {\bibfnamefont {Jose}\ \bibnamefont
  {Carrasco}}, \bibinfo {author} {\bibfnamefont {Barbara}\ \bibnamefont
  {Kraus}}, \ and\ \bibinfo {author} {\bibfnamefont {Peter}\ \bibnamefont
  {Zoller}},\ }\bibfield  {title} {\enquote {\bibinfo {title} {Hardware
  efficient quantum simulation of non-abelian gauge theories with qudits on
  rydberg platforms},}\ }\href {\doibase 10.1103/PhysRevLett.129.160501}
  {\bibfield  {journal} {\bibinfo  {journal} {Phys. Rev. Lett.}\ }\textbf
  {\bibinfo {volume} {129}},\ \bibinfo {pages} {160501} (\bibinfo {year}
  {2022})}\BibitemShut {NoStop}%
\bibitem [{\citenamefont {Kasper}\ \emph {et~al.}(2022)\citenamefont {Kasper},
  \citenamefont {González-Cuadra}, \citenamefont {Hegde}, \citenamefont {Xia},
  \citenamefont {Dauphin}, \citenamefont {Huber}, \citenamefont {Tiemann},
  \citenamefont {Lewenstein}, \citenamefont {Jendrzejewski},\ and\
  \citenamefont {Hauke}}]{kasper2022universal}%
  \BibitemOpen
  \bibfield  {author} {\bibinfo {author} {\bibfnamefont {Valentin}\
  \bibnamefont {Kasper}}, \bibinfo {author} {\bibfnamefont {Daniel}\
  \bibnamefont {González-Cuadra}}, \bibinfo {author} {\bibfnamefont {Apoorva}\
  \bibnamefont {Hegde}}, \bibinfo {author} {\bibfnamefont {Andy}\ \bibnamefont
  {Xia}}, \bibinfo {author} {\bibfnamefont {Alexandre}\ \bibnamefont
  {Dauphin}}, \bibinfo {author} {\bibfnamefont {Felix}\ \bibnamefont {Huber}},
  \bibinfo {author} {\bibfnamefont {Eberhard}\ \bibnamefont {Tiemann}},
  \bibinfo {author} {\bibfnamefont {Maciej}\ \bibnamefont {Lewenstein}},
  \bibinfo {author} {\bibfnamefont {Fred}\ \bibnamefont {Jendrzejewski}}, \
  and\ \bibinfo {author} {\bibfnamefont {Philipp}\ \bibnamefont {Hauke}},\
  }\bibfield  {title} {\enquote {\bibinfo {title} {Universal quantum
  computation and quantum error correction with ultracold atomic mixtures},}\
  }\href {\doibase 10.1088/2058-9565/ac2d39} {\bibfield  {journal} {\bibinfo
  {journal} {Quantum Science and Technology}\ }\textbf {\bibinfo {volume}
  {7}},\ \bibinfo {pages} {015008} (\bibinfo {year} {2022})}\BibitemShut
  {NoStop}%
\bibitem [{\citenamefont {Gimeno}\ \emph {et~al.}(2021)\citenamefont {Gimeno},
  \citenamefont {Urtizberea}, \citenamefont {Román-Roche}, \citenamefont
  {Zueco}, \citenamefont {Camón}, \citenamefont {Alonso}, \citenamefont
  {Roubeau},\ and\ \citenamefont {Luis}}]{Gimeno2021}%
  \BibitemOpen
  \bibfield  {author} {\bibinfo {author} {\bibfnamefont {Ignacio}\ \bibnamefont
  {Gimeno}}, \bibinfo {author} {\bibfnamefont {Ainhoa}\ \bibnamefont
  {Urtizberea}}, \bibinfo {author} {\bibfnamefont {Juan}\ \bibnamefont
  {Román-Roche}}, \bibinfo {author} {\bibfnamefont {David}\ \bibnamefont
  {Zueco}}, \bibinfo {author} {\bibfnamefont {Agustín}\ \bibnamefont
  {Camón}}, \bibinfo {author} {\bibfnamefont {Pablo~J.}\ \bibnamefont
  {Alonso}}, \bibinfo {author} {\bibfnamefont {Olivier}\ \bibnamefont
  {Roubeau}}, \ and\ \bibinfo {author} {\bibfnamefont {Fernando}\ \bibnamefont
  {Luis}},\ }\bibfield  {title} {\enquote {\bibinfo {title} {Broad-band
  spectroscopy of a vanadyl porphyrin: a model electronuclear spin qudit},}\
  }\href {\doibase 10.1039/D1SC00564B} {\bibfield  {journal} {\bibinfo
  {journal} {Chem. Sci.}\ }\textbf {\bibinfo {volume} {12}},\ \bibinfo {pages}
  {5621--5630} (\bibinfo {year} {2021})}\BibitemShut {NoStop}%
\bibitem [{\citenamefont {Chicco}\ \emph {et~al.}(2021)\citenamefont {Chicco},
  \citenamefont {Chiesa}, \citenamefont {Allodi}, \citenamefont {Garlatti},
  \citenamefont {Atzori}, \citenamefont {Sorace}, \citenamefont {De~Renzi},
  \citenamefont {Sessoli},\ and\ \citenamefont {Carretta}}]{Chicco2021}%
  \BibitemOpen
  \bibfield  {author} {\bibinfo {author} {\bibfnamefont {Simone}\ \bibnamefont
  {Chicco}}, \bibinfo {author} {\bibfnamefont {Alessandro}\ \bibnamefont
  {Chiesa}}, \bibinfo {author} {\bibfnamefont {Giuseppe}\ \bibnamefont
  {Allodi}}, \bibinfo {author} {\bibfnamefont {Elena}\ \bibnamefont
  {Garlatti}}, \bibinfo {author} {\bibfnamefont {Matteo}\ \bibnamefont
  {Atzori}}, \bibinfo {author} {\bibfnamefont {Lorenzo}\ \bibnamefont
  {Sorace}}, \bibinfo {author} {\bibfnamefont {Roberto}\ \bibnamefont
  {De~Renzi}}, \bibinfo {author} {\bibfnamefont {Roberta}\ \bibnamefont
  {Sessoli}}, \ and\ \bibinfo {author} {\bibfnamefont {Stefano}\ \bibnamefont
  {Carretta}},\ }\bibfield  {title} {\enquote {\bibinfo {title} {Controlled
  coherent dynamics of [vo(tpp)]{,} a prototype molecular nuclear qudit with an
  electronic ancilla},}\ }\href {\doibase 10.1039/D1SC01358K} {\bibfield
  {journal} {\bibinfo  {journal} {Chem. Sci.}\ }\textbf {\bibinfo {volume}
  {12}},\ \bibinfo {pages} {12046--12055} (\bibinfo {year} {2021})}\BibitemShut
  {NoStop}%
\bibitem [{\citenamefont {Chizzini}\ \emph
  {et~al.}(2022{\natexlab{b}})\citenamefont {Chizzini}, \citenamefont {Crippa},
  \citenamefont {Chiesa}, \citenamefont {Tacchino}, \citenamefont {Petiziol},
  \citenamefont {Tavernelli}, \citenamefont {Santini},\ and\ \citenamefont
  {Carretta}}]{Chizzini2022physrevres}%
  \BibitemOpen
  \bibfield  {author} {\bibinfo {author} {\bibfnamefont {M.}~\bibnamefont
  {Chizzini}}, \bibinfo {author} {\bibfnamefont {L.}~\bibnamefont {Crippa}},
  \bibinfo {author} {\bibfnamefont {A.}~\bibnamefont {Chiesa}}, \bibinfo
  {author} {\bibfnamefont {F.}~\bibnamefont {Tacchino}}, \bibinfo {author}
  {\bibfnamefont {F.}~\bibnamefont {Petiziol}}, \bibinfo {author}
  {\bibfnamefont {I.}~\bibnamefont {Tavernelli}}, \bibinfo {author}
  {\bibfnamefont {P.}~\bibnamefont {Santini}}, \ and\ \bibinfo {author}
  {\bibfnamefont {S.}~\bibnamefont {Carretta}},\ }\bibfield  {title} {\enquote
  {\bibinfo {title} {Molecular nanomagnets with competing interactions as
  optimal units for qudit-based quantum computation},}\ }\href {\doibase
  10.1103/PhysRevResearch.4.043135} {\bibfield  {journal} {\bibinfo  {journal}
  {Phys. Rev. Research}\ }\textbf {\bibinfo {volume} {4}},\ \bibinfo {pages}
  {043135} (\bibinfo {year} {2022}{\natexlab{b}})}\BibitemShut {NoStop}%
\bibitem [{\citenamefont {Koch}\ \emph {et~al.}(2007)\citenamefont {Koch},
  \citenamefont {Yu}, \citenamefont {Gambetta}, \citenamefont {Houck},
  \citenamefont {Schuster}, \citenamefont {Majer}, \citenamefont {Blais},
  \citenamefont {Devoret}, \citenamefont {Girvin},\ and\ \citenamefont
  {Schoelkopf}}]{koch2007chargeinsensitive}%
  \BibitemOpen
  \bibfield  {author} {\bibinfo {author} {\bibfnamefont {Jens}\ \bibnamefont
  {Koch}}, \bibinfo {author} {\bibfnamefont {Terri~M.}\ \bibnamefont {Yu}},
  \bibinfo {author} {\bibfnamefont {Jay}\ \bibnamefont {Gambetta}}, \bibinfo
  {author} {\bibfnamefont {Andrew~A.}\ \bibnamefont {Houck}}, \bibinfo {author}
  {\bibfnamefont {David~I.}\ \bibnamefont {Schuster}}, \bibinfo {author}
  {\bibfnamefont {Johannes}\ \bibnamefont {Majer}}, \bibinfo {author}
  {\bibfnamefont {Alexandre}\ \bibnamefont {Blais}}, \bibinfo {author}
  {\bibfnamefont {Michel~H.}\ \bibnamefont {Devoret}}, \bibinfo {author}
  {\bibfnamefont {Steven~M.}\ \bibnamefont {Girvin}}, \ and\ \bibinfo {author}
  {\bibfnamefont {Robert~J.}\ \bibnamefont {Schoelkopf}},\ }\bibfield  {title}
  {\enquote {\bibinfo {title} {Charge-insensitive qubit design derived from the
  {Cooper} pair box},}\ }\href {\doibase 10.1103/PhysRevA.76.042319} {\bibfield
   {journal} {\bibinfo  {journal} {Physical Review A}\ }\textbf {\bibinfo
  {volume} {76}},\ \bibinfo {pages} {042319} (\bibinfo {year}
  {2007})}\BibitemShut {NoStop}%
\bibitem [{\citenamefont {Blok}\ \emph {et~al.}(2021)\citenamefont {Blok},
  \citenamefont {Ramasesh}, \citenamefont {Schuster}, \citenamefont
  {O’Brien}, \citenamefont {Kreikebaum}, \citenamefont {Dahlen},
  \citenamefont {Morvan}, \citenamefont {Yoshida}, \citenamefont {Yao},\ and\
  \citenamefont {Siddiqi}}]{blok2021quantum}%
  \BibitemOpen
  \bibfield  {author} {\bibinfo {author} {\bibfnamefont {Machiel~S.}\
  \bibnamefont {Blok}}, \bibinfo {author} {\bibfnamefont {Vinay~V.}\
  \bibnamefont {Ramasesh}}, \bibinfo {author} {\bibfnamefont {Thomas}\
  \bibnamefont {Schuster}}, \bibinfo {author} {\bibfnamefont {Kevin}\
  \bibnamefont {O’Brien}}, \bibinfo {author} {\bibfnamefont {John-Mark}\
  \bibnamefont {Kreikebaum}}, \bibinfo {author} {\bibfnamefont {Dar}\
  \bibnamefont {Dahlen}}, \bibinfo {author} {\bibfnamefont {Alexis}\
  \bibnamefont {Morvan}}, \bibinfo {author} {\bibfnamefont {Beni}\ \bibnamefont
  {Yoshida}}, \bibinfo {author} {\bibfnamefont {Norman~Y.}\ \bibnamefont
  {Yao}}, \ and\ \bibinfo {author} {\bibfnamefont {Irfan}\ \bibnamefont
  {Siddiqi}},\ }\bibfield  {title} {\enquote {\bibinfo {title} {Quantum
  {information} {scrambling} on a {superconducting} {qutrit} {processor}},}\
  }\href {\doibase 10.1103/PhysRevX.11.021010} {\bibfield  {journal} {\bibinfo
  {journal} {Physical Review X}\ }\textbf {\bibinfo {volume} {11}},\ \bibinfo
  {pages} {021010} (\bibinfo {year} {2021})}\BibitemShut {NoStop}%
\bibitem [{\citenamefont {Cervera-Lierta}\ \emph {et~al.}(2022)\citenamefont
  {Cervera-Lierta}, \citenamefont {Krenn}, \citenamefont {Aspuru-Guzik},\ and\
  \citenamefont {Galda}}]{cerveralierta2022experimental}%
  \BibitemOpen
  \bibfield  {author} {\bibinfo {author} {\bibfnamefont {Alba}\ \bibnamefont
  {Cervera-Lierta}}, \bibinfo {author} {\bibfnamefont {Mario}\ \bibnamefont
  {Krenn}}, \bibinfo {author} {\bibfnamefont {Alán}\ \bibnamefont
  {Aspuru-Guzik}}, \ and\ \bibinfo {author} {\bibfnamefont {Alexey}\
  \bibnamefont {Galda}},\ }\bibfield  {title} {\enquote {\bibinfo {title}
  {Experimental {high}-{dimensional} {Greenberger}-{Horne}-{Zeilinger}
  {entanglement} with {superconducting} {transmon} {qutrits}},}\ }\href
  {\doibase 10.1103/PhysRevApplied.17.024062} {\bibfield  {journal} {\bibinfo
  {journal} {Physical Review Applied}\ }\textbf {\bibinfo {volume} {17}},\
  \bibinfo {pages} {024062} (\bibinfo {year} {2022})}\BibitemShut {NoStop}%
\bibitem [{\citenamefont {Egger}\ \emph {et~al.}(2019)\citenamefont {Egger},
  \citenamefont {Ganzhorn}, \citenamefont {Salis}, \citenamefont {Fuhrer},
  \citenamefont {Müller}, \citenamefont {Barkoutsos}, \citenamefont {Moll},
  \citenamefont {Tavernelli},\ and\ \citenamefont
  {Filipp}}]{egger2019entanglement}%
  \BibitemOpen
  \bibfield  {author} {\bibinfo {author} {\bibfnamefont {D.J.}\ \bibnamefont
  {Egger}}, \bibinfo {author} {\bibfnamefont {M.}~\bibnamefont {Ganzhorn}},
  \bibinfo {author} {\bibfnamefont {G.}~\bibnamefont {Salis}}, \bibinfo
  {author} {\bibfnamefont {A.}~\bibnamefont {Fuhrer}}, \bibinfo {author}
  {\bibfnamefont {P.}~\bibnamefont {Müller}}, \bibinfo {author} {\bibfnamefont
  {P.Kl.}\ \bibnamefont {Barkoutsos}}, \bibinfo {author} {\bibfnamefont
  {N.}~\bibnamefont {Moll}}, \bibinfo {author} {\bibfnamefont {I.}~\bibnamefont
  {Tavernelli}}, \ and\ \bibinfo {author} {\bibfnamefont {S.}~\bibnamefont
  {Filipp}},\ }\bibfield  {title} {\enquote {\bibinfo {title} {Entanglement
  {generation} in {superconducting} {qubits} {using} {holonomic}
  {operations}},}\ }\href {\doibase 10.1103/PhysRevApplied.11.014017}
  {\bibfield  {journal} {\bibinfo  {journal} {Physical Review Applied}\
  }\textbf {\bibinfo {volume} {11}},\ \bibinfo {pages} {014017} (\bibinfo
  {year} {2019})}\BibitemShut {NoStop}%
\bibitem [{\citenamefont {Nikolaeva}\ \emph
  {et~al.}(2022{\natexlab{b}})\citenamefont {Nikolaeva}, \citenamefont
  {Kiktenko},\ and\ \citenamefont {Fedorov}}]{nikolaeva2022decomposing}%
  \BibitemOpen
  \bibfield  {author} {\bibinfo {author} {\bibfnamefont {Anastasiia~S.}\
  \bibnamefont {Nikolaeva}}, \bibinfo {author} {\bibfnamefont {Evgeniy~O.}\
  \bibnamefont {Kiktenko}}, \ and\ \bibinfo {author} {\bibfnamefont
  {Aleksey~K.}\ \bibnamefont {Fedorov}},\ }\bibfield  {title} {\enquote
  {\bibinfo {title} {Decomposing the generalized {{Toffoli}} gate with
  qutrits},}\ }\href {\doibase 10.1103/PhysRevA.105.032621} {\bibfield
  {journal} {\bibinfo  {journal} {Physical Review A}\ }\textbf {\bibinfo
  {volume} {105}},\ \bibinfo {pages} {032621} (\bibinfo {year}
  {2022}{\natexlab{b}})}\BibitemShut {NoStop}%
\bibitem [{\citenamefont {Elder}\ \emph {et~al.}(2020)\citenamefont {Elder},
  \citenamefont {Wang}, \citenamefont {Reinhold}, \citenamefont {Hann},
  \citenamefont {Chou}, \citenamefont {Lester}, \citenamefont {Rosenblum},
  \citenamefont {Frunzio}, \citenamefont {Jiang},\ and\ \citenamefont
  {Schoelkopf}}]{elder2020highfidelity}%
  \BibitemOpen
  \bibfield  {author} {\bibinfo {author} {\bibfnamefont {Salvatore~S.}\
  \bibnamefont {Elder}}, \bibinfo {author} {\bibfnamefont {Christopher~S.}\
  \bibnamefont {Wang}}, \bibinfo {author} {\bibfnamefont {Philip}\ \bibnamefont
  {Reinhold}}, \bibinfo {author} {\bibfnamefont {Connor~T.}\ \bibnamefont
  {Hann}}, \bibinfo {author} {\bibfnamefont {Kevin~S.}\ \bibnamefont {Chou}},
  \bibinfo {author} {\bibfnamefont {Brian~J.}\ \bibnamefont {Lester}}, \bibinfo
  {author} {\bibfnamefont {Serge}\ \bibnamefont {Rosenblum}}, \bibinfo {author}
  {\bibfnamefont {Luigi}\ \bibnamefont {Frunzio}}, \bibinfo {author}
  {\bibfnamefont {Liang}\ \bibnamefont {Jiang}}, \ and\ \bibinfo {author}
  {\bibfnamefont {Robert~J.}\ \bibnamefont {Schoelkopf}},\ }\bibfield  {title}
  {\enquote {\bibinfo {title} {High-{fidelity} {measurement} of {qubits}
  {encoded} in {multilevel} {superconducting} {circuits}},}\ }\href {\doibase
  10.1103/PhysRevX.10.011001} {\bibfield  {journal} {\bibinfo  {journal}
  {Physical Review X}\ }\textbf {\bibinfo {volume} {10}},\ \bibinfo {pages}
  {011001} (\bibinfo {year} {2020})}\BibitemShut {NoStop}%
\bibitem [{\citenamefont {Jurcevic}\ \emph {et~al.}(2021)\citenamefont
  {Jurcevic}, \citenamefont {Javadi-Abhari}, \citenamefont {Bishop},
  \citenamefont {Lauer}, \citenamefont {Bogorin}, \citenamefont {Brink},
  \citenamefont {Capelluto}, \citenamefont {Günlük}, \citenamefont {Itoko},
  \citenamefont {Kanazawa}, \citenamefont {Kandala}, \citenamefont {Keefe},
  \citenamefont {Krsulich}, \citenamefont {Landers}, \citenamefont
  {Lewandowski}, \citenamefont {McClure}, \citenamefont {Nannicini},
  \citenamefont {Narasgond}, \citenamefont {Nayfeh}, \citenamefont {Pritchett},
  \citenamefont {Rothwell}, \citenamefont {Srinivasan}, \citenamefont
  {Sundaresan}, \citenamefont {Wang}, \citenamefont {Wei}, \citenamefont
  {Wood}, \citenamefont {Yau}, \citenamefont {Zhang}, \citenamefont {Dial},
  \citenamefont {Chow},\ and\ \citenamefont
  {Gambetta}}]{jurcevic2021demonstration}%
  \BibitemOpen
  \bibfield  {author} {\bibinfo {author} {\bibfnamefont {Petar}\ \bibnamefont
  {Jurcevic}}, \bibinfo {author} {\bibfnamefont {Ali}\ \bibnamefont
  {Javadi-Abhari}}, \bibinfo {author} {\bibfnamefont {Lev~S}\ \bibnamefont
  {Bishop}}, \bibinfo {author} {\bibfnamefont {Isaac}\ \bibnamefont {Lauer}},
  \bibinfo {author} {\bibfnamefont {Daniela~F}\ \bibnamefont {Bogorin}},
  \bibinfo {author} {\bibfnamefont {Markus}\ \bibnamefont {Brink}}, \bibinfo
  {author} {\bibfnamefont {Lauren}\ \bibnamefont {Capelluto}}, \bibinfo
  {author} {\bibfnamefont {Oktay}\ \bibnamefont {Günlük}}, \bibinfo {author}
  {\bibfnamefont {Toshinari}\ \bibnamefont {Itoko}}, \bibinfo {author}
  {\bibfnamefont {Naoki}\ \bibnamefont {Kanazawa}}, \bibinfo {author}
  {\bibfnamefont {Abhinav}\ \bibnamefont {Kandala}}, \bibinfo {author}
  {\bibfnamefont {George~A}\ \bibnamefont {Keefe}}, \bibinfo {author}
  {\bibfnamefont {Kevin}\ \bibnamefont {Krsulich}}, \bibinfo {author}
  {\bibfnamefont {William}\ \bibnamefont {Landers}}, \bibinfo {author}
  {\bibfnamefont {Eric~P}\ \bibnamefont {Lewandowski}}, \bibinfo {author}
  {\bibfnamefont {Douglas~T}\ \bibnamefont {McClure}}, \bibinfo {author}
  {\bibfnamefont {Giacomo}\ \bibnamefont {Nannicini}}, \bibinfo {author}
  {\bibfnamefont {Adinath}\ \bibnamefont {Narasgond}}, \bibinfo {author}
  {\bibfnamefont {Hasan~M}\ \bibnamefont {Nayfeh}}, \bibinfo {author}
  {\bibfnamefont {Emily}\ \bibnamefont {Pritchett}}, \bibinfo {author}
  {\bibfnamefont {Mary~Beth}\ \bibnamefont {Rothwell}}, \bibinfo {author}
  {\bibfnamefont {Srikanth}\ \bibnamefont {Srinivasan}}, \bibinfo {author}
  {\bibfnamefont {Neereja}\ \bibnamefont {Sundaresan}}, \bibinfo {author}
  {\bibfnamefont {Cindy}\ \bibnamefont {Wang}}, \bibinfo {author}
  {\bibfnamefont {Ken~X}\ \bibnamefont {Wei}}, \bibinfo {author} {\bibfnamefont
  {Christopher~J}\ \bibnamefont {Wood}}, \bibinfo {author} {\bibfnamefont
  {Jeng-Bang}\ \bibnamefont {Yau}}, \bibinfo {author} {\bibfnamefont {Eric~J}\
  \bibnamefont {Zhang}}, \bibinfo {author} {\bibfnamefont {Oliver~E}\
  \bibnamefont {Dial}}, \bibinfo {author} {\bibfnamefont {Jerry~M}\
  \bibnamefont {Chow}}, \ and\ \bibinfo {author} {\bibfnamefont {Jay~M}\
  \bibnamefont {Gambetta}},\ }\bibfield  {title} {\enquote {\bibinfo {title}
  {Demonstration of quantum volume 64 on a superconducting quantum computing
  system},}\ }\href {\doibase 10.1088/2058-9565/abe519} {\bibfield  {journal}
  {\bibinfo  {journal} {Quantum Science and Technology}\ }\textbf {\bibinfo
  {volume} {6}},\ \bibinfo {pages} {025020} (\bibinfo {year}
  {2021})}\BibitemShut {NoStop}%
\bibitem [{\citenamefont {Shlyakhov}\ \emph {et~al.}(2018)\citenamefont
  {Shlyakhov}, \citenamefont {Zemlyanov}, \citenamefont {Suslov}, \citenamefont
  {Lebedev}, \citenamefont {Paraoanu}, \citenamefont {Lesovik},\ and\
  \citenamefont {Blatter}}]{shlyakhov2018quantum}%
  \BibitemOpen
  \bibfield  {author} {\bibinfo {author} {\bibfnamefont {A.~R.}\ \bibnamefont
  {Shlyakhov}}, \bibinfo {author} {\bibfnamefont {V.~V.}\ \bibnamefont
  {Zemlyanov}}, \bibinfo {author} {\bibfnamefont {M.~V.}\ \bibnamefont
  {Suslov}}, \bibinfo {author} {\bibfnamefont {Andrey~V.}\ \bibnamefont
  {Lebedev}}, \bibinfo {author} {\bibfnamefont {Gheorghe~S.}\ \bibnamefont
  {Paraoanu}}, \bibinfo {author} {\bibfnamefont {Gordey~B.}\ \bibnamefont
  {Lesovik}}, \ and\ \bibinfo {author} {\bibfnamefont {Gianni}\ \bibnamefont
  {Blatter}},\ }\bibfield  {title} {\enquote {\bibinfo {title} {Quantum
  metrology with a transmon qutrit},}\ }\href {\doibase
  10.1103/PhysRevA.97.022115} {\bibfield  {journal} {\bibinfo  {journal}
  {Physical Review A}\ }\textbf {\bibinfo {volume} {97}},\ \bibinfo {pages}
  {022115} (\bibinfo {year} {2018})}\BibitemShut {NoStop}%
\bibitem [{\citenamefont {Egger}\ \emph {et~al.}(2018)\citenamefont {Egger},
  \citenamefont {Werninghaus}, \citenamefont {Ganzhorn}, \citenamefont {Salis},
  \citenamefont {Fuhrer}, \citenamefont {Müller},\ and\ \citenamefont
  {Filipp}}]{egger2018pulsed}%
  \BibitemOpen
  \bibfield  {author} {\bibinfo {author} {\bibfnamefont {Daniel~J.}\
  \bibnamefont {Egger}}, \bibinfo {author} {\bibfnamefont {Max}\ \bibnamefont
  {Werninghaus}}, \bibinfo {author} {\bibfnamefont {Marc}\ \bibnamefont
  {Ganzhorn}}, \bibinfo {author} {\bibfnamefont {Gian}\ \bibnamefont {Salis}},
  \bibinfo {author} {\bibfnamefont {Andreas}\ \bibnamefont {Fuhrer}}, \bibinfo
  {author} {\bibfnamefont {Peter}\ \bibnamefont {Müller}}, \ and\ \bibinfo
  {author} {\bibfnamefont {Stefan}\ \bibnamefont {Filipp}},\ }\bibfield
  {title} {\enquote {\bibinfo {title} {Pulsed reset protocol for
  fixed-frequency superconducting qubits},}\ }\href {\doibase
  10.1103/PhysRevApplied.10.044030} {\bibfield  {journal} {\bibinfo  {journal}
  {Physical Review Applied}\ }\textbf {\bibinfo {volume} {10}},\ \bibinfo
  {pages} {044030} (\bibinfo {year} {2018})}\BibitemShut {NoStop}%
\bibitem [{\citenamefont {Roy}\ \emph {et~al.}(2022)\citenamefont {Roy},
  \citenamefont {Li}, \citenamefont {Kapit},\ and\ \citenamefont
  {Schuster}}]{roy2022realization}%
  \BibitemOpen
  \bibfield  {author} {\bibinfo {author} {\bibfnamefont {Tanay}\ \bibnamefont
  {Roy}}, \bibinfo {author} {\bibfnamefont {Ziqian}\ \bibnamefont {Li}},
  \bibinfo {author} {\bibfnamefont {Eliot}\ \bibnamefont {Kapit}}, \ and\
  \bibinfo {author} {\bibfnamefont {David~I.}\ \bibnamefont {Schuster}},\
  }\bibfield  {title} {\enquote {\bibinfo {title} {Realization of two-qutrit
  quantum algorithms on a programmable superconducting processor},}\ }\href
  {http://arxiv.org/abs/2211.06523} {\bibfield  {journal} {\bibinfo  {journal}
  {arXiv:2211.06523}\ } (\bibinfo {year} {2022})}\BibitemShut {NoStop}%
\bibitem [{\citenamefont {Kiktenko}\ \emph {et~al.}(2015)\citenamefont
  {Kiktenko}, \citenamefont {Fedorov}, \citenamefont {Man'ko},\ and\
  \citenamefont {Man'ko}}]{kiktenko2015multilevel}%
  \BibitemOpen
  \bibfield  {author} {\bibinfo {author} {\bibfnamefont {Evgeniy.~O.}\
  \bibnamefont {Kiktenko}}, \bibinfo {author} {\bibfnamefont {Aleksey.~K.}\
  \bibnamefont {Fedorov}}, \bibinfo {author} {\bibfnamefont {Olga~V.}\
  \bibnamefont {Man'ko}}, \ and\ \bibinfo {author} {\bibfnamefont
  {Vladimir~I.}\ \bibnamefont {Man'ko}},\ }\bibfield  {title} {\enquote
  {\bibinfo {title} {Multilevel superconducting circuits as two-qubit systems:
  {operations}, state preparation, and entropic inequalities},}\ }\href
  {\doibase 10.1103/PhysRevA.91.042312} {\bibfield  {journal} {\bibinfo
  {journal} {Physical Review A}\ }\textbf {\bibinfo {volume} {91}},\ \bibinfo
  {pages} {042312} (\bibinfo {year} {2015})}\BibitemShut {NoStop}%
\bibitem [{\citenamefont {Cao}\ \emph {et~al.}(2023)\citenamefont {Cao},
  \citenamefont {Bakr}, \citenamefont {Campanaro}, \citenamefont {Fasciati},
  \citenamefont {Wills}, \citenamefont {Lall}, \citenamefont {Shteynas},
  \citenamefont {Chidambaram}, \citenamefont {Rungger},\ and\ \citenamefont
  {Leek}}]{cao2023emulatinga}%
  \BibitemOpen
  \bibfield  {author} {\bibinfo {author} {\bibfnamefont {Shuxiang}\
  \bibnamefont {Cao}}, \bibinfo {author} {\bibfnamefont {Mustafa}\ \bibnamefont
  {Bakr}}, \bibinfo {author} {\bibfnamefont {Giulio}\ \bibnamefont
  {Campanaro}}, \bibinfo {author} {\bibfnamefont {Simone~D.}\ \bibnamefont
  {Fasciati}}, \bibinfo {author} {\bibfnamefont {James}\ \bibnamefont {Wills}},
  \bibinfo {author} {\bibfnamefont {Deep}\ \bibnamefont {Lall}}, \bibinfo
  {author} {\bibfnamefont {Boris}\ \bibnamefont {Shteynas}}, \bibinfo {author}
  {\bibfnamefont {Vivek}\ \bibnamefont {Chidambaram}}, \bibinfo {author}
  {\bibfnamefont {Ivan}\ \bibnamefont {Rungger}}, \ and\ \bibinfo {author}
  {\bibfnamefont {Peter}\ \bibnamefont {Leek}},\ }\bibfield  {title} {\enquote
  {\bibinfo {title} {Emulating two qubits with a four-level transmon qudit for
  variational quantum algorithms},}\ }\href {http://arxiv.org/abs/2303.04796}
  {\bibfield  {journal} {\bibinfo  {journal} {arXiv:2303.04796}\ } (\bibinfo
  {year} {2023})}\BibitemShut {NoStop}%
\bibitem [{\citenamefont {Seifert}\ \emph {et~al.}(2023)\citenamefont
  {Seifert}, \citenamefont {Li}, \citenamefont {Roy}, \citenamefont {Schuster},
  \citenamefont {Chong},\ and\ \citenamefont {Baker}}]{seifert2023exploring}%
  \BibitemOpen
  \bibfield  {author} {\bibinfo {author} {\bibfnamefont {Lennart~Maximilian}\
  \bibnamefont {Seifert}}, \bibinfo {author} {\bibfnamefont {Ziqian}\
  \bibnamefont {Li}}, \bibinfo {author} {\bibfnamefont {Tanay}\ \bibnamefont
  {Roy}}, \bibinfo {author} {\bibfnamefont {David~I.}\ \bibnamefont
  {Schuster}}, \bibinfo {author} {\bibfnamefont {Frederic~T.}\ \bibnamefont
  {Chong}}, \ and\ \bibinfo {author} {\bibfnamefont {Jonathan~M.}\ \bibnamefont
  {Baker}},\ }\bibfield  {title} {\enquote {\bibinfo {title} {Exploring
  {{Ququart Computation}} on a {{Transmon}} using {{Optimal Control}}},}\
  }\href {http://arxiv.org/abs/2304.11159} {\bibfield  {journal} {\bibinfo
  {journal} {arXiv:2304.11159}\ } (\bibinfo {year} {2023})}\BibitemShut
  {NoStop}%
\bibitem [{\citenamefont {Liu}\ \emph {et~al.}(2023)\citenamefont {Liu},
  \citenamefont {Wang}, \citenamefont {Zhang}, \citenamefont {Zhang},
  \citenamefont {Cai}, \citenamefont {Xu}, \citenamefont {Li}, \citenamefont
  {Han}, \citenamefont {Li}, \citenamefont {Xue}, \citenamefont {Liu},
  \citenamefont {You}, \citenamefont {Jin},\ and\ \citenamefont
  {Yu}}]{liu2023performing}%
  \BibitemOpen
  \bibfield  {author} {\bibinfo {author} {\bibfnamefont {Pei}\ \bibnamefont
  {Liu}}, \bibinfo {author} {\bibfnamefont {Ruixia}\ \bibnamefont {Wang}},
  \bibinfo {author} {\bibfnamefont {Jing-Ning}\ \bibnamefont {Zhang}}, \bibinfo
  {author} {\bibfnamefont {Yingshan}\ \bibnamefont {Zhang}}, \bibinfo {author}
  {\bibfnamefont {Xiaoxia}\ \bibnamefont {Cai}}, \bibinfo {author}
  {\bibfnamefont {Huikai}\ \bibnamefont {Xu}}, \bibinfo {author} {\bibfnamefont
  {Zhiyuan}\ \bibnamefont {Li}}, \bibinfo {author} {\bibfnamefont {Jiaxiu}\
  \bibnamefont {Han}}, \bibinfo {author} {\bibfnamefont {Xuegang}\ \bibnamefont
  {Li}}, \bibinfo {author} {\bibfnamefont {Guangming}\ \bibnamefont {Xue}},
  \bibinfo {author} {\bibfnamefont {Weiyang}\ \bibnamefont {Liu}}, \bibinfo
  {author} {\bibfnamefont {Li}~\bibnamefont {You}}, \bibinfo {author}
  {\bibfnamefont {Yirong}\ \bibnamefont {Jin}}, \ and\ \bibinfo {author}
  {\bibfnamefont {Haifeng}\ \bibnamefont {Yu}},\ }\bibfield  {title} {\enquote
  {\bibinfo {title} {Performing $\mathrm{SU}(d)$ operations and rudimentary
  algorithms in a superconducting transmon qudit for $d=3$ and $d=4$},}\ }\href
  {\doibase 10.1103/PhysRevX.13.021028} {\bibfield  {journal} {\bibinfo
  {journal} {Phys. Rev. X}\ }\textbf {\bibinfo {volume} {13}},\ \bibinfo
  {pages} {021028} (\bibinfo {year} {2023})}\BibitemShut {NoStop}%
\bibitem [{\citenamefont {Rigetti}\ and\ \citenamefont
  {Devoret}(2010)}]{rigetti2010fully}%
  \BibitemOpen
  \bibfield  {author} {\bibinfo {author} {\bibfnamefont {Chad}\ \bibnamefont
  {Rigetti}}\ and\ \bibinfo {author} {\bibfnamefont {Michel}\ \bibnamefont
  {Devoret}},\ }\bibfield  {title} {\enquote {\bibinfo {title} {Fully
  microwave-tunable universal gates in superconducting qubits with linear
  couplings and fixed transition frequencies},}\ }\href {\doibase
  10.1103/PhysRevB.81.134507} {\bibfield  {journal} {\bibinfo  {journal}
  {Physical Review B}\ }\textbf {\bibinfo {volume} {81}},\ \bibinfo {pages}
  {134507} (\bibinfo {year} {2010})}\BibitemShut {NoStop}%
\bibitem [{\citenamefont {Bravyi}\ \emph {et~al.}(2022)\citenamefont {Bravyi},
  \citenamefont {Dial}, \citenamefont {Gambetta}, \citenamefont {Gil},\ and\
  \citenamefont {Nazario}}]{bravyi2022future}%
  \BibitemOpen
  \bibfield  {author} {\bibinfo {author} {\bibfnamefont {Sergey}\ \bibnamefont
  {Bravyi}}, \bibinfo {author} {\bibfnamefont {Oliver}\ \bibnamefont {Dial}},
  \bibinfo {author} {\bibfnamefont {Jay~M}\ \bibnamefont {Gambetta}}, \bibinfo
  {author} {\bibfnamefont {Dario}\ \bibnamefont {Gil}}, \ and\ \bibinfo
  {author} {\bibfnamefont {Zaira}\ \bibnamefont {Nazario}},\ }\bibfield
  {title} {\enquote {\bibinfo {title} {The future of quantum computing with
  superconducting qubits},}\ }\href {https://doi.org/10.1063/5.0082975}
  {\bibfield  {journal} {\bibinfo  {journal} {Journal of Applied Physics}\
  }\textbf {\bibinfo {volume} {132}},\ \bibinfo {pages} {160902} (\bibinfo
  {year} {2022})}\BibitemShut {NoStop}%
\bibitem [{\citenamefont {Wack}\ \emph {et~al.}(2021)\citenamefont {Wack},
  \citenamefont {Paik}, \citenamefont {Javadi-Abhari}, \citenamefont
  {Jurcevic}, \citenamefont {Faro}, \citenamefont {Gambetta},\ and\
  \citenamefont {Johnson}}]{wack2021quality}%
  \BibitemOpen
  \bibfield  {author} {\bibinfo {author} {\bibfnamefont {Andrew}\ \bibnamefont
  {Wack}}, \bibinfo {author} {\bibfnamefont {Hanhee}\ \bibnamefont {Paik}},
  \bibinfo {author} {\bibfnamefont {Ali}\ \bibnamefont {Javadi-Abhari}},
  \bibinfo {author} {\bibfnamefont {Petar}\ \bibnamefont {Jurcevic}}, \bibinfo
  {author} {\bibfnamefont {Ismael}\ \bibnamefont {Faro}}, \bibinfo {author}
  {\bibfnamefont {Jay~M.}\ \bibnamefont {Gambetta}}, \ and\ \bibinfo {author}
  {\bibfnamefont {Blake~R.}\ \bibnamefont {Johnson}},\ }\bibfield  {title}
  {\enquote {\bibinfo {title} {Quality, {speed}, and {scale}: three key
  attributes to measure the performance of near-term quantum computers},}\
  }\href {http://arxiv.org/abs/2110.14108} {\bibfield  {journal} {\bibinfo
  {journal} {arXiv:2110.14108}\ } (\bibinfo {year} {2021})}\BibitemShut
  {NoStop}%
\bibitem [{\citenamefont {Place}\ \emph {et~al.}(2021)\citenamefont {Place},
  \citenamefont {Rodgers}, \citenamefont {Mundada}, \citenamefont {Smitham},
  \citenamefont {Fitzpatrick}, \citenamefont {Leng}, \citenamefont {Premkumar},
  \citenamefont {Bryon}, \citenamefont {Vrajitoarea}, \citenamefont {Sussman},
  \citenamefont {Cheng}, \citenamefont {Madhavan}, \citenamefont {Babla},
  \citenamefont {Le}, \citenamefont {Gang}, \citenamefont {Jäck},
  \citenamefont {Gyenis}, \citenamefont {Yao}, \citenamefont {Cava},
  \citenamefont {de~Leon},\ and\ \citenamefont {Houck}}]{place2021new}%
  \BibitemOpen
  \bibfield  {author} {\bibinfo {author} {\bibfnamefont {Alexander P.~M.}\
  \bibnamefont {Place}}, \bibinfo {author} {\bibfnamefont {Lila V.~H.}\
  \bibnamefont {Rodgers}}, \bibinfo {author} {\bibfnamefont {Pranav}\
  \bibnamefont {Mundada}}, \bibinfo {author} {\bibfnamefont {Basil~M.}\
  \bibnamefont {Smitham}}, \bibinfo {author} {\bibfnamefont {Mattias}\
  \bibnamefont {Fitzpatrick}}, \bibinfo {author} {\bibfnamefont {Zhaoqi}\
  \bibnamefont {Leng}}, \bibinfo {author} {\bibfnamefont {Anjali}\ \bibnamefont
  {Premkumar}}, \bibinfo {author} {\bibfnamefont {Jacob}\ \bibnamefont
  {Bryon}}, \bibinfo {author} {\bibfnamefont {Andrei}\ \bibnamefont
  {Vrajitoarea}}, \bibinfo {author} {\bibfnamefont {Sara}\ \bibnamefont
  {Sussman}}, \bibinfo {author} {\bibfnamefont {Guangming}\ \bibnamefont
  {Cheng}}, \bibinfo {author} {\bibfnamefont {Trisha}\ \bibnamefont
  {Madhavan}}, \bibinfo {author} {\bibfnamefont {Harshvardhan~K.}\ \bibnamefont
  {Babla}}, \bibinfo {author} {\bibfnamefont {Xuan~Hoang}\ \bibnamefont {Le}},
  \bibinfo {author} {\bibfnamefont {Youqi}\ \bibnamefont {Gang}}, \bibinfo
  {author} {\bibfnamefont {Berthold}\ \bibnamefont {Jäck}}, \bibinfo {author}
  {\bibfnamefont {András}\ \bibnamefont {Gyenis}}, \bibinfo {author}
  {\bibfnamefont {Nan}\ \bibnamefont {Yao}}, \bibinfo {author} {\bibfnamefont
  {Robert~J.}\ \bibnamefont {Cava}}, \bibinfo {author} {\bibfnamefont
  {Nathalie~P.}\ \bibnamefont {de~Leon}}, \ and\ \bibinfo {author}
  {\bibfnamefont {Andrew~A.}\ \bibnamefont {Houck}},\ }\bibfield  {title}
  {\enquote {\bibinfo {title} {New material platform for superconducting
  transmon qubits with coherence times exceeding 0.3 milliseconds},}\ }\href
  {\doibase 10.1038/s41467-021-22030-5} {\bibfield  {journal} {\bibinfo
  {journal} {Nature Communications}\ }\textbf {\bibinfo {volume} {12}},\
  \bibinfo {pages} {1779} (\bibinfo {year} {2021})}\BibitemShut {NoStop}%
\bibitem [{\citenamefont {McKay}\ \emph {et~al.}(2017)\citenamefont {McKay},
  \citenamefont {Wood}, \citenamefont {Sheldon}, \citenamefont {Chow},\ and\
  \citenamefont {Gambetta}}]{mckay2017efficient}%
  \BibitemOpen
  \bibfield  {author} {\bibinfo {author} {\bibfnamefont {David~C.}\
  \bibnamefont {McKay}}, \bibinfo {author} {\bibfnamefont {Christopher~J.}\
  \bibnamefont {Wood}}, \bibinfo {author} {\bibfnamefont {Sarah}\ \bibnamefont
  {Sheldon}}, \bibinfo {author} {\bibfnamefont {Jerry~M.}\ \bibnamefont
  {Chow}}, \ and\ \bibinfo {author} {\bibfnamefont {Jay~M.}\ \bibnamefont
  {Gambetta}},\ }\bibfield  {title} {\enquote {\bibinfo {title} {Efficient {Z}
  gates for quantum computing},}\ }\href {\doibase 10.1103/PhysRevA.96.022330}
  {\bibfield  {journal} {\bibinfo  {journal} {Physical Review A}\ }\textbf
  {\bibinfo {volume} {96}},\ \bibinfo {pages} {022330} (\bibinfo {year}
  {2017})}\BibitemShut {NoStop}%
\bibitem [{\citenamefont {Schirmer}\ \emph {et~al.}(2002)\citenamefont
  {Schirmer}, \citenamefont {Greentree}, \citenamefont {Ramakrishna},\ and\
  \citenamefont {Rabitz}}]{schirmer2002constructive}%
  \BibitemOpen
  \bibfield  {author} {\bibinfo {author} {\bibfnamefont {Sophie~G.}\
  \bibnamefont {Schirmer}}, \bibinfo {author} {\bibfnamefont {Andrew~D.}\
  \bibnamefont {Greentree}}, \bibinfo {author} {\bibfnamefont {Viswanath}\
  \bibnamefont {Ramakrishna}}, \ and\ \bibinfo {author} {\bibfnamefont
  {Herschel}\ \bibnamefont {Rabitz}},\ }\bibfield  {title} {\enquote {\bibinfo
  {title} {Constructive control of quantum systems using factorization of
  unitary operators},}\ }\href {\doibase 10.1088/0305-4470/35/39/313}
  {\bibfield  {journal} {\bibinfo  {journal} {Journal of Physics A:
  Mathematical and General}\ }\textbf {\bibinfo {volume} {35}},\ \bibinfo
  {pages} {8315--8339} (\bibinfo {year} {2002})}\BibitemShut {NoStop}%
\bibitem [{\citenamefont {Brennen}\ \emph
  {et~al.}(2005{\natexlab{a}})\citenamefont {Brennen}, \citenamefont
  {O’Leary},\ and\ \citenamefont {Bullock}}]{brennen2005criteria}%
  \BibitemOpen
  \bibfield  {author} {\bibinfo {author} {\bibfnamefont {Gavin}\ \bibnamefont
  {Brennen}}, \bibinfo {author} {\bibfnamefont {Dianne}\ \bibnamefont
  {O’Leary}}, \ and\ \bibinfo {author} {\bibfnamefont {Stephen}\ \bibnamefont
  {Bullock}},\ }\bibfield  {title} {\enquote {\bibinfo {title} {Criteria for
  exact qudit universality},}\ }\href {\doibase 10.1103/PhysRevA.71.052318}
  {\bibfield  {journal} {\bibinfo  {journal} {Physical Review A}\ }\textbf
  {\bibinfo {volume} {71}},\ \bibinfo {pages} {052318} (\bibinfo {year}
  {2005}{\natexlab{a}})}\BibitemShut {NoStop}%
\bibitem [{\citenamefont {Patterson}\ \emph {et~al.}(2019)\citenamefont
  {Patterson}, \citenamefont {Rahamim}, \citenamefont {Tsunoda}, \citenamefont
  {Spring}, \citenamefont {Jebari}, \citenamefont {Ratter}, \citenamefont
  {Mergenthaler}, \citenamefont {Tancredi}, \citenamefont {Vlastakis},
  \citenamefont {Esposito},\ and\ \citenamefont
  {Leek}}]{patterson2019calibration}%
  \BibitemOpen
  \bibfield  {author} {\bibinfo {author} {\bibfnamefont {A.D.}\ \bibnamefont
  {Patterson}}, \bibinfo {author} {\bibfnamefont {J.}~\bibnamefont {Rahamim}},
  \bibinfo {author} {\bibfnamefont {T.}~\bibnamefont {Tsunoda}}, \bibinfo
  {author} {\bibfnamefont {P.A.}\ \bibnamefont {Spring}}, \bibinfo {author}
  {\bibfnamefont {S.}~\bibnamefont {Jebari}}, \bibinfo {author} {\bibfnamefont
  {K.}~\bibnamefont {Ratter}}, \bibinfo {author} {\bibfnamefont
  {M.}~\bibnamefont {Mergenthaler}}, \bibinfo {author} {\bibfnamefont
  {G.}~\bibnamefont {Tancredi}}, \bibinfo {author} {\bibfnamefont
  {B.}~\bibnamefont {Vlastakis}}, \bibinfo {author} {\bibfnamefont
  {M.}~\bibnamefont {Esposito}}, \ and\ \bibinfo {author} {\bibfnamefont
  {P.J.}\ \bibnamefont {Leek}},\ }\bibfield  {title} {\enquote {\bibinfo
  {title} {Calibration of a {{Cross-Resonance Two-Qubit Gate Between Directly
  Coupled Transmons}}},}\ }\href {\doibase 10.1103/PhysRevApplied.12.064013}
  {\bibfield  {journal} {\bibinfo  {journal} {Physical Review Applied}\
  }\textbf {\bibinfo {volume} {12}},\ \bibinfo {pages} {064013} (\bibinfo
  {year} {2019})}\BibitemShut {NoStop}%
\bibitem [{\citenamefont {Magesan}\ and\ \citenamefont
  {Gambetta}(2020)}]{magesan2020effective}%
  \BibitemOpen
  \bibfield  {author} {\bibinfo {author} {\bibfnamefont {Easwar}\ \bibnamefont
  {Magesan}}\ and\ \bibinfo {author} {\bibfnamefont {Jay~M.}\ \bibnamefont
  {Gambetta}},\ }\bibfield  {title} {\enquote {\bibinfo {title} {Effective
  {Hamiltonian} models of the cross-resonance gate},}\ }\href {\doibase
  10.1103/PhysRevA.101.052308} {\bibfield  {journal} {\bibinfo  {journal}
  {Physical Review A}\ }\textbf {\bibinfo {volume} {101}},\ \bibinfo {pages}
  {052308} (\bibinfo {year} {2020})}\BibitemShut {NoStop}%
\bibitem [{\citenamefont {Tripathi}\ \emph {et~al.}(2019)\citenamefont
  {Tripathi}, \citenamefont {Khezri},\ and\ \citenamefont
  {Korotkov}}]{tripathi2019operation}%
  \BibitemOpen
  \bibfield  {author} {\bibinfo {author} {\bibfnamefont {Vinay}\ \bibnamefont
  {Tripathi}}, \bibinfo {author} {\bibfnamefont {Mostafa}\ \bibnamefont
  {Khezri}}, \ and\ \bibinfo {author} {\bibfnamefont {Alexander~N.}\
  \bibnamefont {Korotkov}},\ }\bibfield  {title} {\enquote {\bibinfo {title}
  {Operation and intrinsic error budget of a two-qubit cross-resonance gate},}\
  }\href {\doibase 10.1103/PhysRevA.100.012301} {\bibfield  {journal} {\bibinfo
   {journal} {Physical Review A}\ }\textbf {\bibinfo {volume} {100}},\ \bibinfo
  {pages} {012301} (\bibinfo {year} {2019})}\BibitemShut {NoStop}%
\bibitem [{\citenamefont {Malekakhlagh}\ \emph {et~al.}(2020)\citenamefont
  {Malekakhlagh}, \citenamefont {Magesan},\ and\ \citenamefont
  {McKay}}]{malekakhlagh2020firstprinciples}%
  \BibitemOpen
  \bibfield  {author} {\bibinfo {author} {\bibfnamefont {Moein}\ \bibnamefont
  {Malekakhlagh}}, \bibinfo {author} {\bibfnamefont {Easwar}\ \bibnamefont
  {Magesan}}, \ and\ \bibinfo {author} {\bibfnamefont {David~C.}\ \bibnamefont
  {McKay}},\ }\bibfield  {title} {\enquote {\bibinfo {title} {First-principles
  analysis of cross-resonance gate operation},}\ }\href {\doibase
  10.1103/PhysRevA.102.042605} {\bibfield  {journal} {\bibinfo  {journal}
  {Physical Review A}\ }\textbf {\bibinfo {volume} {102}},\ \bibinfo {pages}
  {042605} (\bibinfo {year} {2020})}\BibitemShut {NoStop}%
\bibitem [{\citenamefont {Sheldon}\ \emph {et~al.}(2016)\citenamefont
  {Sheldon}, \citenamefont {Magesan}, \citenamefont {Chow},\ and\ \citenamefont
  {Gambetta}}]{sheldon2016procedure}%
  \BibitemOpen
  \bibfield  {author} {\bibinfo {author} {\bibfnamefont {Sarah}\ \bibnamefont
  {Sheldon}}, \bibinfo {author} {\bibfnamefont {Easwar}\ \bibnamefont
  {Magesan}}, \bibinfo {author} {\bibfnamefont {Jerry~M.}\ \bibnamefont
  {Chow}}, \ and\ \bibinfo {author} {\bibfnamefont {Jay~M.}\ \bibnamefont
  {Gambetta}},\ }\bibfield  {title} {\enquote {\bibinfo {title} {Procedure for
  systematically tuning up cross-talk in the cross-resonance gate},}\ }\href
  {\doibase 10.1103/PhysRevA.93.060302} {\bibfield  {journal} {\bibinfo
  {journal} {Physical Review A}\ }\textbf {\bibinfo {volume} {93}},\ \bibinfo
  {pages} {060302} (\bibinfo {year} {2016})}\BibitemShut {NoStop}%
\bibitem [{\citenamefont {Sundaresan}\ \emph {et~al.}(2020)\citenamefont
  {Sundaresan}, \citenamefont {Lauer}, \citenamefont {Pritchett}, \citenamefont
  {Magesan}, \citenamefont {Jurcevic},\ and\ \citenamefont
  {Gambetta}}]{sundaresan2020reducing}%
  \BibitemOpen
  \bibfield  {author} {\bibinfo {author} {\bibfnamefont {Neereja}\ \bibnamefont
  {Sundaresan}}, \bibinfo {author} {\bibfnamefont {Isaac}\ \bibnamefont
  {Lauer}}, \bibinfo {author} {\bibfnamefont {Emily}\ \bibnamefont
  {Pritchett}}, \bibinfo {author} {\bibfnamefont {Easwar}\ \bibnamefont
  {Magesan}}, \bibinfo {author} {\bibfnamefont {Petar}\ \bibnamefont
  {Jurcevic}}, \ and\ \bibinfo {author} {\bibfnamefont {Jay~M.}\ \bibnamefont
  {Gambetta}},\ }\bibfield  {title} {\enquote {\bibinfo {title} {Reducing
  {unitary} and {spectator} {errors} in {cross} {resonance} with {optimized}
  {rotary} {echoes}},}\ }\href {\doibase 10.1103/PRXQuantum.1.020318}
  {\bibfield  {journal} {\bibinfo  {journal} {PRX Quantum}\ }\textbf {\bibinfo
  {volume} {1}},\ \bibinfo {pages} {020318} (\bibinfo {year}
  {2020})}\BibitemShut {NoStop}%
\bibitem [{\citenamefont {Schutjens}\ \emph {et~al.}(2013)\citenamefont
  {Schutjens}, \citenamefont {Dagga}, \citenamefont {Egger},\ and\
  \citenamefont {Wilhelm}}]{schutjens2013singlequbit}%
  \BibitemOpen
  \bibfield  {author} {\bibinfo {author} {\bibfnamefont {Ron}\ \bibnamefont
  {Schutjens}}, \bibinfo {author} {\bibfnamefont {Abu}\ \bibnamefont {Dagga}},
  \bibinfo {author} {\bibfnamefont {Daniel~J.}\ \bibnamefont {Egger}}, \ and\
  \bibinfo {author} {\bibfnamefont {Frank~K.}\ \bibnamefont {Wilhelm}},\
  }\bibfield  {title} {\enquote {\bibinfo {title} {Single-qubit gates in
  frequency-crowded transmon systems},}\ }\href {\doibase
  10.1103/PhysRevA.88.052330} {\bibfield  {journal} {\bibinfo  {journal}
  {Physical Review A}\ }\textbf {\bibinfo {volume} {88}},\ \bibinfo {pages}
  {052330} (\bibinfo {year} {2013})}\BibitemShut {NoStop}%
\bibitem [{\citenamefont {Vesterinen}\ \emph {et~al.}(2014)\citenamefont
  {Vesterinen}, \citenamefont {Saira}, \citenamefont {Bruno},\ and\
  \citenamefont {DiCarlo}}]{vesterinen2014mitigating}%
  \BibitemOpen
  \bibfield  {author} {\bibinfo {author} {\bibfnamefont {Visa}\ \bibnamefont
  {Vesterinen}}, \bibinfo {author} {\bibfnamefont {Olli-Pentti}\ \bibnamefont
  {Saira}}, \bibinfo {author} {\bibfnamefont {Alessandro}\ \bibnamefont
  {Bruno}}, \ and\ \bibinfo {author} {\bibfnamefont {Leonardo}\ \bibnamefont
  {DiCarlo}},\ }\bibfield  {title} {\enquote {\bibinfo {title} {Mitigating
  information leakage in a crowded spectrum of weakly anharmonic qubits},}\
  }\href {http://arxiv.org/abs/1405.0450} {\bibfield  {journal} {\bibinfo
  {journal} {arXiv:1405.0450}\ } (\bibinfo {year} {2014})}\BibitemShut
  {NoStop}%
\bibitem [{\citenamefont {Quantum}(accessed Oct. 2022)}]{IBMQuantum}%
  \BibitemOpen
  \bibfield  {author} {\bibinfo {author} {\bibfnamefont {IBM}\ \bibnamefont
  {Quantum}},\ }\href@noop {} {}\bibinfo {howpublished}
  {\url{https://quantum-computing.ibm.com/}} (\bibinfo {year} {accessed Oct.
  2022})\BibitemShut {NoStop}%
\bibitem [{\citenamefont {Werninghaus}\ \emph {et~al.}(2021)\citenamefont
  {Werninghaus}, \citenamefont {Egger}, \citenamefont {Roy}, \citenamefont
  {Machnes}, \citenamefont {Wilhelm},\ and\ \citenamefont
  {Filipp}}]{werninghaus2021leakage}%
  \BibitemOpen
  \bibfield  {author} {\bibinfo {author} {\bibfnamefont {Max}\ \bibnamefont
  {Werninghaus}}, \bibinfo {author} {\bibfnamefont {Daniel~J.}\ \bibnamefont
  {Egger}}, \bibinfo {author} {\bibfnamefont {Federico}\ \bibnamefont {Roy}},
  \bibinfo {author} {\bibfnamefont {Shai}\ \bibnamefont {Machnes}}, \bibinfo
  {author} {\bibfnamefont {Frank~K.}\ \bibnamefont {Wilhelm}}, \ and\ \bibinfo
  {author} {\bibfnamefont {Stefan}\ \bibnamefont {Filipp}},\ }\bibfield
  {title} {\enquote {\bibinfo {title} {Leakage reduction in fast
  superconducting qubit gates via optimal control},}\ }\href {\doibase
  10.1038/s41534-020-00346-2} {\bibfield  {journal} {\bibinfo  {journal} {npj
  Quantum Information}\ }\textbf {\bibinfo {volume} {7}},\ \bibinfo {pages}
  {14} (\bibinfo {year} {2021})}\BibitemShut {NoStop}%
\bibitem [{\citenamefont {Motzoi}\ \emph {et~al.}(2009)\citenamefont {Motzoi},
  \citenamefont {Gambetta}, \citenamefont {Rebentrost},\ and\ \citenamefont
  {Wilhelm}}]{motzoi2009simple}%
  \BibitemOpen
  \bibfield  {author} {\bibinfo {author} {\bibfnamefont {Felix}\ \bibnamefont
  {Motzoi}}, \bibinfo {author} {\bibfnamefont {Jay~M.}\ \bibnamefont
  {Gambetta}}, \bibinfo {author} {\bibfnamefont {Patrick}\ \bibnamefont
  {Rebentrost}}, \ and\ \bibinfo {author} {\bibfnamefont {Frank~K.}\
  \bibnamefont {Wilhelm}},\ }\bibfield  {title} {\enquote {\bibinfo {title}
  {Simple {pulses} for {elimination} of {leakage} in {weakly} {nonlinear}
  {qubits}},}\ }\href {\doibase 10.1103/PhysRevLett.103.110501} {\bibfield
  {journal} {\bibinfo  {journal} {Physical Review Letters}\ }\textbf {\bibinfo
  {volume} {103}},\ \bibinfo {pages} {110501} (\bibinfo {year}
  {2009})}\BibitemShut {NoStop}%
\bibitem [{\citenamefont {Seifert}\ \emph {et~al.}(2022)\citenamefont
  {Seifert}, \citenamefont {Chadwick}, \citenamefont {Litteken}, \citenamefont
  {Chong},\ and\ \citenamefont {Baker}}]{seifert2022timeefficient}%
  \BibitemOpen
  \bibfield  {author} {\bibinfo {author} {\bibfnamefont {Lennart~Maximilian}\
  \bibnamefont {Seifert}}, \bibinfo {author} {\bibfnamefont {Jason}\
  \bibnamefont {Chadwick}}, \bibinfo {author} {\bibfnamefont {Andrew}\
  \bibnamefont {Litteken}}, \bibinfo {author} {\bibfnamefont {Frederic~T.}\
  \bibnamefont {Chong}}, \ and\ \bibinfo {author} {\bibfnamefont {Jonathan~M.}\
  \bibnamefont {Baker}},\ }\bibfield  {title} {\enquote {\bibinfo {title}
  {Time-efficient qudit gates through incremental pulse re-seeding},}\ }in\
  \href {\doibase 10.1109/QCE53715.2022.00051} {\emph {\bibinfo {booktitle}
  {2022 IEEE International Conference on Quantum Computing and Engineering
  (QCE)}}}\ (\bibinfo {year} {2022})\ pp.\ \bibinfo {pages}
  {304--313}\BibitemShut {NoStop}%
\bibitem [{\citenamefont {Simm}\ \emph {et~al.}(2023)\citenamefont {Simm},
  \citenamefont {Machnes},\ and\ \citenamefont {Wilhelm}}]{simm2023two}%
  \BibitemOpen
  \bibfield  {author} {\bibinfo {author} {\bibfnamefont {Alexander}\
  \bibnamefont {Simm}}, \bibinfo {author} {\bibfnamefont {Shai}\ \bibnamefont
  {Machnes}}, \ and\ \bibinfo {author} {\bibfnamefont {Frank~K}\ \bibnamefont
  {Wilhelm}},\ }\bibfield  {title} {\enquote {\bibinfo {title} {Two qubits in
  one transmon--qec without ancilla hardware},}\ }\href
  {https://arxiv.org/abs/2302.14707} {\bibfield  {journal} {\bibinfo  {journal}
  {arXiv:2302.14707}\ } (\bibinfo {year} {2023})}\BibitemShut {NoStop}%
\bibitem [{\citenamefont {Peterer}\ \emph {et~al.}(2015)\citenamefont
  {Peterer}, \citenamefont {Bader}, \citenamefont {Jin}, \citenamefont {Yan},
  \citenamefont {Kamal}, \citenamefont {Gudmundsen}, \citenamefont {Leek},
  \citenamefont {Orlando}, \citenamefont {Oliver},\ and\ \citenamefont
  {Gustavsson}}]{peterer2015coherence}%
  \BibitemOpen
  \bibfield  {author} {\bibinfo {author} {\bibfnamefont {Michael~J.}\
  \bibnamefont {Peterer}}, \bibinfo {author} {\bibfnamefont {Samuel~J.}\
  \bibnamefont {Bader}}, \bibinfo {author} {\bibfnamefont {Xiaoyue}\
  \bibnamefont {Jin}}, \bibinfo {author} {\bibfnamefont {Fei}\ \bibnamefont
  {Yan}}, \bibinfo {author} {\bibfnamefont {Archana}\ \bibnamefont {Kamal}},
  \bibinfo {author} {\bibfnamefont {Theodore~J.}\ \bibnamefont {Gudmundsen}},
  \bibinfo {author} {\bibfnamefont {Peter~J.}\ \bibnamefont {Leek}}, \bibinfo
  {author} {\bibfnamefont {Terry~P.}\ \bibnamefont {Orlando}}, \bibinfo
  {author} {\bibfnamefont {William~D.}\ \bibnamefont {Oliver}}, \ and\ \bibinfo
  {author} {\bibfnamefont {Simon}\ \bibnamefont {Gustavsson}},\ }\bibfield
  {title} {\enquote {\bibinfo {title} {Coherence and {decay} of {higher}
  {energy} {levels} of a {superconducting} {transmon} {qubit}},}\ }\href
  {\doibase 10.1103/PhysRevLett.114.010501} {\bibfield  {journal} {\bibinfo
  {journal} {Physical Review Letters}\ }\textbf {\bibinfo {volume} {114}},\
  \bibinfo {pages} {010501} (\bibinfo {year} {2015})}\BibitemShut {NoStop}%
\bibitem [{\citenamefont {Miao}\ \emph {et~al.}(2022)\citenamefont {Miao},
  \citenamefont {McEwen}, \citenamefont {Atalaya}, \citenamefont {Kafri},
  \citenamefont {Pryadko}, \citenamefont {Bengtsson}, \citenamefont {Opremcak},
  \citenamefont {Satzinger}, \citenamefont {Chen}, \citenamefont {Klimov},
  \citenamefont {Quintana} \emph {et~al.}}]{miao2022overcoming}%
  \BibitemOpen
  \bibfield  {author} {\bibinfo {author} {\bibfnamefont {Kevin~C.}\
  \bibnamefont {Miao}}, \bibinfo {author} {\bibfnamefont {Matt}\ \bibnamefont
  {McEwen}}, \bibinfo {author} {\bibfnamefont {Juan}\ \bibnamefont {Atalaya}},
  \bibinfo {author} {\bibfnamefont {Dvir}\ \bibnamefont {Kafri}}, \bibinfo
  {author} {\bibfnamefont {Leonid~P.}\ \bibnamefont {Pryadko}}, \bibinfo
  {author} {\bibfnamefont {Andreas}\ \bibnamefont {Bengtsson}}, \bibinfo
  {author} {\bibfnamefont {Alex}\ \bibnamefont {Opremcak}}, \bibinfo {author}
  {\bibfnamefont {Kevin~J.}\ \bibnamefont {Satzinger}}, \bibinfo {author}
  {\bibfnamefont {Zijun}\ \bibnamefont {Chen}}, \bibinfo {author}
  {\bibfnamefont {Paul~V.}\ \bibnamefont {Klimov}}, \bibinfo {author}
  {\bibfnamefont {Chris}\ \bibnamefont {Quintana}},  \emph {et~al.},\
  }\bibfield  {title} {\enquote {\bibinfo {title} {Overcoming leakage in
  scalable quantum error correction},}\ }\href
  {http://arxiv.org/abs/2211.04728} {\bibfield  {journal} {\bibinfo  {journal}
  {arXiv:2211.04728}\ } (\bibinfo {year} {2022})}\BibitemShut {NoStop}%
\bibitem [{\citenamefont {Li}\ \emph {et~al.}(2019)\citenamefont {Li},
  \citenamefont {Ding},\ and\ \citenamefont {Xie}}]{li2019tackling}%
  \BibitemOpen
  \bibfield  {author} {\bibinfo {author} {\bibfnamefont {Gushu}\ \bibnamefont
  {Li}}, \bibinfo {author} {\bibfnamefont {Yufei}\ \bibnamefont {Ding}}, \ and\
  \bibinfo {author} {\bibfnamefont {Yuan}\ \bibnamefont {Xie}},\ }\bibfield
  {title} {\enquote {\bibinfo {title} {Tackling the {{Qubit Mapping Problem}}
  for {{NISQ-Era Quantum Devices}}},}\ }\href {http://arxiv.org/abs/1809.02573}
  {\bibfield  {journal} {\bibinfo  {journal} {arXiv:1809.02573}\ } (\bibinfo
  {year} {2019})}\BibitemShut {NoStop}%
\bibitem [{\citenamefont {Earnest}\ \emph {et~al.}(2021)\citenamefont
  {Earnest}, \citenamefont {Tornow},\ and\ \citenamefont
  {Egger}}]{earnest2021pulse}%
  \BibitemOpen
  \bibfield  {author} {\bibinfo {author} {\bibfnamefont {Nathan}\ \bibnamefont
  {Earnest}}, \bibinfo {author} {\bibfnamefont {Caroline}\ \bibnamefont
  {Tornow}}, \ and\ \bibinfo {author} {\bibfnamefont {Daniel~J}\ \bibnamefont
  {Egger}},\ }\bibfield  {title} {\enquote {\bibinfo {title} {Pulse-efficient
  circuit transpilation for quantum applications on cross-resonance-based
  hardware},}\ }\href {https://doi.org/10.1103/PhysRevResearch.3.043088}
  {\bibfield  {journal} {\bibinfo  {journal} {Physical Review Research}\
  }\textbf {\bibinfo {volume} {3}},\ \bibinfo {pages} {043088} (\bibinfo {year}
  {2021})}\BibitemShut {NoStop}%
\bibitem [{\citenamefont {Miller}\ \emph {et~al.}(2022)\citenamefont {Miller},
  \citenamefont {Fischer}, \citenamefont {Sokolov}, \citenamefont
  {Barkoutsos},\ and\ \citenamefont {Tavernelli}}]{miller2022hardwaretailored}%
  \BibitemOpen
  \bibfield  {author} {\bibinfo {author} {\bibfnamefont {Daniel}\ \bibnamefont
  {Miller}}, \bibinfo {author} {\bibfnamefont {Laurin~E.}\ \bibnamefont
  {Fischer}}, \bibinfo {author} {\bibfnamefont {Igor~O.}\ \bibnamefont
  {Sokolov}}, \bibinfo {author} {\bibfnamefont {Panagiotis~Kl.}\ \bibnamefont
  {Barkoutsos}}, \ and\ \bibinfo {author} {\bibfnamefont {Ivano}\ \bibnamefont
  {Tavernelli}},\ }\bibfield  {title} {\enquote {\bibinfo {title}
  {Hardware-{Tailored} {Diagonalization} {Circuits}},}\ }\href
  {http://arxiv.org/abs/2203.03646} {\bibfield  {journal} {\bibinfo  {journal}
  {arXiv:2203.03646}\ } (\bibinfo {year} {2022})}\BibitemShut {NoStop}%
\bibitem [{\citenamefont {Brylinski}\ and\ \citenamefont
  {Brylinski}(2002)}]{brylinski2002universal}%
  \BibitemOpen
  \bibfield  {author} {\bibinfo {author} {\bibfnamefont {Jean-Luc}\
  \bibnamefont {Brylinski}}\ and\ \bibinfo {author} {\bibfnamefont {Ranee}\
  \bibnamefont {Brylinski}},\ }\bibfield  {title} {\enquote {\bibinfo {title}
  {Universal quantum gates},}\ }in\ \href
  {https://doi.org/10.1201/9781420035377} {\emph {\bibinfo {booktitle}
  {Mathematics of quantum computation}}}\ (\bibinfo  {publisher} {Chapman and
  Hall/CRC},\ \bibinfo {year} {2002})\ pp.\ \bibinfo {pages}
  {117--134}\BibitemShut {NoStop}%
\bibitem [{\citenamefont {Brennen}\ \emph
  {et~al.}(2005{\natexlab{b}})\citenamefont {Brennen}, \citenamefont
  {Bullock},\ and\ \citenamefont {O'Leary}}]{brennen2005efficient}%
  \BibitemOpen
  \bibfield  {author} {\bibinfo {author} {\bibfnamefont {Gavin~K.}\
  \bibnamefont {Brennen}}, \bibinfo {author} {\bibfnamefont {Stephen~S.}\
  \bibnamefont {Bullock}}, \ and\ \bibinfo {author} {\bibfnamefont {Dianne~P.}\
  \bibnamefont {O'Leary}},\ }\bibfield  {title} {\enquote {\bibinfo {title}
  {Efficient {circuits} for {exact}-{universal} {computations} with
  {qudits}},}\ }\href {https://dl.acm.org/doi/abs/10.5555/2012086.2012095}
  {\bibfield  {journal} {\bibinfo  {journal} {Quantum Information \&
  Computation}\ }\textbf {\bibinfo {volume} {6}},\ \bibinfo {pages} {436 --
  454} (\bibinfo {year} {2005}{\natexlab{b}})}\BibitemShut {NoStop}%
\bibitem [{\citenamefont {Shende}\ \emph {et~al.}(2006)\citenamefont {Shende},
  \citenamefont {Bullock},\ and\ \citenamefont {Markov}}]{shende2006synthesis}%
  \BibitemOpen
  \bibfield  {author} {\bibinfo {author} {\bibfnamefont {V.V.}\ \bibnamefont
  {Shende}}, \bibinfo {author} {\bibfnamefont {S.S.}\ \bibnamefont {Bullock}},
  \ and\ \bibinfo {author} {\bibfnamefont {I.L.}\ \bibnamefont {Markov}},\
  }\bibfield  {title} {\enquote {\bibinfo {title} {Synthesis of quantum-logic
  circuits},}\ }\href {\doibase 10.1109/TCAD.2005.855930} {\bibfield  {journal}
  {\bibinfo  {journal} {IEEE Transactions on Computer-Aided Design of
  Integrated Circuits and Systems}\ }\textbf {\bibinfo {volume} {25}},\
  \bibinfo {pages} {1000--1010} (\bibinfo {year} {2006})}\BibitemShut {NoStop}%
\bibitem [{\citenamefont {Di}\ and\ \citenamefont
  {Wei}(2013)}]{di2013synthesis}%
  \BibitemOpen
  \bibfield  {author} {\bibinfo {author} {\bibfnamefont {Yao-Min}\ \bibnamefont
  {Di}}\ and\ \bibinfo {author} {\bibfnamefont {Hai-Rui}\ \bibnamefont {Wei}},\
  }\bibfield  {title} {\enquote {\bibinfo {title} {Synthesis of multivalued
  quantum logic circuits by elementary gates},}\ }\href {\doibase
  10.1103/PhysRevA.87.012325} {\bibfield  {journal} {\bibinfo  {journal}
  {Physical Review A}\ }\textbf {\bibinfo {volume} {87}},\ \bibinfo {pages}
  {012325} (\bibinfo {year} {2013})}\BibitemShut {NoStop}%
\bibitem [{\citenamefont {{Qiskit contributors}}(2023)}]{Qiskit}%
  \BibitemOpen
  \bibfield  {author} {\bibinfo {author} {\bibnamefont {{Qiskit
  contributors}}},\ }\href {\doibase 10.5281/zenodo.7757946} {\enquote
  {\bibinfo {title} {Qiskit: An open-source framework for quantum computing},}\
  } (\bibinfo {year} {2023})\BibitemShut {NoStop}%
\bibitem [{\citenamefont {Iten}\ \emph {et~al.}(2016)\citenamefont {Iten},
  \citenamefont {Colbeck}, \citenamefont {Kukuljan}, \citenamefont {Home},\
  and\ \citenamefont {Christandl}}]{iten2016quantum}%
  \BibitemOpen
  \bibfield  {author} {\bibinfo {author} {\bibfnamefont {Raban}\ \bibnamefont
  {Iten}}, \bibinfo {author} {\bibfnamefont {Roger}\ \bibnamefont {Colbeck}},
  \bibinfo {author} {\bibfnamefont {Ivan}\ \bibnamefont {Kukuljan}}, \bibinfo
  {author} {\bibfnamefont {Jonathan}\ \bibnamefont {Home}}, \ and\ \bibinfo
  {author} {\bibfnamefont {Matthias}\ \bibnamefont {Christandl}},\ }\bibfield
  {title} {\enquote {\bibinfo {title} {Quantum circuits for isometries},}\
  }\href {\doibase 10.1103/PhysRevA.93.032318} {\bibfield  {journal} {\bibinfo
  {journal} {Physical Review A}\ }\textbf {\bibinfo {volume} {93}},\ \bibinfo
  {pages} {032318} (\bibinfo {year} {2016})}\BibitemShut {NoStop}%
\bibitem [{\citenamefont {Chen}\ and\ \citenamefont
  {Wang}(2013)}]{chen2013qcompiler}%
  \BibitemOpen
  \bibfield  {author} {\bibinfo {author} {\bibfnamefont {Y.G.}\ \bibnamefont
  {Chen}}\ and\ \bibinfo {author} {\bibfnamefont {J.B.}\ \bibnamefont {Wang}},\
  }\bibfield  {title} {\enquote {\bibinfo {title} {Qcompiler: {{Quantum}}
  compilation with the {{CSD}} method},}\ }\href {\doibase
  10.1016/j.cpc.2012.10.019} {\bibfield  {journal} {\bibinfo  {journal}
  {Computer Physics Communications}\ }\textbf {\bibinfo {volume} {184}},\
  \bibinfo {pages} {853--865} (\bibinfo {year} {2013})}\BibitemShut {NoStop}%
\bibitem [{\citenamefont {Goss}\ \emph {et~al.}(2022)\citenamefont {Goss},
  \citenamefont {Morvan}, \citenamefont {Marinelli}, \citenamefont {Mitchell},
  \citenamefont {Nguyen}, \citenamefont {Naik}, \citenamefont {Chen},
  \citenamefont {J{\"u}nger}, \citenamefont {Kreikebaum}, \citenamefont
  {Santiago}, \citenamefont {Wallman},\ and\ \citenamefont
  {Siddiqi}}]{goss2022highfidelity}%
  \BibitemOpen
  \bibfield  {author} {\bibinfo {author} {\bibfnamefont {Noah}\ \bibnamefont
  {Goss}}, \bibinfo {author} {\bibfnamefont {Alexis}\ \bibnamefont {Morvan}},
  \bibinfo {author} {\bibfnamefont {Brian}\ \bibnamefont {Marinelli}}, \bibinfo
  {author} {\bibfnamefont {Bradley~K.}\ \bibnamefont {Mitchell}}, \bibinfo
  {author} {\bibfnamefont {Long~B.}\ \bibnamefont {Nguyen}}, \bibinfo {author}
  {\bibfnamefont {Ravi~K.}\ \bibnamefont {Naik}}, \bibinfo {author}
  {\bibfnamefont {Larry}\ \bibnamefont {Chen}}, \bibinfo {author}
  {\bibfnamefont {Christian}\ \bibnamefont {J{\"u}nger}}, \bibinfo {author}
  {\bibfnamefont {John~Mark}\ \bibnamefont {Kreikebaum}}, \bibinfo {author}
  {\bibfnamefont {David~I.}\ \bibnamefont {Santiago}}, \bibinfo {author}
  {\bibfnamefont {Joel~J.}\ \bibnamefont {Wallman}}, \ and\ \bibinfo {author}
  {\bibfnamefont {Irfan}\ \bibnamefont {Siddiqi}},\ }\bibfield  {title}
  {\enquote {\bibinfo {title} {High-fidelity qutrit entangling gates for
  superconducting circuits},}\ }\href {\doibase 10.1038/s41467-022-34851-z}
  {\bibfield  {journal} {\bibinfo  {journal} {Nature Communications}\ }\textbf
  {\bibinfo {volume} {13}},\ \bibinfo {pages} {7481} (\bibinfo {year}
  {2022})}\BibitemShut {NoStop}%
\bibitem [{\citenamefont {Blais}\ \emph {et~al.}(2021)\citenamefont {Blais},
  \citenamefont {Grimsmo}, \citenamefont {Girvin},\ and\ \citenamefont
  {Wallraff}}]{blais2021circuit}%
  \BibitemOpen
  \bibfield  {author} {\bibinfo {author} {\bibfnamefont {Alexandre}\
  \bibnamefont {Blais}}, \bibinfo {author} {\bibfnamefont {Arne~L.}\
  \bibnamefont {Grimsmo}}, \bibinfo {author} {\bibfnamefont {Steven~M.}\
  \bibnamefont {Girvin}}, \ and\ \bibinfo {author} {\bibfnamefont {Andreas}\
  \bibnamefont {Wallraff}},\ }\bibfield  {title} {\enquote {\bibinfo {title}
  {Circuit quantum electrodynamics},}\ }\href {\doibase
  10.1103/RevModPhys.93.025005} {\bibfield  {journal} {\bibinfo  {journal}
  {Reviews of Modern Physics}\ }\textbf {\bibinfo {volume} {93}},\ \bibinfo
  {pages} {025005} (\bibinfo {year} {2021})}\BibitemShut {NoStop}%
\bibitem [{\citenamefont {Knill}\ and\ \citenamefont
  {Laflamme}(1997)}]{KnillLaflamme}%
  \BibitemOpen
  \bibfield  {author} {\bibinfo {author} {\bibfnamefont {Emanuel}\ \bibnamefont
  {Knill}}\ and\ \bibinfo {author} {\bibfnamefont {Raymond}\ \bibnamefont
  {Laflamme}},\ }\bibfield  {title} {\enquote {\bibinfo {title} {Theory of
  quantum error-correcting codes},}\ }\href {\doibase 10.1103/PhysRevA.55.900}
  {\bibfield  {journal} {\bibinfo  {journal} {Physical Review A}\ }\textbf
  {\bibinfo {volume} {55}},\ \bibinfo {pages} {900--911} (\bibinfo {year}
  {1997})}\BibitemShut {NoStop}%
\bibitem [{\citenamefont {Temme}\ \emph {et~al.}(2017)\citenamefont {Temme},
  \citenamefont {Bravyi},\ and\ \citenamefont {Gambetta}}]{temme2017error}%
  \BibitemOpen
  \bibfield  {author} {\bibinfo {author} {\bibfnamefont {Kristan}\ \bibnamefont
  {Temme}}, \bibinfo {author} {\bibfnamefont {Sergey}\ \bibnamefont {Bravyi}},
  \ and\ \bibinfo {author} {\bibfnamefont {Jay~M}\ \bibnamefont {Gambetta}},\
  }\bibfield  {title} {\enquote {\bibinfo {title} {Error mitigation for
  short-depth quantum circuits},}\ }\href
  {https://doi.org/10.1103/PhysRevLett.119.180509} {\bibfield  {journal}
  {\bibinfo  {journal} {Physical review letters}\ }\textbf {\bibinfo {volume}
  {119}},\ \bibinfo {pages} {180509} (\bibinfo {year} {2017})}\BibitemShut
  {NoStop}%
\bibitem [{\citenamefont {Gambetta}(2013)}]{gambetta2013quantum}%
  \BibitemOpen
  \bibfield  {author} {\bibinfo {author} {\bibfnamefont {Jay~M.}\ \bibnamefont
  {Gambetta}},\ }\href {https://juser.fz-juelich.de/record/153195} {\emph
  {\bibinfo {title} {Quantum {information} {processing} - {lecture} {notes} of
  the 44th {IFF} {spring} {school} 2013}}},\ edited by\ \bibinfo {editor}
  {\bibfnamefont {David~P.}\ \bibnamefont {DiVincenzo}}\ (\bibinfo  {publisher}
  {Forschungszentrum Jülich, Zentralbibliothek},\ \bibinfo {year}
  {2013})\BibitemShut {NoStop}%
\bibitem [{\citenamefont {Johansson}\ \emph {et~al.}(2012)\citenamefont
  {Johansson}, \citenamefont {Nation},\ and\ \citenamefont
  {Nori}}]{johansson2012qutip}%
  \BibitemOpen
  \bibfield  {author} {\bibinfo {author} {\bibfnamefont {Robert}\ \bibnamefont
  {Johansson}}, \bibinfo {author} {\bibfnamefont {Paul~D.}\ \bibnamefont
  {Nation}}, \ and\ \bibinfo {author} {\bibfnamefont {Franco}\ \bibnamefont
  {Nori}},\ }\bibfield  {title} {\enquote {\bibinfo {title} {{QuTiP}: {an}
  open-source {Python} framework for the dynamics of open quantum systems},}\
  }\href {\doibase 10.1016/j.cpc.2012.02.021} {\bibfield  {journal} {\bibinfo
  {journal} {Computer Physics Communications}\ }\textbf {\bibinfo {volume}
  {183}},\ \bibinfo {pages} {1760--1772} (\bibinfo {year} {2012})}\BibitemShut
  {NoStop}%
\bibitem [{\citenamefont {Nielsen}(2002)}]{nielsen2002simple}%
  \BibitemOpen
  \bibfield  {author} {\bibinfo {author} {\bibfnamefont {Michael~A.}\
  \bibnamefont {Nielsen}},\ }\bibfield  {title} {\enquote {\bibinfo {title} {A
  simple formula for the average gate fidelity of a quantum dynamical
  operation},}\ }\href {\doibase 10.1016/S0375-9601(02)01272-0} {\bibfield
  {journal} {\bibinfo  {journal} {Physics Letters A}\ }\textbf {\bibinfo
  {volume} {303}},\ \bibinfo {pages} {249--252} (\bibinfo {year}
  {2002})}\BibitemShut {NoStop}%
\bibitem [{\citenamefont {Nielsen}\ and\ \citenamefont
  {Chuang}(2010)}]{nielsen_chuang_2010}%
  \BibitemOpen
  \bibfield  {author} {\bibinfo {author} {\bibfnamefont {Michael~A.}\
  \bibnamefont {Nielsen}}\ and\ \bibinfo {author} {\bibfnamefont {Isaac~L.}\
  \bibnamefont {Chuang}},\ }\href {\doibase 10.1017/CBO9780511976667} {\emph
  {\bibinfo {title} {Quantum Computation and Quantum Information: 10th
  Anniversary Edition}}}\ (\bibinfo  {publisher} {Cambridge University Press},\
  \bibinfo {address} {Cambridge},\ \bibinfo {year} {2010})\BibitemShut {NoStop}%
\end{thebibliography}
\end{document}